\newif\ifloguseIEEEConf
\ifloguseIEEEConf
    \documentclass[10pt,conference]{IEEEtran}
\else
    \documentclass{article}
\fi
\usepackage{color}
\usepackage{graphicx}
\usepackage{ algorithmic,algorithm}
\usepackage{lscape}
\usepackage{amssymb}
\usepackage[letterpaper,top=3cm,bottom=3.25cm,left=2.75cm, right=2.75cm]{geometry}
\usepackage{amsfonts,amsmath, amssymb, graphics,geometry,framed,enumitem}
\usepackage{xcolor}
\usepackage{caption}
\usepackage{subcaption}
\usepackage{tikz}

\usepackage[framemethod=TikZ]{mdframed}
\definecolor{shadecolor}{RGB}{180,180,180}
\mdfdefinestyle{MyFrame}{%
    linecolor=black,
    outerlinewidth=1pt,
    roundcorner=20pt,
    innertopmargin= 2pt, 
    innerbottommargin=2pt,
    innerrightmargin=20pt,
    innerleftmargin=20pt,
    needspace=10pt,
    backgroundcolor=gray!10!white}

\begin{document}

\newtheorem{theorem}{Theorem}
\newtheorem{corollary}{Corollary}
\newtheorem{lemma}{Lemma}
\newtheorem{example}{Example}
\newtheorem{definition}{Definition}
\newtheorem{proposition}{Proposition}
\newtheorem{observation}{Observation}
\newtheorem{conjecture}{Conjecture}
\newtheorem{remark}{Remark}
\newcommand{\N}{\mathbb{N}}
\newcommand*{\Line}[3][]{\tikz \draw[#1] #2 -- #3;}%
\newcommand{\qed}{\hfill $\diamondsuit$}
\newcommand{\noam}{Noam Presman\ }
\newcommand{\bhat}{Bhattacharyya }
\newcommand{ \etld } {\tilde{\epsilon}}
\newcommand{ \ALGOSTEP } {$\rhd$ }
\newcommand{ \ALGOIN } {$\left[\blacktriangleright\blacktriangleright\blacktriangleright\right]$ }
\newcommand{ \ALGOOUT } {$\left[\blacktriangleleft\blacktriangleleft\blacktriangleleft\right]$ }
\newcommand{ \ALGOLINE }{
\Line[blue, thick, dotted]{(0,0)}{(5.00in,0in)}}
\newcommand{ \He } {\mathcal{H}}
\newcommand{\rulesep}{\unskip\ \vrule\ }

\newlength\myindent 
\setlength\myindent{6em} 
\newcommand\bindent{%
  \begingroup 
  \setlength{\itemindent}{\myindent} 
  \addtolength{\algorithmicindent}{\myindent} 
}
\newcommand\eindent{\endgroup} 

\ifloguseIEEEConf

\else
  \newcommand{\QED}{\hfill $\diamondsuit$}
  \newcommand{\proof}{\noindent {\bf Proof}\ }
\fi

\definecolor{sourcecolour}{HTML}{F0F0F0}

\let\oldalgorithm\algorithm
\let\endoldalgorithm\endalgorithm
\renewenvironment{algorithm}[1][htbp]{
  \let\graphicsformat\justifiedandcolored
  \oldalgorithm[#1]
}%
  {\endoldalgorithm}

\def\justifiedandcolored#1{%
  \setlength\fboxrule{0pt}%
  \setlength\fboxsep{0pt}%
  \kern-2pt
  \colorbox{red}{%
    \hbox to\linewidth{#1}%
  }%
  \par
  \kern-2pt
}

\let\graphicsformat\centering

\let\graphicsformat\centering

\title{ Recursive Descriptions of Polar Codes }

\author{Noam Presman and Simon Litsyn\thanks{Noam Presman and Simon Litsyn are with the
   the School of Electrical Engineering,
   Tel Aviv University, Ramat Aviv 69978 Israel.
   (e-mails:
 \{presmann, litsyn\}@eng.tau.ac.il.).}}
\date{}

\maketitle

\begin{abstract}
Polar codes are  recursive general concatenated codes. This property  motivates a recursive formalization of the known decoding algorithms: Successive Cancellation, Successive Cancellation with Lists and Belief Propagation. Using such  description allows an easy development of these algorithms for arbitrary polarizing kernels. Hardware architectures for these decoding algorithms are also described in a recursive way, both for Arikan's standard polar codes and for arbitrary polarizing kernels.
\end{abstract}
\section{Introduction}
Polar codes were introduced by Arikan \cite{Arikan} and provided
a scheme for achieving the symmetric capacity of binary
memoryless channels (B-MC) with polynomial encoding and decoding
complexities. Arikan used  the so-called $(u+v,v)$ construction, which is based on the
following linear kernel

$$
{\bf G}_2 = \left[
      \begin{array}{cc}
        1 & 0 \\
        1 & 1 \\
      \end{array}
    \right].
$$
In this scheme,  a $2^n\times2^n$ matrix, ${\bf G}_2^{\bigotimes n}$, is generated by performing the  Kronecker power on ${\bf G}_2$. An input vector $\bf u$ of length   $N=2^n$  is transformed into an $N$ length vector $\bf x$ by multiplying a certain permutation of the vector $\bf u$ by ${\bf G}_2^{\bigotimes n}$. The vector $\bf x$ is  transmitted  over  $N$ independent copies of the memoryless channel, $\mathcal{W}$.  This results in new $N$ (dependent) channels between the individual components of $\bf u$  and the outputs of the channels. Arikan showed that these
channels exhibit the phenomenon of polarization under Successive
Cancellation (SC) decoding. This means that as $n$ grows, there is a
proportion of $I(\mathcal{W})$ (the symmetric channel capacity) of the
channels that become clean channels (i.e. having the capacity
approaching $1$) and the rest of the channels become completely
noisy (i.e. with the capacity approaching $0$). Arikan showed that the  SC decoding algorithm has an algorithmic time and space complexity which is $O(N\cdot \log(N))$  (the same asymptotic complexities apply also for the encoding algorithm). Furthermore, it was shown \cite{Arikan2} that asymptotically in the block length $N$, the block error probability of this scheme decays to zero like $O(2^{-\sqrt{N}})$.

Generalizations of Arikan's code structures were soon to follow. Korada \textit{et al.} considered binary and linear kernels \cite{Korada}. They showed that a binary linear kernel is polarizing if and only if   there does not exist a column permutation of its generating matrix which is upper-triangular, and analyzed its rate of polarization, by introducing the notion of the kernel exponent.  Mori and Tanaka considered the general case of a mapping $g(\cdot)$, which is not necessarily linear and binary,
as a basis for channel polarization constructions \cite{Mori2010}. They gave
sufficient conditions for polarization and generalized the exponent
for these cases. They further showed examples of linear and non-binary  Reed-Solomon codes and Algebraic
Geometry codes  with exponents that are far better
than the exponents of the known binary kernels  \cite{MoriandTanka3}. The authors of this correspondence gave examples of binary but non-linear kernels having the optimal exponent per their kernel dimensions \cite{PrShLi2,Presman2014}.

All of the aforementioned polar code  structures have homogenous kernels, meaning that the  alphabet of their inputs and their outputs are the same. The authors of this correspondence considered the case that some of the inputs of a kernel may have different alphabet than the rest of the inputs \cite{Presman2011}. This results in the so-called mixed-kernels structure, that have demonstrated good performance for finite length codes in many cases. A further generalization of the polar code structure was suggested by Trifonov \cite{Trifonov2011}, in which the outer polar codes were replaced by suitable codes  along with their appropriate decoding algorithms. We note here, that the representation of polar codes as instances of general concatenated codes (GCC) is fundamental to this correspondence, and we elaborate on it in the sequel.

Generalizations and alternatives to SC as the decoding algorithm were also extensively studied.  Tal and Vardy introduced the Successive Cancellation List (SCL) decoder \cite{Tal11,Tal2012}. In this algorithm, the decoder considers up to $L$  concurrent decoding alternatives on each one of its stages, where $L$ is the size of the list. At the final stage of the algorithm,  the most likely result is selected from the list. The asymptotic time and space complexities of this decoder are the same as those of the standard SC algorithm, multiplied by $L$. Furthermore, incorporation of a cyclic redundancy check code (CRC) as an outer-code, results in a scheme with an excellent error-correcting performance, which in many cases is comparable with state of the art schemes (see e.g. \cite[Section V]{Tal2012}).

 Belief-Propagation is an alternative to the SC decoding algorithm. This is a message passing iterative decoding algorithm that operates on the normal factor graph representation of the code. It is known to outperform SC over the Binary Erasure Channel (BEC) \cite{Hussami2009}  and seems to have good performance on other channels as well \cite{Hussami2009,Arikan3}.

Leroux \textit{et al.} considered efficient hardware implementations for the SC decoder of the $(u+v,v)$ polar code \cite{Leroux10,Leroux2012}. They gave an explicit design of a "line decoder" with $N/2$ processing elements and $O(N)$ memory elements. Their work, contains an efficient approximate min-sum decoder, and a discussion on a fixed point implementation. Their design is verified by an ASIC synthesis. Efficient limited parallelism decoders were considered by Leroux \textit{et al.} \cite{Leroux2013} and by Pamuk and Arikan \cite{Pamuk2013}. Hardware implementation of SCL decoder was discussed in Balatsoukas-Stimming \textit{et al.} papers \cite{Balatsoukas-Stimming2013, Balatsoukas-Stimming2013a}.   Pamuk considered a hardware design of BP decoder tailored for an FPGA
implementation \cite{Pamuk2011}.

The  goal of this paper is to emphasize the formalization of  polar codes as recursive GCCs and the implication of this property on the encoding and decoding algorithms. The main contributions of this manuscript are as follows: 1) Formalizing Tal and Vardy's SCL as a recursive algorithm, and thereby generalizing it to arbitrary kernels. 2) Formalizing Leroux \textit{et al.} SC line decoder and generalizing it to arbitrary kernels. 3) Defining a BP decoder with GCC schedule, and suggesting a BP line architecture for it.

The paper is organized as follows. In Section \ref{sec:Prelim} we  describe  polar code kernels as the generating building blocks of polar codes. We then elaborate on the fact that polar codes are examples of recursive GCC structures. This fundamental notion, is the motivation for formalizing the encoding and decoding algorithms in a recursive fashion in Sections \ref{sec:RecEnccOfDecAlgor} and \ref{sec:RecDescOfDecAlgor}, respectively. In particular, we study the standard SC, the SCL (both for  Arikan's kernels and arbitrary ones) and BP (for linear lower triangular kernels) decoding algorithms. These formalizations lay the ground for schematic architectures of the decoding algorithms in Subsection \ref{sec:HrdwreArikConstr}. Specifically, we restate Leroux \textit{et al.} SC pipeline and  SC line decoders, and introduce a line decoder for the GCC schedule of the BP algorithm. Finally, in Subsection \ref{sec:HardArchiForOthKer}, we consider generalizations of these architectures for arbitrary kernels.

\section{Preliminaries}\label{sec:Prelim}

Throughout we use the following notations. For a natural number ${\ell}$, we denote
$[{\ell}]=\left\{1,2,3,...,{\ell}\right\}$ and $[{\ell}]_{-}=\left\{0,1,2,...,{\ell}-1\right\}$. We represent vectors by bold letters.  For $i\geq j$, let ${\bf
u}^i_j=\left[u_j \,\,\, u_{j+1} \ldots \,\,\,\, u_i\right]$ be the sub-vector of ${\bf u}$ of length
$i-j+1$ (if $i<j$ we say that ${\bf
u}^i_j=[\,\,\,]$, the empty vector, and its length is $0$). For two vectors $\bf u$ and $\bf v$ of lengths $n_u$ and $n_v$, we denote the $n_u+n_v$ length vector which is the concatenation of $\bf u$ to $\bf v$ by $[{\bf u}, {\bf v}]$ or $[{\bf u}  \,\,\,{\bf v}]$ or ${\bf u}\bullet{\bf v}$ or just ${\bf u}{\bf v}$. For a scalar $x$, the  $n_u+1$  length vector ${\bf u}\bullet x$, is just the concatenation of the vector ${\bf u}$ with the length one vector containing $x$. Matrices are denoted by boldface capital letters. We denote the set of all the matrices of $n_1$ rows and $n_2$ columns over a field $F$ by $ F^{n_1\times n_2}$. Let  ${\bf A}\in F^{n_1\times n_2}$. We denote row $i$ (column $j$) of the matrix by ${\bf A}_{i \rightarrow }$ (${\bf A}_{ \downarrow j}$). The element at row $i$ and column $j$ is denoted by $A_{i,j}$.  The sub-matrix  containing only rows $i_1\leq i\leq i_2$ and columns $j_1\leq j\leq j_2$ is denoted as ${\bf A}_{i_1:i_2,j_1:j_2}$.

In this paper we consider kernels that are based on bijective
transformations over a field $F$.
A channel polarization kernel of ${\ell}$ dimensions, denoted by $g(\cdot)$, is a mapping
$$
g:F^{{\ell}}\rightarrow F^{{\ell}}.
$$
This means that $g({\bf u})={\bf x}, \,\,\,\, {\bf u}, {\bf x}
\in F^{{\ell}}$.

 We refer to this type of kernel as a \textit{homogeneous kernel}, because its $\ell$ input coordinates and $\ell$ output coordinate are from the same alphabet $F$. Symbols from an alphabet $F$ are called $F$-symbols in this paper. The homogenous kernel $g(\cdot)$ may generate a polar  code of length $\ell^m$ $F$-sybmols by inducing a larger mapping from it, in the following way \cite{Mori2010}.
\begin{definition}[Homogenous Polar Code Generation]\label{def:constructG2}
Given an ${\ell}$ dimensions transformation $g(\cdot)$, we
construct a mapping $g^{(n)}(\cdot)$ of $N={\ell}^n$  dimensions (i.e.
$g^{(n)}(\cdot):F^{{\ell}^n}\rightarrow F^{{\ell}^n}$)
in the following recursive fashion.
$$g^{(1)}({\bf u}_0^{\ell-1})=g({\bf u}_0^{\ell-1})\,\,\,;$$

$$\text{for } n>1, \,\,\,\, g^{(n)}=\Big[ g\left(\gamma_{0,0}, \gamma_{1,0}, \gamma_{2,0}, \ldots, \gamma_{\ell-1,0}\right),$$
$$\,\,\,\,\,\,\,g\left(\gamma_{0,1}, \gamma_{1,1}, \gamma_{2,1}, \ldots, \gamma_{\ell-1,1}\right),\ldots,$$
$$
\,\,\,\,\,\,\,g\left(\gamma_{0, N/{\ell}-1}, \gamma_{1,  N/{\ell}-1}, \gamma_{2,  N/{\ell}-1}, \ldots, \gamma_{\ell-1, N/{\ell}-1}\right)  \Big],
$$
\normalsize
where
$$
\left[\gamma_{i,j}\right]_{j=0}^{j=N/\ell-1}=g^{(n-1)}\left({\bf u}_{ i \cdot (N/\ell)} ^{(i+1)\cdot (N/\ell) -1}\right),
\,\,\,\,\,\,\,\,  i\in\left[\ell\right]_{-}.
$$
\end{definition}

\subsection{Polar Codes as Recursive General Concatenated Codes}\label{subsect:kernelAndCodeDecompose}
 General Concatenated Codes (GCC)\footnote{The construction of the GCCs is a generalization of  Forney's code concatenation method \cite{Forney1966}.} are error correcting codes that are constructed by a technique, which was introduced by Blokh and Zyabolov \cite{Blokh1974} and Zinoviev \cite{Zinoviev1976}. In this construction, we have $\ell$ outer-codes $\left\{\mathcal{C}_i\right\}_{i=0}^{\ell-1}$, where $\mathcal{C}_i$ is an $N_{out}$ length code of size $M_i$ over alphabet $F_i$. We also have an inner-code of length $N_{in}$ and size $\prod_{i=0}^{\ell-1}|F_i|$ over alphabet $F$, with a nested encoding function $\phi : F_0\times F_1 \times ... \times F_{\ell-1} \rightarrow F^{N_{in}}$. The GCC that is generated by these components is a code of length  $N_{out}\cdot N_{in}$ symbols and of size  $\prod_{i=0}^{\ell-1}M_i$. It is created by taking an $\ell\times N_{out}$ matrix, in which the $i^{\text{th}}$ row is a codeword from $\mathcal{C}_i$, and applying the inner mapping $\phi$ on each of the $N_{out}$ columns of the matrix. As Dumer describes in his survey \cite{DumerConcatCodes},   GCCs can give good code parameters  for short length codes when using   appropriate combinations of outer-codes and a nested inner-code. In fact, some of them give the best parameters known. Moreover,  decoding algorithms may utilize their structure by performing local decoding steps  on the outer-codes and utilizing the inner-code layer for exchanging decisions between the outer-codes.

  As Arikan already noted, polar codes are instances of recursive GCCs \cite[Section I.D]{Arikan}. This observation is useful as it allows to formalize the construction of large length polar code as a concatenation of several smaller length polar codes (outer-codes) by using a kernel mapping (an inner-code). Therefore, applying this notion to Definition \ref{def:constructG2}, we observe that a polar code of length $\ell^m$ symbols, may be regarded as a collection of $\ell$ outer polar codes of length $\ell^{m-1}$ (the $i^{th}$ outer-code  is $\left[\gamma_{i,j}\right]_{j=0}^{j=N/\ell-1}=g^{(n-1)}\left({\bf u}_{i\cdot N/{\ell}}^{(i+1)\cdot N/\ell-1}\right)$ for $i\in \left[\ell\right]_{-}$). These codes are then joined together by employing an inner-code (defined by the kernel function $g(\cdot)$) on the outputs of these mappings. There are $N/\ell$ instances of the inner-mapping, such that instance number $j\in \left[N/{\ell}\right]_{-}$ is applied on the   $j^{th}$ symbol from each outer-code.

\begin{figure}
\center
  \includegraphics[scale = 0.13]{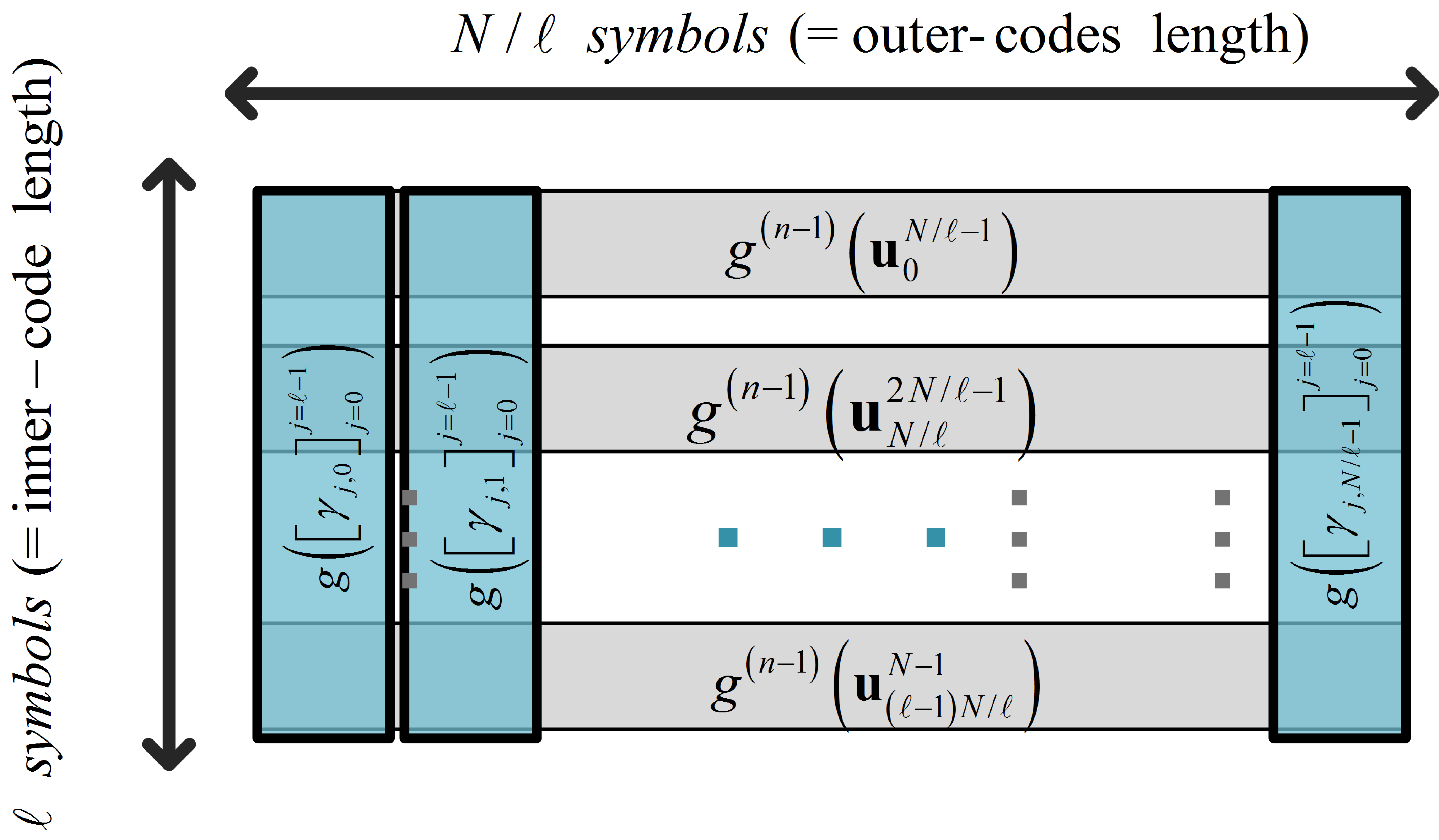}\\
  \caption{A GCC representation  of a polar code of length $\ell^n$ symbols constructed by a homogenous kernel according to Definition \ref{def:constructG2}   }\label{fig: def2GCC}
\end{figure}

  The above GCC formalization is illustrated in Figure \ref{fig: def2GCC}. In this figure, we see the $\ell$ outer-code codewords of length $\ell^{m-1}$ depicted as  gray horizontal rectangles (similar to rows of a matrix). The instances of the inner-codeword mapping are depicted as  vertical rectangles that are located on top of the gray outer-codes rows (resembling columns of a matrix). This is appropriate,  as this mapping operates on columns  of the matrix which rows are the outer-code codewords. Note that for brevity we only drew three instances of the inner mapping, but there should be $\ell^{m-1}$ instances of it, one for each column of this matrix. In the homogenous case, the outer-codes themselves are constructed in the same manner. Note, however, that even though these outer-codes have the same structure, they form different codes in the general case. The reason is that they may have different sets of frozen symbols.

\begin{figure}
\center
  \includegraphics[scale = 0.23]{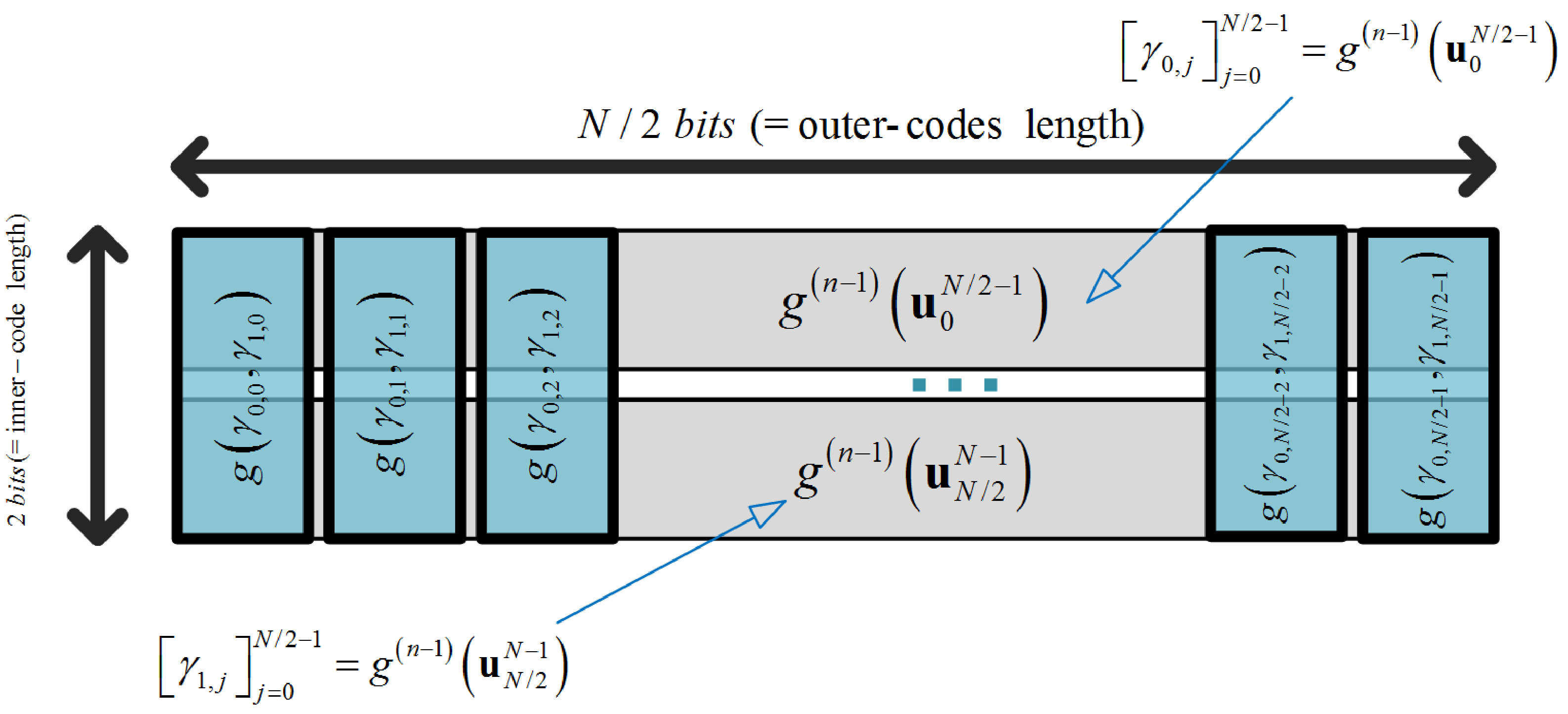}\\
  \caption{Example \ref{ex:UVV}'s GCC representation  (Arikan's construction)  }\label{fig:gccUVV}
\end{figure}

\begin{example}[Arikan's Construction]\label{ex:UVV}
Let $g(u_0,u_1) = [u_0\,\,\, u_1]\cdot {\bf G}_2$. Let ${\bf u}$ be an $N=2^n$ length binary vector.  The vector $\bf u$ is transformed into an $N$ length vector $\bf x$ by using a bijective mapping $g^{(n)}(\cdot):\{0,1\}^{N}\rightarrow \{0,1\}^{N}$. The transformation is defined recursively as
$$
\text{for } n=1\,\,\,\,\, g^{(1)}({\bf u})=g({\bf u})=\left[u_0+u_1,u_1\right],
$$
\begin{equation}\label{eq:Constr}
\text{for }  n>1\,\,\,\,\, g^{(n)}({\bf u})={\bf x}_0^{N-1}\,\,\,\,,
\end{equation}
where $\left[x_{2j}, \,\,\, x_{2j+1}\right] = \left[\gamma_{0,j}+\gamma_{1,j},\,\,\,\,\,\gamma_{1,j}\right]$ for $j\in [N/2]_{-}$, and $\left[\gamma_{0,j} \right]_{j=0}^{N/2-1}=g^{(n-1)}\left({\bf u}_0^{N/2-1}\right)$, $\left[\gamma_{1,j} \right]_{j=0}^{N/2-1}=g^{(n-1)}\left({\bf u}_{N/2}^{N-1}\right)$ are the two outer-codes (each one of length $N/2$ bits). Figure \ref{fig:gccUVV} depicts the GCC block diagram for this example.
\end{example}

The GCC structure of polar codes can be also  represented by a layered\footnote{In a layered graph, the vertices set can be partitioned into a sequence of sub-sets called layers and denoted by $L_0,L_1,\cdots,L_{k-1}$. The edges of the graph connect only vertices within the layer or in successive layers.} Forney's normal factor graph \cite{Forney2001}.  Layer $\# 0$ of this graph contains the inner mappings (represented as sets of vertices), and therefore we refer to it as the \textit{inner-layer}. Layer $\#1$ contains the vertices of the inner layers of all the outer-codes that are concatenated by layer $\#0$. We may continue and generate layer $\#i$ by considering the outer-codes that are concatenated by layer $\#(i-1)$ and include in this layer all the vertices describing their inner mappings. This recursive construction process may continue until we reach to outer-codes that cannot be decomposed into non-trivial inner-codes and outer-codes.  Edges (representing variables) connect between outputs of the outer-codes to the input of the inner mappings. This representation can be viewed as observing the GCC structure in Figure \ref{fig: def2GCC} from its side.

\begin{figure}
\center
  \includegraphics[scale = 0.2]{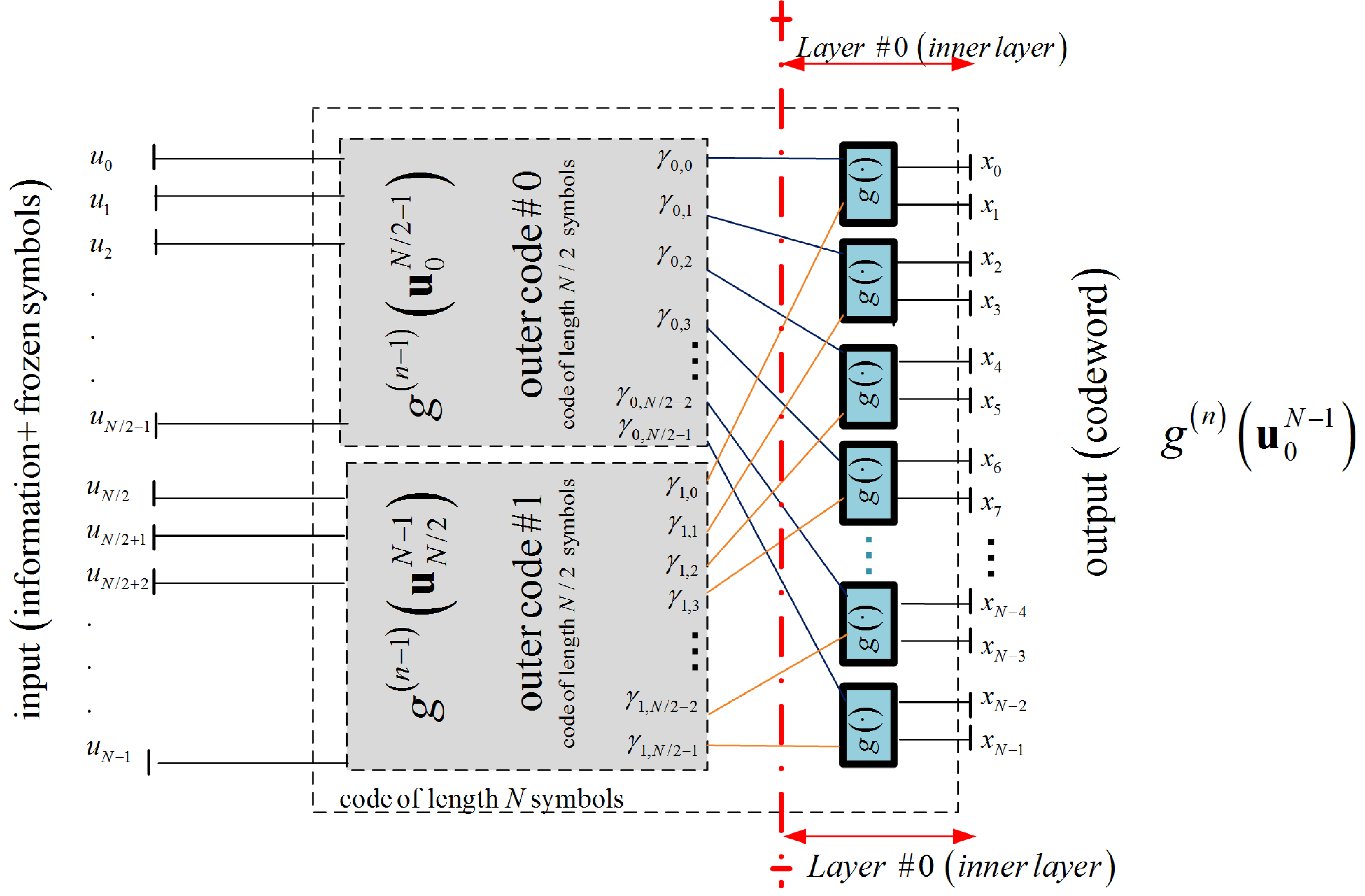}\\
  \caption{ Representation of a polar code with kernel of $\ell = 2$  dimensions   as a layered factor graph    }\label{fig:GCCWithLayer0}
\end{figure}

\begin{figure}
\center
  \includegraphics[scale = 0.2]{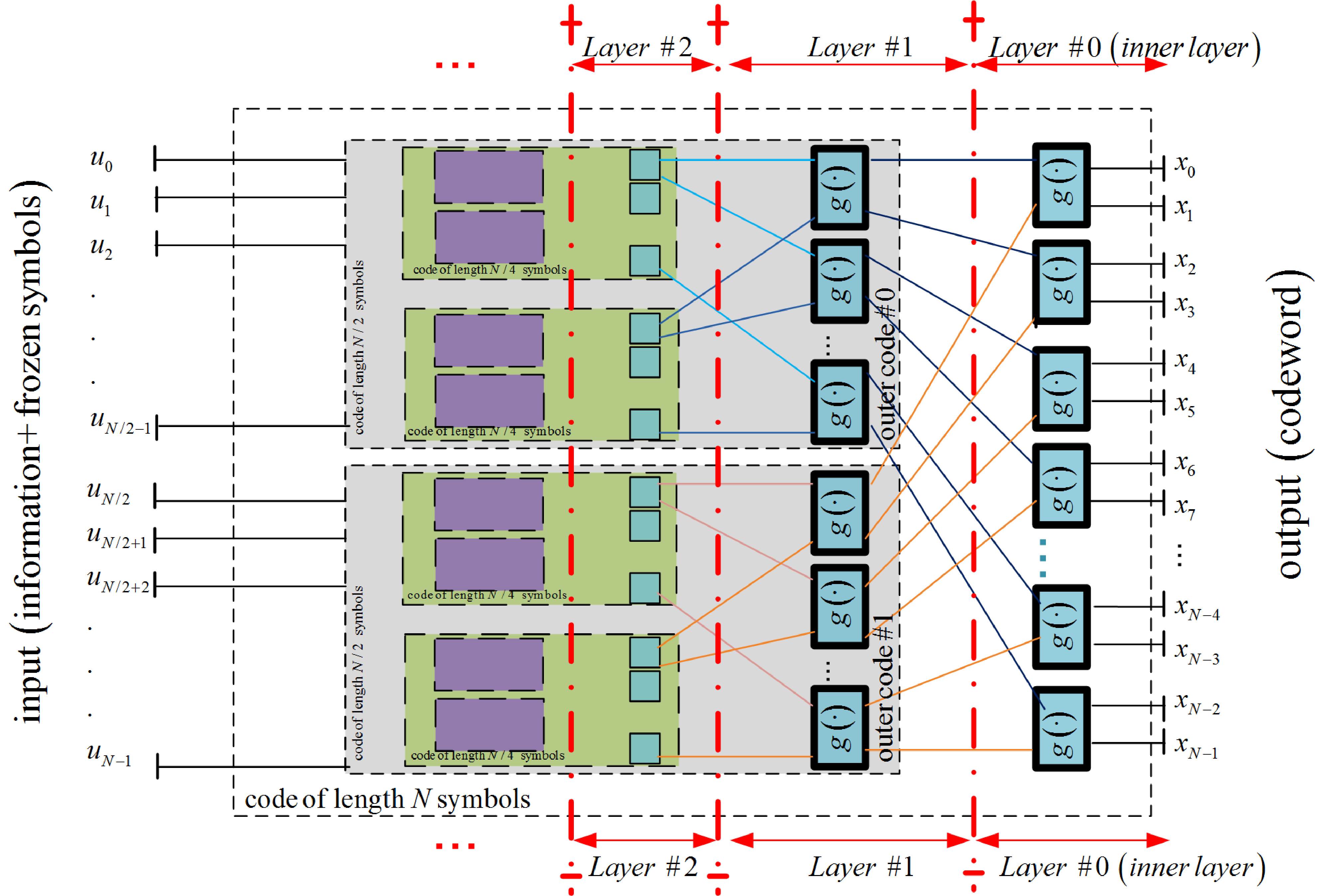}\\
  \caption{ Representation of a polar code with kernel of $\ell = 2$  dimensions as a layered factor graph (detailed version of Figure \ref{fig:GCCWithLayer0} - recursion unfolded)    }\label{fig:GCCWithLayer}
\end{figure}

\begin{figure}
\center
  \includegraphics[scale = 0.15]{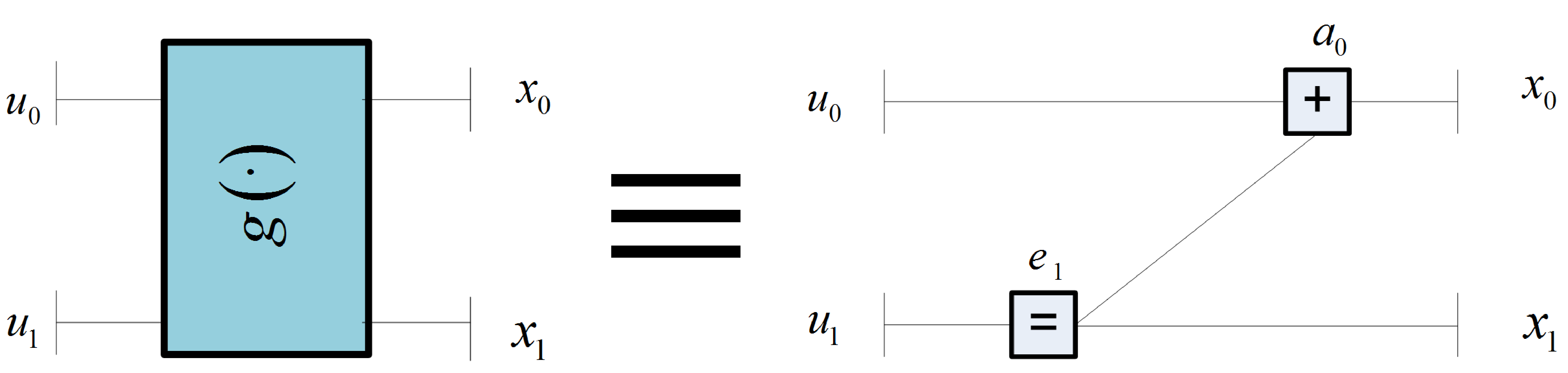}\\
  \caption{ Normal factor graph representation of the $g(\cdot)$ block from Figures \ref{fig:GCCWithLayer0} and \ref{fig:GCCWithLayer} for Arikan's $(u+v,v)$ construction   }\label{fig:GCCWithLayerUVVPart}
\end{figure}

\begin{example}[Layered Normal Factor Graph for Arikan's Construction]\label{ex:ArikanLayeredFactorGraph}
Figures and \ref{fig:GCCWithLayer0} and \ref{fig:GCCWithLayer} depict a layered factor graph representation for length $N=2^n$ symbols polar code with kernel of $\ell = 2$ dimensions. Figure \ref{fig:GCCWithLayer0} gives only a block structure of the graph, in which we have the two outer-codes of length $N/2$ that are connected by the inner layer (note the similarities to the GCC block diagram in Figure \ref{fig:gccUVV}). Half edges represent the inputs ${\bf u}_{0}^{N-1}$ and the outputs  ${\bf x}_{0}^{N-1}$ of the transformation. The edges (denoted by $\gamma_{i,j},\,\,\,\,  j\in\left[N/2\right]_{-}, i\in[2]_{-}$) connect the outputs of the two outer-codes to the inputs of the inner mapping blocks,  $g(\cdot)$. A more elaborated version of this figure is given in Figure \ref{fig:GCCWithLayer}, in which we unfolded the recursive construction.

Strictly speaking, the green blocks that represent the $g(\cdot)$ inner-mapping are themselves factor graphs (i.e.  collections of vertices and edges). An example of a normal factor graph specifying such a block is given in Figure \ref{fig:GCCWithLayerUVVPart} for  Arikan's $(u+v,v)$ construction (see Example \ref{ex:UVV}). Vertex $a_0$ represents a parity constraint and vertex $e_1$ represents an equivalence constraint.  The half edges $u_0,u_1$ represent the inputs of the mapping, and the half edges $x_0,x_1$ represent its outputs. This graphical structures is probably the most popular visual representation of polar codes (see e.g. \cite[Figure 12]{Arikan} and \cite[Figure 5.2]{KoradaThesis} ) and is also known as the "butterflies" graph because of the edges arrangement  in Figure \ref{fig:GCCWithLayer}.
\end{example}


\subsection{Mixed-Kernels Polar Codes}\label{sec:mixedKernels}
Thus far, we described homogenous kernels constructions in which a single kernel and code alphabet is used for generating the polar codes structures. It may be advantageous in terms of error-correction performance and complexity to combine two types of kernels (each one over different alphabet) into one structure. Such constructions are called mixed-kernels structures  \cite{Presman2011,Presman2011b}. In order to have a more comprehensive introduction to the notion of mixed-kernels we give an example of the structure (taken from \cite{Presman2011,Presman2011b}).

\begin{example}[Mixed-Kernels Construction]\label{ex:MxdKernelsExample}
Let $g(\cdot)$ be a four dimensions binary mapping defined as $g({\bf u}) = {\bf u} \cdot {\bf G}_2^{\otimes 2}$.
Using $g(\cdot)$ we define an additional kernel
\begin{equation}
g_0(u_0,u_{(1,2)},u_3) \triangleq  g(u_0,u_1,u_2,u_3), \text{where $u_{(1,2)} \triangleq [u_1,u_2]\in \{0,1\}^2$}.
\end{equation}
In other words we take the $u_1$ and $u_2$ binary inputs to $g(\cdot)$ and combine them into a single quaternary entity  $u_{(1,2)}$. We informally say that $u_1$ and $u_2$ were glued together generating  $u_{(1,2)}$.

Let $g_1(\cdot):\left(\{0,1\}^2 \right)^4\rightarrow \left(\{0,1\}^2 \right)^4$ be a polarizing kernel over the quaternary alphabet. For example, $g_1(\cdot)$ can be a kernel, based on the extended Reed-Solomon code of length four, $G_{RS}(4)$ that was proven by Mori and Tanaka \cite[Example 20]{MoriandTanka} to be a polarizing kernel. The homogenous polar code generated by $g_1(\cdot)$ is dubbed the RS4 polar code. Using $g_1(\cdot)$, we can extend the mapping  of $g_0(\cdot)$ to a length $N=4^n$  bits code. Both $g_0(\cdot)$ and $g_1(\cdot)$ are referred to as the \textit{constituent} kernels of the construction. Note that $g_1(\cdot)$ is introduced in order to handle the glued bits $u_{(1,2)}$ of the input of $g_0(\cdot)$ and therefore is also referred to as the \textit{auxiliary kernel} of the construction. The standard Arikan's  construction (based on the Kronecker power) does not suffice here,
because of the glued bits $u_{(1,2)}$, that need to be jointly treated as a quaternary symbol.

The mixed-kernels construction can be readily explained in terms of GCC structure. Let $g^{(1)}(\cdot)=g_0(\cdot)$.
In order to extend this construction to a mapping $g^{(n)}\left({\bf u}_0^{4^n-1}\right)$, $n>1$ for which some of the inputs are glued, we suggest the following recursive GCC  construction. We define three outer-code:
\begin{description}
  \item[outer-code  $\#0$:] $\left[\gamma_{0,j}\right]_{j=0}^{j=N/4-1}=g^{(n-1)}\left({\bf u}_0^{N/4-1}\right),\,\,\,\, u_j,\gamma_{0,j}\in \{0,1\} ,\,\,\,\,j\in [N/4]_{-}$;
        \item[outer-code  $\#1$:] $\left[\gamma_{1,j}\right]_{j=0}^{j=N/4-1}=g^{(n-1)}_2\left(\left[{ u}_{\left(N/4+2j,N/4+2j+1\right)} \right]_{j=0}^{j=N/4-1}\right),\,\,\,\, u_{\left(N/4+2j,N/4+2j+1\right)}, \gamma_{1,j} \in \{0,1\}^2 ,\,\,\,\,j\in [N/4]_{-}$;
  \item[outer-code  $\#2$:] $\left[\gamma_{2,j}\right]_{j=0}^{j=N/4-1}=g^{(n-1)}\left({\bf u}_{3N/4}^{N-1}\right),\,\,\,\,u_{j+3N/4},\gamma_{2,j}\in \{0,1\}  ,\,\,\,\,j\in [N/4]_{-}$,
\end{description}

where $u_{(i,j)}$ means that the items of sub-vector ${\bf u}_{i}^{j}$ were glued together, generating an element from the larger alphabet $\{0,1\}^{j-i+1}$.
Note that outer-codes $\#0$ and $\#2$ are just mixed-kernels constructions of length $N/4$ bits. The output of these outer-codes are binary vectors, but the input is a mixture of binary and quaternary symbols (generated by bits that were glued together). Outer-code $\#1$ is a  homogenous polar code construction of length $N/4$ quaternary symbols, that  all of its input symbols and output symbols are bits that were glued together in pairs. Finally, these three outer-codes are combined together using the $g_0(\cdot)$ inner mapping.
$$g^{(n)}=\Big[ g_0\left(\gamma_{0,0}, \gamma_{1,0}, \gamma_{2,0}\right),\,\,g_0\left(\gamma_{0,1}, \gamma_{1,1}, \gamma_{2,1}\right),\ldots,$$
$$
\,\,\,\,\,\,\,g_0\left(\gamma_{0,N/4-1}, \gamma_{1, N/4-1}, \gamma_{2, N/4-1}\right)  \Big].
$$
Figure \ref{fig: illustrationOfExampleMixed} depicts this GCC construction. Note that outer-code $\#1$ was drawn as a rectangle having the same width of  outer-code $\#0$ (or $\#2$). This property symbolizes that all the outer-codes have the same length in terms of \textit{symbols}. On the other hand, the height of the rectangle of outer-code $\# 1$ is twice the height of each of the rectangles of the  other two outer-codes. This property indicates that the symbols alphabet size of  outer-code $\#1$  is twice  the size of the symbols alphabet  of the   other  outer-codes (for which the symbols are bits). This is because outer-code $\#1$ is a quaternary mapping in which both the input symbols and the output symbols are pairs of glued bits.

\begin{figure}
\center
  \includegraphics[scale = 0.07]{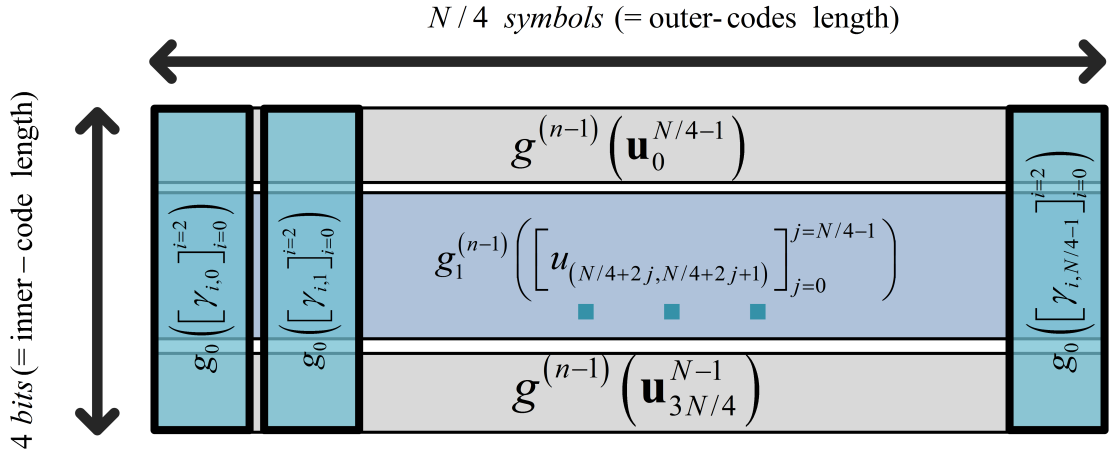}\\
  \caption{A GCC representation of the length $N=4^n$ bits mixed-kernels polar code $g^{(n)}(\cdot)$  described in Example \ref{ex:MxdKernelsExample}   }\label{fig: illustrationOfExampleMixed}
\end{figure}

\end{example}

The recursive GCC structure of polar codes enables recursive formalizations of the algorithms associated with them. These algorithms benefit from   simple and clear descriptions, which support elegant analysis.  Furthermore, in some cases it allows reuse of resources and indicates which operations may be done in parallel. The essence of the recursive encoding algorithm has already been described in Definition \ref{def:constructG2}. In Section  \ref{sec:RecEnccOfDecAlgor} we formalize these ideas and in addition describe an algorithm for systematic encoding of polar codes.   Afterwards, we consider the decoding algorithms of polar codes, giving them a recursive formulation in Section \ref{sec:RecDescOfDecAlgor}.

\section{Recursive Descriptions of Polar Codes Encoding Algorithms}\label{sec:RecEnccOfDecAlgor}
In this section we discuss encoding algorithms for polar codes. We begin in Subsection \ref{sec:nonSysEnc} by describing a non-systematic encoding algorithm that is a direct consequence of the GCC structure discussed in Subsection \ref{subsect:kernelAndCodeDecompose}. Subsection \ref{sec:sysEnc} considers systematic encoding algorithm of linear polar codes with lower triangular kernel generating matrix.
\subsection{Non-Systematic Encoding}\label{sec:nonSysEnc}
In this subsection we consider a non-systematic  recursive encoding algorithm that is based on the recursive GCC structures of polar codes. Let us begin by describing the algorithm for Arikan's $(u+v,v)$ polar code.
Let $\bf u$ be an $N$ length binary vector, serving as  the encoder input. Each polar code of length $N$ is defined by its $N$ length frozen-indicator vector $\bf z$, such that $z_i =1$ if and only if the $i^{th}$ input of the encoder is frozen (i.e. fixed and known to both of the encoder and the decoder) and $z_i=0$ otherwise. For a $(u+v,v)$ polar code of  dimension $k$, we have that $w_H({\bf z}) = N-k$, where $w_H(\cdot)$ is the Hamming weight of the vector. Given an information vector $\breve {\bf u}\in \{0,1\}^k$, it is the role of the encoder to output a binary codeword ${\bf x}\in \{0,1\}^N$ representing its corresponding codeword. Given an information vector $\breve{\bf u}$ and    $\bf z$, it is easy to generate ${\bf u} \in \{0,1\}^n$, the encoder input, such that values of $\tilde{\bf u}$ are sequentially assigned to the non-frozen components of $\bf u$ and elements corresponding to frozen indices are set to a predetermined value (here, we arbitrarily decided  to set the frozen values of $\bf u$ to zero), i.e.
\begin{equation}
\text{if } z_i=1 \text{ then }  u_i = 0,\,\,\,\,\,\,\, \forall i\in [N]_{-} ;
\end{equation}
$$
u_{\theta_i}  =  \breve{u}_{i},\,\,\,\,\, \forall i\in [k]_{-},
$$
where  $\boldsymbol \theta$ is a $k$ length vector, such that $\theta_i$ is the  $i^{\text{th}}$ index of $\bf z$ corresponding to zero value (i.e. indicating a non-frozen input symbol). The signature of the non-systematic encoding algorithm for $N=2^n$ code is
\begin{equation}\label{eq:nonSysEnc}
{\bf x} = \text{NonSysEncoder}\left({\bf u}\right).
\end{equation}

Algorithm \ref{algo:nonSysEncUVV} describes a recursive implementation of the encoder. Note that for a scalar input $u$ to (\ref{eq:nonSysEnc}) (i.e. $n=0$) we have the output $x$ equal to $u$.

\begin{algorithm}
\caption{Non-Systematic Encoder for $(u+v,v)$ Polar Code, of Length $N=2^n$ Bits, $n\geq 1$ }          
\label{algo:nonSysEncUVV}
  \begin{description}
\item[\ALGOIN \underline{Input:} ] $\bf u$.
\item[//Initialization: ] \hfill \\
\ALGOSTEP Allocate two binary vectors ${\bf x}^{(0)}$ and ${\bf x}^{(1)}$ each one of length $N/2$.
 \item[//Encode the Outer-Codes:] \hfill \\
    \ALGOSTEP Encode the two outer-codes of length $N/2$ using the information sub-vectors ${\bf u}_{0}^{N/2}$ and  ${\bf u}_{N/2}^{N-1}$:
    \begin{equation}\label{eq:OuterCodeUVV}
        {\bf x}^{(i)} = \text{NonSysEncoder}\left({\bf u}_{i\cdot N/2}^{{(i+1)\cdot N/2}-1}\right),\,\,\,\,\,\,\, \forall i\in\{0,1\}.
    \end{equation}
 \item[//Encode the Inner-Code:] \hfill \\
\ALGOSTEP Apply the inner-code $(u+v,v)$ on the pairs $\left[{ x}^{(0)}_j,\,\,\,\,\,{ x}^{(1)}_j\right]$:
    \begin{equation}\label{eq:InnerCodeUVV}
        {\bf x}_{2j}^{2j+1} = \left[ x^{(0)}_{j}+x^{(1)}_{j}, \,\,\,\,\,\, x^{(1)}_{j}\right], \,\,\,\,\,\,\forall j\in\left[N/2\right]_{-}.
    \end{equation}
\item[\ALGOOUT \underline{Output:} ] ${\bf x}$.
  \end{description}
\end{algorithm}

Let us now consider a general kernel $g(\cdot)$ of $\ell$ dimensions over a field $F$, i.e. $g:F^{\ell}\rightarrow F^{\ell}$. The signature of the encoder remains the same, only that both $\bf u$ and $\bf x$ are in $F^{N}$.
Algorithm \ref{algo:nonSysGenHomKernel} describes the encoding procedure for this case. Similarly  to the $(u+v,v)$ case, the function has its output equal to its its input for scalar inputs.

\begin{algorithm}
\caption{Non-Systematic Encoder for Homogenous Polar Code of Length $N=\ell^n$ $F$-Symbols, Based on Kernel $g(\cdot)$,  $n\geq 1$ }          
\label{algo:nonSysGenHomKernel}
  \begin{description}
\item[\ALGOIN \underline{Input:} ] $\bf u$.
\item[//Initialization: ] \hfill \\
\ALGOSTEP Allocate $\ell$ vectors $\left\{{\bf x}^{(i)}\right\}_{i=0}^{\ell}$,  each one of length $N/\ell$ $F$-symbols.
 \item[//Encode the Outer-Codes:] \hfill \\
    \ALGOSTEP Encode the $\ell$ outer-codes of length $N/\ell$ using the information sub-vectors $\left\{{\bf u}_{i\cdot N/\ell}^{(i+1)\cdot N/\ell-1}\right\}_{i\in[\ell]_{-}}$:
    \begin{equation}\label{eq:OuterCodeUVV}
        {\bf x}^{(i)} = \text{NonSysEncoder}\left({\bf u}_{i\cdot N/\ell}^{(i+1)\cdot N/\ell-1}\right),\,\,\,\,\,\,\, \forall i\in \left[\ell\right]_{-}.
    \end{equation}
 \item[//Encode the Inner-Code:] \hfill \\
\ALGOSTEP Apply the inner-code $g(\cdot)$ on the sub-vectors $\left[{ x}^{(i)}_j\right]_{i\in[\ell]_{-}}$, $\forall j\in\left[N/\ell\right]_{-}$:
    \begin{equation}\label{eq:InnerCodeUVV}
        {\bf x}_{j\cdot \ell}^{(j+1)\cdot \ell-1} = g\left({ x}^{(0)}_j,{ x}^{(1)}_j,\ldots,{ x}^{(\ell-1)}_j\right), \,\,\,\,\,\,\forall j\in\left[N/\ell\right]_{-}.
    \end{equation}
\item[\ALGOOUT \underline{Output:} ] ${\bf x}$.
  \end{description}
\end{algorithm}

Encoding of mixed-kernels is performed in a similar fashion. The difference is that in the outer-code encoding phase we need to provide  information sub-vectors of different lengths. Let us consider the mixed-kernels instance given in Example \ref{ex:MxdKernelsExample}. We have three computations of outer-codes
    \begin{equation}\label{eq:OuterCodeUVV}
        {\bf x}^{(0)} = \text{NonSysEncoder}\left({\bf u}_{0}^{ N/4-1}\right);\,\,\,{\bf x}^{(1)} = \text{NonSysEncoder}^{(RS4)}\left({\bf u}_{N/4}^{ 3 \cdot N/4-1}\right); \,\,\,{\bf x}^{(2)} = \text{NonSysEncoder}\left({\bf u}_{3N/4}^{ N-1}\right),
    \end{equation}
where ${\bf x}^{(0)},{\bf x}^{(2)}\in \{0,1\}^{N/4} $ and ${\bf x}^{(1)} \in \{0,1\}^{N/2}$. The function $\text{NonSysEncoder}^{(RS4)}$ is the encoding procedure of the homogenous $RS4$ code which input and output are  $GF(4)$ vectors. The elements of $GF(4)$  are represented by their binary vector (cartesian) form.

\subsection{Systematic Encoding} \label{sec:sysEnc}
In this subsection we consider systematic encoding of polar codes. A systematic encoder has the property that the non-frozen symbols of the encoder input vector, $\bf u$, appear explicitly in their corresponding codeword, $\bf x$. Formally speaking, for a length $N$ code, we define a bijective mapping function ${\mathrm{m}_N(\cdot): [N]_{-}\rightarrow [N]_{-}}$, such that a systematic encoder corresponding to $\mathrm{m}_N(\cdot)$ outputs $\bf x$, satisfying $u_t = x_{m_{N}(t)}$ for all non-frozen indices $t\in [N]_{-}$ (i.e. $z_t = 0$). A systematic encoder is advantageous because it facilitates retrieval of  the user information without performing a decoding first (assuming no errors occurred in the received codeword). Furthermore,  Arikan demonstrated by simulations   systematic coding systems  having better  BER performance  compared to  non-systematic  coding systems using the same  $(u+v,v)$  polar code \cite{Arikan2011}.

In this paper we consider systematic encoders for linear kernels having a lower triangular generating matrix ${\bf G}\in F^{\ell\times\ell}$. The signature of the systematic encoder is defined as follows:
\begin{equation}\label{eq:sysEnc}
\left[{\bf x},\,\,\,\, \tilde{{\bf u}} \right]= \text{SysEncoder}\left({\bf u},\,\,\,{\bf z}\right),
\end{equation}
where the vectors ${\bf x},{\bf u}$ and ${\bf z}$ were defined before in Subsection \ref{sec:nonSysEnc} and $\bf x$ is a systematic encoding of $\bf u$.  The vector $\tilde{{\bf u}}\in F^{N}$ is the input for the non-systematic encoder that results in $\bf x$, i.e. ${\bf x} = \text{NonSysEncoder}\left({\tilde{\bf u}}\right)$ and   $\tilde{u}_t = 0$ if $z_t = 1$ for all $t\in[N]_{-}$. While not being a necessary output of the algorithm, $\tilde {\bf u}$ is used here to enable a more comprehensible description of the systematic encoder. Indeed, the   systematic encoder may be understood as an algorithm for finding the vector ${\tilde{\bf u}}$ meeting these requirements.

Let us first consider the  $N=\ell$ $F$-symbols case. In this case we have to find $\tilde{\bf u}$ such that $\tilde{\bf u}  \cdot {\bf G} = {\bf x}$, and
\begin{equation}\label{eq:sysConditionForBaseCase}
\forall t\in [\ell]_{-},\,\, \,\, \,\, \,\,  u_t=x_{\mathrm{m}_{\ell}(t)} \text{ if } z_t=0 \text{ and otherwise } \tilde{u}_t =0.\,\,
\end{equation}
For this base case we take $\mathrm{m}_{\ell}(\cdot)$ to be the identity function, i.e. $\mathrm{m}_{\ell}(t) = t,\,\,\,\,\, \forall t\in [\ell]_{-}$. Algorithm \ref{algo:SysGenHomKernelBaseCase} describes the systematic encoding procedure for this case. It can be easily shown by induction on the \textit{for-loop} variable $j$ (beginning with $\ell-1$ and ending with $0$) that on each step  condition  (\ref{eq:sysConditionForBaseCase})  is met.
\begin{algorithm}

\caption{Systematic Encoder for Homogenous Length $N=\ell$ $F$-Symbols  Polar Code, Based on Lower Triangular Kernel ${\bf G}\in F^{(\ell\times\ell)}$}          
\label{algo:SysGenHomKernelBaseCase}
  \begin{description}
\item[\ALGOIN \underline{Input:} ] $\bf u$; $\bf z$.
\item[//Initializations: ] \hfill \\
\ALGOSTEP Allocate two vectors $\tilde{{\bf u}},{\bf x}\in F^{\ell}$. Initialize ${\bf x} = {\bf 0}$.
 \item[//Successively encode $\bf u$:] \hfill \\
  \ALGOSTEP \textbf{For} $j = \ell-1$ to $0$ \textbf{Do}
        \begin{itemize}
        \item \textbf{If} $z_j==0$ \textbf{Then} set $\tilde{u}_{j} = G_{j,j}^{-1}\cdot \left(u_j - x_j\right)$ ; \textbf{Else} set $\tilde{u}_{j} = 0$;
        \item Set ${\bf x} =  {\bf x}  + \tilde{u}_{j}\cdot {\bf G}_{j \rightarrow}$;
        \end{itemize}
\item[\ALGOOUT \underline{Output:} ] $\bf x$; $\tilde{\bf u}$.
  \end{description}
\end{algorithm}

For the general $N=\ell^n$ case where $n>1$ we  utilize the GCC structure of the polar code in order to perform   systematic encoding. Let us first describe the indices mapping function $\mathrm{m}_{N}(\cdot)$. As was already noted in the GCC discussion, and was exemplified in the non-systematic encoder (Algorithm \ref{algo:nonSysGenHomKernel}), the input sub-vector ${\bf u}_{i\cdot N/{\ell}}^{(i+1)N/{\ell}-1}$ is also the input of outer-code $\mathcal{C}_i$ for all $i\in [\ell]_{-}$. The following requirement of the mapping function will prove useful in the recursive implementation.
\begin{equation}\label{eq:MappingReq}
\mathrm{m}_{N}(t) \equiv \left\lfloor \frac{t}{N/{\ell}}\right\rfloor \,\,\,\,\,\, (mod\,\,\,\,\, \ell)\,\,\,\,\,\, \forall t\in [N]_{-},\,\,\,\, \forall N=\ell^n.
\end{equation}
The implication of (\ref{eq:MappingReq}) is that non-frozen symbols placed at index $t$, such that $b\cdot N/{\ell} \leq t < (b+1)\cdot N/{\ell}$, where $b\in [\ell]_{-}$ should appear at the output $x_{\tau}$ where $\tau = a\cdot \ell +b$ and $a$ is some number in $\left[N/{\ell}\right]_{-}$. Note that index $t$ of the input corresponds to the inputs of outer-code $\mathcal{C}_b$. Furthermore, if ${\bf x}^{(i)}$ is the outer-code codeword of $\mathcal{C}_i$ we have $x_{\tau} = \sum_{i=b}^{\ell-1}G_{i,b}\cdot x^{(i)}_{a}$ (see Figure \ref{fig: def2GCC}). This connection is useful, because if we already systematically encoded all the inputs $t'$ such that  $(b+1)\cdot N/{\ell} \leq t'$ (corresponding to outer-codes $\mathcal{C}_{b'}$ where $b'\geq b+1$), by appropriately  calling the systematic encoder of $\mathcal{C}_b$ we can ensure that indeed $x_{\tau} = u_t$.

It can be  proven by induction that a mapping function implementing the following recursion formula indeed satisfies (\ref{eq:MappingReq}):
\begin{equation}\label{eq:RecursionFormMapping}
\text{for } n>1, \,\,\,\,\,\,\mathrm{m}_{\ell^n}\left(t\right) = \ell \cdot \mathrm{m}_{\ell^{n-1}}\left(\mathrm{R}_{\ell^{n-1}}\left[t\right]\right)+ \left\lfloor \frac{t}{\ell^{n-1}}\right\rfloor,\,\,\,\,\,\,\, \forall t\in \left[\ell^n\right]_{-};
\end{equation}
$$
\mathrm{m}_{\ell}\left(t\right) = t, \,\,\,\,\,\,\, \forall t\in \left[\ell\right]_{-},
$$
where $\mathrm{R}_{\beta}(\alpha)$ is the remainder of $\alpha$ divided by $\beta$.   Note that according to this definition, $\mathrm{m}_{\ell}(\cdot)$ is a base $\ell$ reversal function, i.e.  $\mathrm{m}_{\ell}(t)$ has its base $\ell$ representation being equal to the base $\ell$ representation of $t$ given in reverse order (for $\ell=2$ this transformation is also known as the reverse shuffle operation).

Algorithm \ref{algo:SysGenHomKernel} describes the recursive algorithm for the general $N=\ell^n$ case (for $n>1$).
The algorithm can also be easily adapted for mixed-kernels.  Let us  prove that the algorithm meets the systematic encoding requirement.

\begin{observation}\label{obsrv:sysEncObv1}
After round $i$ of the for-loop (beginning with $\ell-1$ and ending with $0$), codeword components $x_{\tau}$ such that $\mathrm{R}_{\ell}\left[{\tau}\right]\geq i$ are not changed anymore by Algorithm   \ref{algo:SysGenHomKernel}.
\end{observation}
\proof Equation (\ref{eq:SysEncEncInnerCode})  updates   vector $\bf x$ at the end of round $i$. Since $\bf G$ is lower triangular, we always have that $\forall i'\in[\ell]_{-}$,  ${G}_{i',j'}=0$ for $j'>i'$. Therefore all the updates for rounds $i'<i$ of the for-loop will have zeros in the vector $G_{i'\rightarrow}$ in places corresponding to $x_{\tau}$ where $\mathrm{R}_{\ell}\left[{\tau}\right]\geq i$. \qed

 \begin{observation}\label{obsrv:sysEncObv2}
After round $i$ of the for-loop (beginning with $\ell-1$ and ending with $0$), we have $x_{\mathrm{m}_{N}\left(t\right)}= u_t  $ for all non-frozen components $u_t$ such that $\mathrm{R}_{\ell}\left[\mathrm{m}_{N}\left(t\right)\right] = i$,  where $i\in [\ell]_{-}$ and $t\in \left[N\right]_{-}$.
\end{observation}
\proof Let $t$ be such that $\mathrm{R}_{\ell}\left[\mathrm{m}_{N}\left(t\right)\right] = i$. Following (\ref{eq:MappingReq}) we have that $\frac{N}{\ell}\cdot i \leq t \leq \frac{N}{\ell}\cdot ( i +1)-1$. Consequently,  it is encoded at round $i$ of the for-loop, dedicated for encoding $\mathcal{C}_i$. Assume that $t = \frac{N}{\ell}\cdot i+r$ where $r = \mathrm{R}_{N/\ell}\left[t\right]$. In (\ref{eq:SysEncLabelPreparations}) we have $ \tilde{u}'_r =  G_{i,i}^{-1}\cdot\left(u_{ t }-x_{\mathrm{m}_N\left(t\right)}\right)$.  After applying the systematic encoder in  (\ref{eq:SysEncCall}), we have that $\tilde{x}_{\mathrm{m}_{\frac{N}{\ell}}\left(r\right)} = \tilde{u}'_r$. Following the execution of (\ref{eq:SysEncEncInnerCode}), we have $x_{\mathrm{m}_{N}\left(t\right)} = x_{\mathrm{m}_{N}\left(t\right)} + \tilde{x}_{\left\lfloor\mathrm{m}_{N}\left(t\right)/\ell\right\rfloor}\cdot G_{i,i}$. However, due to (\ref{eq:RecursionFormMapping}), we have $\left\lfloor\mathrm{m}_{N}\left(t\right)/\ell\right\rfloor = \mathrm{m}_{N/{\ell}}\left(\mathrm{R}_{N/{\ell}}\left[t\right]\right)=  \mathrm{m}_{N/{\ell}}\left(r\right) $. Therefore, we have $x_{\mathrm{m}_{N}\left(t\right)} = x_{\mathrm{m}_{N}\left(t\right)} +\tilde{u}'_r\cdot G_{i,i}=u_t$. Owing to Observation \ref{obsrv:sysEncObv1}, the value of $x_{\mathrm{m}_{N}\left(t\right)}$ will not further change in the algorithm, which proves the statement. \qed

\begin{algorithm}
\caption{Systematic Encoder for Homogenous Length $N=\ell^n$ $F$-Symbols  Polar Code, Based on Lower Triangular Kernel ${\bf G}\in F^{(\ell\times\ell)}$}          
\label{algo:SysGenHomKernel}
  \begin{description}
\item[\ALGOIN \underline{Input:} ] $\bf u$; $\bf z$.
\item[//Initializations: ] \hfill \\
\ALGOSTEP Allocate $\ell$ vectors $\left\{\tilde{{\bf u}}^{(i)}\right\}_{i\in[\ell]_{-}}$ of length $\frac{N}{\ell}$ $F$-symbols.

\ALGOSTEP Allocate two vectors $\tilde{\bf u}',\tilde{\bf x}\in F^{N/{\ell}}$.

\ALGOSTEP Allocate two vectors $\tilde{\bf u},{\bf x}\in F^{N}$. Initialize ${\bf x} = {\bf 0}$.

 \item[//Successively encode $\bf u$:] \hfill \\
  \ALGOSTEP \textbf{For} $i = \ell-1$ to $0$ \textbf{Do}  //encode $\mathcal{C}_i$
        \begin{itemize}
        \item Prepare vector $\tilde{\bf u}'$ which serves as the modified input to the encoder of $\mathcal{C}_i$:
\begin{equation}\label{eq:SysEncLabelPreparations}
\tilde{u}'_{r} = \left\{
                  \begin{array}{ll}
                    0, & \hbox{$z_{i\cdot \frac{N}{\ell}  +r } = 1$;} \\
                    G_{i,i}^{-1}\cdot\left(u_{ i \cdot \frac{N}{\ell} +r}-x_{\mathrm{m}_N\left( i \cdot \frac{N}{\ell} +r\right)}\right), & \hbox{otherwise.}
                  \end{array}
                \right.\,\,\,\,\,\,\,\,\,\,\,\,\ \forall r\in\left[\frac{N}{\ell}\right]_{-};
\end{equation}
        \item Run  $\mathcal{C}_i$ systematic encoder:
\begin{equation}\label{eq:SysEncCall}
\left[\tilde{{\bf x}},\,\,\,\, \tilde{{\bf u}}^{(i)} \right]= \text{SysEncoder}\left(\tilde{{\bf u}}',\,\,\,\,\,{\bf z}_{i\cdot \frac{N}{\ell} }^{{(i+1)\cdot \frac{N}{\ell} -1}}\right);
\end{equation}
        \item Update the encoded vector $\bf x$
\begin{equation}\label{eq:SysEncEncInnerCode}
  {\bf x}_{r'\cdot\ell}^{(r'+1)\cdot\ell-1}  =   {\bf x}_{r'\cdot\ell}^{(r'+1)\cdot\ell-1}   + \tilde{x}_{r'}\cdot {\bf G}_{i \rightarrow},\,\,\,\,\,\,\,\,\,\, \forall r'\in \left[\frac{N}{\ell}\right]_{-} ;
\end{equation}
        \end{itemize}

\item[\ALGOOUT \underline{Output:} ]\begin{itemize}
\item ${\bf x}$;
\item $\tilde {\bf u} = \left[\tilde{{\bf u}}^{(0)}, \,\,\, \tilde{{\bf u}}^{(1)},\ldots ,\tilde{{\bf u}}^{(\ell-1)} \right]$;
\end{itemize}

\end{description}

\end{algorithm}

\section{Recursive Descriptions of Polar Codes Decoding Algorithms}\label{sec:RecDescOfDecAlgor}
In this section we describe decoding algorithms for polar codes in a recursive framework that is induced from their recursive GCC structures.
Roughly speaking, all the algorithms we consider here have a similar format. Consider the GCC structure of Figure \ref{fig: def2GCC}. In this construction we have a length $N$ symbols code that is composed of $\ell$ outer-codes, denoted by  $\left\{\mathcal{C}_i\right\}_{i=0}^{\ell-1}$, each one of length $N/\ell$ symbols. The  decoding algorithms that are considered here  are composed of $\ell$ pairs of steps.  The $i^{th}$ pair is dedicated to decoding $\mathcal{C}_i$ as described in Algorithm \ref{algo:schemDescCrFormat}.

\begin{algorithm}
\caption{Decoding Outer-code $\mathcal{C}_i,\,\,\,\, i\in [\ell]_{-}$ }          
\label{algo:schemDescCrFormat}                           
  \begin{description}
  \item[//STEP $2\cdot i$:] \hfill \\
    \ALGOSTEP Using the previous steps, prepare the inputs to the decoder of outer-code $\mathcal{C}_i$.
   \item[//STEP $2\cdot i+1$:] \hfill \\
   \ALGOSTEP Run the decoder of code $\mathcal{C}_i$ on the inputs you prepared.\\
   \ALGOSTEP  Process the output of this decoder, together with the outputs of the previous steps.
    \end{description}
\end{algorithm}

Typically, the codes  $\left\{\mathcal{C}_i\right\}_{i=0}^{\ell-1}$ are polar codes of length $N/\ell$ symbols, thereby creating the recursive structure of the decoding algorithm.

Note that the decoding algorithm structure in Algorithm \ref{algo:schemDescCrFormat} is quite typical for decoding algorithms of GCCs. As an example, see the decoding algorithms in Dumer's survey on GCCs \cite{DumerConcatCodes}. In addition, the recursive decoding algorithms for Reed-Muller (RM) codes, utilizing their Plotkin $(u+v,v)$ recursive GCC structure were extensively studied by Dumer \cite{Dumer2006,Dumer2006b} and are closely related to the algorithms we present here. Actually, Dumer's simplified decoding algorithm for RM codes \cite[Section IV]{Dumer2006b} is the SC decoding for  Arikan's structure, we describe in Subsection \ref{sec:recSCDec}.

The algorithms we describe in a recursive fashion are the SC (Subsection \ref{sec:recSCDec}), Tal and Vardy's SCL (Subsection \ref{sec:SCListDecoding}) and BP (Subsection \ref{sec:BP}). For all of these algorithms, we first consider Arikan's $(u+v,v)$ code and then provide generalizations for other kernels, both  homogenous and mixed.
We note, that when possible, we prefer that the inputs to the algorithm and the internal computations are interpreted as log likelihood ratios (LLRs). Consequently,  the SC algorithm and  BP are described in such manner. In SCL, however, we need to be able to decide among different simultaneous decoding   option, therefore we use log-likelihoods (LLs) instead of LLRs.

Furthermore, in our discussion we do not consider how to efficiently compute these quantities. In some cases, especially with large kernels or with large alphabet size, these calculations pose a computational challenge. Approaches to adhere this challenge, are efficient decoding algorithms (such as variants of Viterbi algorithms) or approximations of the computations (for example, the min-sum approximation that Leroux \textit{et al.} used \cite{Leroux2012} or the near Maximum Likelihood  (ML) decoding algorithms that were used by Trifonov \cite{Trifonov2011}).

\begin{remark}[SCL Decoding with LLRs]\label{rem:SCLWithLLRs}
 Balatsoukas-Stimming \textit{et al.} presented an LLR based SCL decoder \cite{Balatsoukas-Stimming2013a} in which the decoding options are selected based on a measure called the path-metric (PM). PM is a function of the computed LLRs and the already decided information symbols. It can be easily   seen that tracking the PM measure can also be integrated into the recursive description given in Subsection \ref{sec:SCListDecoding}. This can be achieved by introducing an additional data-structure to hold its computations.
\end{remark}

\subsection{A Recursive Description of the SC Algorithm}\label{sec:recSCDec}

We begin by considering the SC decoder for Arikan's $(u+v,v)$ construction. Description of the algorithm for   generalized  arbitrary kernels then follows.
The inputs of the SC algorithm for Arikan's construction are listed below.
\begin{itemize}
\item An $N$ length vector of input LLRs, $\boldsymbol \lambda$, such that $\lambda_j = \ln \frac{\Pr \left(Y_j = y_j | X_j = 0\right) }{\Pr \left(Y_j = y_j | X_j = 1\right) }$ for $j\in [N]_{-}$,
     where ${  Y}_j$ is the measurement of the $j^{\text{th}}$ channel ${X }_j \rightarrow {  Y}_j $.
\item Vector indicator ${\bf z} \in \{0,1\}^N$, in which $z_i=1$ if and only if  element number $i$ of the information vector $\bf u$ is frozen.
\end{itemize}
The algorithm outputs the following structures.
\begin{itemize}
\item An $N$ length binary vector $\hat {\bf u} $ containing the information word that the decoder estimated. This vector includes the frozen symbols placed in their appropriate positions.
\item An $N$ length binary vector $\hat{\bf x}$ which is the codeword corresponding to   $\hat{\bf u}$.
\end{itemize}
The SC function signature is defined as
\begin{equation}\label{eq:SCSignature}
\left[\hat{{\bf u}},\,\,\, \hat{{\bf x}} \right] = \text{SCDecoder}\left({\boldsymbol \lambda},\,\,\,\, {\bf z}  \right).
\end{equation}

First, let us describe the decoding algorithm for length $N=2$ bits code, i.e. for the basic kernel $g^{(1)}(u,v)=(u+v,v)$. We get as input ${\boldsymbol \lambda}=[\lambda_0,\lambda_1]$ which are the LLRs of the output of the channel ($\lambda_{0}$ corresponds to the first output of the channel and $\lambda_{1}$ corresponds to the second output). The procedure has four steps as described in Algorithm \ref{algo:SCRecsDescUVV}.
\begin{algorithm}
\caption{SC of the $(u+v,v)$ Kernel }          
\label{algo:SCRecsDescUVV}
 \bindent
\begin{description}
\item[\ALGOIN Input: ]  ${\boldsymbol \lambda};\,\,\,\, {\bf z}.$
   \item[//STEP 0:] \hfill \\
    \ALGOSTEP Compute the LLR of $u$: $\hat{\lambda} = 2\tanh^{-1}\left(\tanh(\lambda_0/2)\tanh(\lambda_1/2)\right)$.
   \item[//STEP 1:] \hfill \\
       \ALGOSTEP  Decide on $u$, (denote the decision by $\hat{u}$).
   \item[//STEP 2:] \hfill \\
       \ALGOSTEP  Compute the LLR of $v$  (given the estimate of $\hat{u}$): $\hat{\lambda} = (-1)^{\hat{u}}\cdot\lambda_0+\lambda_1$.
   \item[//STEP 3:] \hfill \\
      \ALGOSTEP  Decide on $v$, (denote the decision by $\hat{v}$).
\item[\ALGOOUT Output: ]
     \begin{itemize}
     \item $\hat{\bf u} = \left[\hat{u}, \hat{v}\right]$;
      \item $\hat{\bf x} = \left[\hat{u}+\hat{v}, \hat{v}\right]$.
     \end{itemize}
 \end{description}
   \eindent

\end{algorithm}
 Note that steps $1$ and $3$, may be done based on the LLRs computed on steps $0$ and $2$, respectively (i.e. by their sign), or by using an additional side information (for example, if $u$ is frozen, then the decision is based on  its known value).
%
 A decoder for length $N$ polar code is described in Algorithm \ref{algo:SCRecsDescUVVN}.

 \begin{algorithm}
\caption{SC Recursive Description for Length $N=2^n$ Bits  $(u+v,v)$ Polar Code }          
\label{algo:SCRecsDescUVVN}
\begin{description}
\item[\ALGOIN Input: ]  ${\boldsymbol \lambda};\,\,\,\, {\bf z}.$
   \item[//STEP 0:] \hfill \\
       \ALGOSTEP  Compute the
 LLR input vector, ${ \hat{\boldsymbol \lambda} }_0^{N/2-1}$, for the first outer-code such that $$\hat{\lambda}_i= 2\tanh^{-1}\left(\tanh(\lambda_{2i}/2)\tanh(\lambda_{2i+1}/2)\right),\,\,\,\, \forall i\in \left[N/2\right]_{-}.$$
   \item[//STEP 1:] \hfill \\
       \ALGOSTEP  Give the vector ${\hat {\boldsymbol\lambda}}$ as an input to the  polar code decoder of length $N/2$. Also provide to the decoding algorithm, the indices of the frozen bits from the first half of the codeword (corresponding to the first outer-code), i.e. run
       \begin{equation}\label{eq:UVVSCCall0}
            \left[\hat{\bf u}^{(0)},\,\,\, \hat{\bf x}^{(0)} \right] = \text{SCDecoder}\left({\hat{\boldsymbol \lambda}},\,\,\,\, {\bf z}_{0}^{N/2-1}  \right).
       \end{equation}
   According to  (\ref{eq:SCSignature}),  $\hat{\bf u}^{(0)} $ is the information word estimation for the first outer-code, and $\hat{\bf x}^{(0)}$  is its corresponding codeword.
   \item[//STEP 2:] \hfill \\
      \ALGOSTEP  Using $\boldsymbol \lambda$ and  $\hat{\bf x}^{(0)}$, prepare the LLR input vector, ${\hat{\boldsymbol \lambda}}_0^{N/2-1}$, for the second outer-code, such that $$\hat{\lambda}_i = (-1)^{\hat{x}^{(0)}_i}\cdot \lambda_{2i}+\lambda_{2i+i},\,\,\,\, \forall i\in \left[N/2\right]_{-}.$$
   \item[//STEP 3:] \hfill \\
      \ALGOSTEP  Give the vector ${\hat {\boldsymbol\lambda}}$ as an input to the polar code decoder of length $N/2$. In addition, provide  the indices of the frozen bits from the second half of the codeword (corresponding to the second outer-code), i.e. run
       \begin{equation}\label{eq:UVVSCCall0}
            \left[\hat{\bf u}^{(1)},\,\,\, \hat{\bf x}^{(1)} \right] = \text{SCDecoder}\left({\hat{\boldsymbol \lambda}},\,\,\,\, {\bf z}_{N/2}^{N-1}  \right),
       \end{equation}
       where  $\hat{\bf u}^{(1)}$  and $\hat{\bf x}^{(1)}$ are the estimations of the information word and its corresponding codeword of the second outer-code.
\item[\ALGOOUT Output: ]
     \begin{itemize}
     \item $\hat{\bf u} = \left[\hat{{\bf u}}^{(0)}, \hat{{\bf u}}^{(1)}\right]$;
      \item $\hat{\bf x} =\left[\hat{x}^{(0)}_i+\hat{x}^{(1)}_i,\hat{x}^{(1)}_i\right]_{i=0}^{N/2-1}$.
     \end{itemize}
 \end{description}
\end{algorithm}

Let us now generalize this decoding algorithm for a GCC homogenous scheme with   general kernel.
In this case for length $N$ $F$-symbols code, we have an $\ell$ length mapping $g({\bf u})={\bf x}$ over the $F$ alphabet, i.e. $g(\cdot):F^{\ell}\rightarrow F^{\ell}$.  The inputs and the outputs of the decoding algorithm are the same as in the $(u+v,v)$ case, except that here the LLRs may correspond to non-binary alphabet. As a consequence, we need to have $|F|-1$ LLR input vectors $\left\{\boldsymbol \lambda^{(t)}\right\}_{t\in F\backslash \{0\}}$ each one of length $N$ and defined such that
\begin{equation}\label{eq:llrDefinitionF}
\lambda_j^{(t)} = \ln \frac{\Pr \left(Y_j = y_j | X_j = 0\right) }{\Pr \left(Y_j = y_j | X_j = t\right) }
\end{equation}
for $j\in [N]_{-}$,  where ${  Y}_j$ is the measurement of the $j^{\text{th}}$ channel ${X }_j \rightarrow {  Y}_j $. Furthermore $\bf u$ and $\bf x$ are in $F^{N}$. Note that we always have $\lambda_j^{(0)}=0$ and therefore it doesn't have to be calculated. The following is the signature for the general SC decoder
\begin{equation}\label{eq:SCGenSignature}
\left[\hat{{\bf u}},\,\,\, \hat{{\bf x}} \right] = \text{SCDecoder}\left(\left\{{\boldsymbol \lambda}^{(t)}\right\}_{t\in F\backslash \{0\}},\,\,\,\, {\bf z}  \right).
\end{equation}

 In the GCC structure of this polar code there exist at most $\ell$ outer-codes $\left\{\mathcal{C}_i\right\}$, each one of length $N/\ell$ symbols. We may have less than $\ell$ outer-codes, in case some of the inputs are glued (which results in a mixed-kernels construction). In such cases, the outer-code corresponding to the glued inputs is considered to be over a larger size input alphabet.
We assume that each outer-code has a decoding algorithm associated with it. This decoding algorithm is assumed to receive as input the "channel" observations on the outer-code symbols (usually manifested as probabilities matrices, or LLR vectors).
If the outer-code is a polar code, then this algorithm should also receive the indices of the frozen symbols of the outer-code. We require that the algorithm  outputs its estimation on the information vector and its corresponding outer-code codeword.

Let us first consider an $\ell$ length code generated by a single application of the kernel i.e. ${\bf x} = g\left({\bf u}\right)$. Note that  this is the base case of the recursion. Assuming that we already decided on symbols ${\bf u}_{0}^{i-1}$ (denote this decision by $\hat{\bf u}_{0}^{i-1}$), computing the LLR vector $\hat{ \boldsymbol\lambda}^{(t)}$ corresponding to  the $i^{th}$ input of the transformation (i.e. $u_{i}$) is done according to the following rule
\begin{equation}\label{eq:genSCRule}
\hat{\lambda}^{(t)}=\ln\left(\frac{\sum_{{\bf u}_{i+1}^{\ell-1}\in F^{\ell-i-1}}  R_g\left(\hat{\bf u}_{0}^{i-1},0,{\bf u}_{i+1}^{\ell-1}\right)}{\sum_{{\bf u}_{i+1}^{\ell-1}\in F^{\ell-i-1}} R_g\left(\hat{\bf u}_{0}^{i-1},t,{\bf u}_{i+1}^{\ell-1}\right)}\right),
\end{equation}
where
\begin{equation}\label{eq:llrSingleCw}
R_g({\bf u}_{0}^{\ell-1}) =\exp\left(-\sum_{r=0}^{\ell-1} \lambda_r^{\left(x_r\right) }\right),\,\,\,\, \text{ such that } {\bf x}=g({\bf u}).
\end{equation}
Consequently, SC decoding for the $\ell$ length polar code includes sequential calculations of the likelihood values $\left\{\hat{\lambda}^{(t)} \right\}_{t\in F\backslash\{0\}}$ corresponding to non-frozen $u_i$ according to (\ref{eq:genSCRule}) followed by a decision on $u_i$  (denoted by $\hat{u}_i$)  for $i\in \left[\ell\right]_{-}$. If $u_i$ is frozen, we set $\hat{u}_i$ to be equal to its predetermined value. Finally in (\ref{eq:SCGenSignature}) we output $\hat{\bf u} = \left[\hat{u}_0\,\,\, \hat{u}_1\,\,\, \ldots \hat{u}_{\ell-1}  \right]$, and  $\hat{\bf x} = g\left(\hat{\bf u}\right)$.

We now turn to describe the SC decoding algorithm for length $N>\ell$ homogenous polar code over $F$ based on the same kernel $g(\cdot)$. As we already   mentioned, due to the code structure, the decoding algorithm is composed of pairs of steps, such that the $i^{th}$ pair  deals with the  $i^{th}$ outer-code, where $i\in[\ell]_{-}$.

We denote the information word that was estimated by the decoder of the $m^{\text{th}}$ outer-code   by $\hat{\bf u}^{(m)}$ and its corresponding codeword by $\hat{\bf x}^{(m)}$, both of them are of length $N/{\ell}$ symbols. Algorithm \ref{algo:GenDetDescSC} describes the pair of steps of the SC algorithm $i\in [\ell]_{-}$ and Algorithm \ref{algo:SCDecodingOutput} specifies its output generation.

\begin{algorithm}
\caption{ SC Decoder  Steps Dedicated for Outer-Code $\mathcal{C}_i,\,\,\,\, i\in [\ell]_{-}$ }          
\label{algo:GenDetDescSC}

\begin{description}

\item [//STEP $2\cdot i$:] \hfill \\
\ALGOSTEP  Prepare $|F|-1$ LLR input vectors $\left\{\hat{\boldsymbol \lambda}^{(t)}\right\}_{t\in F\backslash\{0\}}$ each one of length $N/\ell$ using (\ref{eq:genSCRule2}), i.e
\begin{equation}\label{eq:genSCRule2}
\hat{\lambda}^{(t)}_{j}=\ln\left(\frac{\sum_{{\bf w}_{i+1}^{\ell-1}\in F^{\ell-i-1}}  R_g\left(\hat{x}_{j}^{(0)},\hat{x}_{j}^{(1)},\ldots,\hat{x}_{j}^{(i-1)}, 0,{\bf w}_{i+1}^{\ell-1}\right)}{\sum_{{\bf w}_{i+1}^{\ell-1}\in F^{\ell-i+1}} R_g\left(\hat{x}_{j}^{(0)},\hat{x}_{j}^{(1)},\ldots,\hat{x}_{j}^{(i-1)}, t,{\bf w}_{i+1}^{\ell-1}\right)}\right),\,\,\,  \forall t\in F\backslash \{0\}  \text{ and } \forall j\in \left[N/\ell\right]_{-},
\end{equation}
where  $\hat{\bf x}^{(m)}$ is the estimated codeword of outer-code $\mathcal{C}^{(m)}$ that was computed at the previous steps,  $m\in \left[i\right]_{-}$. Note that for the LLR calculation of   $\hat{\lambda}^{(t)}_{j}$  in (\ref{eq:genSCRule2}) we use  input LLRs corresponding to channel indices $j\cdot\ell, j\cdot\ell+1,\ldots,(j+1)\cdot \ell-1$.
%
%
   \item [//STEP $2\cdot i+1$:] \hfill \\
     \ALGOSTEP Decode    the $i^{th}$ outer-code using the computed LLR vectors $\left\{\hat{\boldsymbol \lambda}^{(t)}\right\}_{t\in F\backslash\{0\}}$, i.e.
   \begin{equation} \label{eq:SCLGenStepr}
   \left[\hat{\bf u}^{(i)},\,\,\, \hat{\bf x}^{(i)}\right] = \text{SCDecoder}\left(\left\{\hat{\boldsymbol \lambda}^{(t)}\right\}_{t\in F\backslash\{0\}},\,\,\,\, {\bf z}_{i\cdot N/\ell}^{(i+1)\cdot N/\ell-1}  \right).
   \end{equation}
\end{description}
\end{algorithm}

\begin{algorithm}
\caption{SC Decoder Output Generation }          
\label{algo:SCDecodingOutput}                           
\begin{description}
\item[\ALGOOUT Output: ] (occurs after applying Algorithm \ref{algo:GenDetDescSC} for all $i\in [\ell]_{-}$)\hfill\\
\begin{itemize}
\item $\hat{\bf u} = \left[\hat{\bf u}^{(0)},\,\,\,\hat{\bf u}^{(1)}, \ldots, \hat{\bf u}^{(\ell-1)}  \right] $;
\item $\hat{\bf x}_{j\cdot \ell}^{(j+1)\cdot {\ell}-1} = g\left(\hat{x}^{(0)}_j,\hat{x}^{(1)}_j,\ldots,\hat{x}^{(\ell-1)}_j\right),\,\,\,\,\,\,\,\forall j\in \left[N/\ell\right]_{-}$.
\end{itemize}
\end{description}
\end{algorithm}

\begin{remark}[LLR Calculations Simplification for Linear Kernels]
Let us assume that $g(\cdot)$ is an $\ell$ dimensions linear kernel, having a generating matrix $\bf G\in F^{\ell\times\ell}$, such that ${\bf x} = g({\bf  u}) = {\bf u}\cdot {\bf G}$.
It can be easily seen that if ${\hat{\bf x} = \hat{\bf u}_{0}^{i-1}\cdot {\bf G}_{0:i-1,0:\ell-1}}$, then (\ref{eq:genSCRule}) is equivalent to

\begin{equation}\label{eq:genLinKernelLLR}
\hat{\lambda}^{(t)}=
\ln\left(\frac{\sum_{{\bf x} \in \Gamma_i+\hat{\bf x}}  \exp\left(-\sum_{r=0}^{\ell-1} \lambda_r^{\left( {x}_r \right) }\right)}{\sum_{{\bf x} \in \Gamma_i+\hat{\bf x}+t\cdot {\bf G}_{i\rightarrow}} \exp\left(-\sum_{r=0}^{\ell-1} \lambda_r^{\left( {x}_r \right) }\right) }\right),\,\,\,\,\, t\in F\backslash \{0\},
\end{equation}

where $\Gamma_i = \left\{ {\bf v} \left|  {\bf v} = {\bf w} \cdot {\bf G}_{\left((i+1):(\ell-1)\right),\left(0:(\ell-1)\right)}, \,\,\,\, {\bf v}\in F^{\ell},{\bf w}\in F^{\ell-1-i}  \right. \right\}$. Note that $\Gamma_i$  is the linear code induced by the last $\ell-1-i$ rows of the generating matrix ${\bf G}$. Furthermore  $\Gamma_i+\hat{\bf x}$ is the coset of the linear code $\Gamma_i$ that is induced by the coset vector $\hat{\bf x}$.

The calculation method in (\ref{eq:genLinKernelLLR}) is attractive because it implements the enumeration of the cosets members as a summation of  $\hat{\bf x}$ (the estimated coset vector, computed throughout the algorithm) with predetermined sets $\Gamma_i+t\cdot {\bf G}_{ i \rightarrow}$ (the cosets of $\Gamma_i$ in $\Gamma_{i-1}$). Therefore,  efficient ways to calculate (\ref{eq:genLinKernelLLR})  for the case of ${\hat{\bf  x}} = \bf 0$ (e.g. using trellis decoding by employing the dual code of $\Gamma_i$)  can be easily utilized for calculating  (\ref{eq:genLinKernelLLR}) for non-zero ${\hat{\bf  x}}$. This can be done by appropriately modifying the input LLR vector reflecting the notion that all the possible enumerated codewords are members of the cosets, considered in the case of ${\hat{\bf  x}} = \bf 0$, shifted by the constant vector $\hat{\bf x}$. As a consequence, using as inputs the LLRs of a modified channel generated by adding the known vector $\hat{\bf x}$ to the original channel output will allow to employ the computations of the zero case for general cases.

Algorithm \ref{algo:GenDetDescSC} can be adapted to support the computation technique suggested here. First, we initialize the vector $\hat{\bf x}$ (later given as output) to be the all-zeros vector. Secondly, we replace (\ref{eq:genSCRule2}) by the following calculation
\begin{equation}\label{eq:genLinKernelLLR2}
\hat{\lambda}^{(t)}_j=
\ln\left(\frac{\sum_{{\bf x} \in \Gamma_i+\hat{\bf x}_{j\cdot\ell}^{(j+1)\ell-1}}  \exp\left(-\sum_{r=0}^{\ell-1} \lambda_{j\cdot \ell+r}^{\left( {x}_r \right) }\right)}{\sum_{{\bf x} \in \Gamma_i+\hat{\bf x}_{j\cdot\ell}^{(j+1)\ell-1}+t\cdot {\bf G}_{i\rightarrow}} \exp\left(-\sum_{r=0}^{\ell-1} \lambda_{j\cdot \ell+r}^{\left( {x}_r \right) }\right) }\right),\,\,\,\,\, t\in F\backslash \{0\}.
\end{equation}
Thirdly, after estimating ${\hat {\bf x}}^{(i)}$, the outer-code codeword of $\mathcal{C}_i$  in (\ref{eq:SCLGenStepr}), we need to update the coset vector $\hat{\bf x}$ by calculating
\begin{equation}\label{eq:reEncodingGenSC}
\hat{\bf x}_{j\cdot \ell}^{(j+1)\cdot {\ell}-1} = \hat{\bf x}_{j\cdot \ell}^{(j+1)\cdot {\ell}-1} + \hat{x}^{(i)}_j\cdot {\bf G}_{i\rightarrow},\,\,\,\, \forall j\in \left[N/\ell\right]_{-}.
\end{equation}
As a consequence, in Algorithm \ref{algo:SCDecodingOutput} the decoder can just output $\hat{\bf x}$ that was calculated throughout the odd steps of the  algorithm. This simplification is used in our suggested schematic implementation in Subsection \ref{sec:SCLineForGeneralKernel}.
\end{remark}

In case we have a mixed-kernels construction, the generalization is quite easy. In order to illustrate this we consider an example of $\ell$ dimensions kernel in which we have glued the symbols ${  u}_1$ and ${  u}_2$  to a new symbol ${  u}_{1,2} \in F^{2}$ (see Example \ref{ex:MxdKernelsExample} for an instance of such structure). In this case, we treat these two symbols as  one entity, and consider the outer-code associated with them, denoted as ${\mathcal{C}}_{1,2}$, as an $N/{\ell}$ length code over the alphabet $F^{2}$. The only change we have in the decoding algorithm is for the pair of decoding steps of Algorithm  \ref{algo:GenDetDescSC} corresponding to this "glued" symbols outer-code. For the first step in the pair, we need to compute $|F|^2-1$ LLR vectors $\left\{\hat{\boldsymbol \lambda}^{(t_0,t_1)}\right\}_{\left(t_0,t_1\right)\in F^2\backslash\{(0,0)\}}$ each one of length $N/\ell$. These vectors serve as an input to the the decoder of ${\mathcal{C}}_{1,2}$. In this case, each LLR component in the vector, is a function of both ${ u}_1$ and ${ u}_2$ inputs to the kernel. Equation (\ref{eq:genSCRule2}) is therefore updated as follows:
\begin{equation}\label{eq:genSCRuleMixed}
\hat{\lambda}^{(t_1,t_2)}_{j}=\ln\left(\frac{\sum_{{\bf w}_{3}^{\ell-1}\in F^{\ell-3}}  R_g\left(\hat{x}_{j}^{(0)},0, 0, {\bf w}_{3}^{\ell-1}\right)}{\sum_{{\bf w}_{3}^{\ell-1}\in F^{\ell-3}} R_g\left(\hat{x}_{j}^{(0)},t_1,t_2, {\bf w}_{3}^{\ell-1}\right)}\right),\,\,\, \forall j\in \left[N/\ell\right]_{-}.
\end{equation}
The second step of the pair in Algorithm \ref{algo:GenDetDescSC} remains unchanged.

\subsection{A Recursive Description of the SCL  Algorithm} \label{sec:SCListDecoding}
In this subsection we provide a recursive description of the SCL decoder, originally introduced by Tal and Vardy \cite{Tal2012}.
 Each stage of the SCL algorithm involves comparisons of  likelihoods of different SC decoding possibilities (resulting from keeping more than one decision option at the previous decoding stages). Therefore, we  assume that the inputs to the algorithm as well as its internal computation values are interpreted as likelihoods, instead of LLRs\footnote {The notion of likelihoods normalization that  was considered by Tal and Vardy  \cite[Algorithm 14]{Tal2012} to avoid floating-point or fixed-point underflows is also applicable here and should be employed for numerical stability.}(note, however, that LLRs can be used as well, see Remark \ref{rem:SCLWithLLRs}). Note, that if the decoding list is of size $1$, then the formulation given below is of the SC decoder described in Subsection \ref{sec:recSCDec} (with the only difference that likelihoods are employed instead of LLRs).

The SCL algorithm, described in this subsection, returns as output a list of decoding possibilities. The most likely element of this list should be given as output.

\subsubsection{Sequential Decoders  as Path Traversal Algorithms in  Decoding Trees}
Before dwelling into the details of  SCL  let us discuss the general idea that this algorithm entails. Sequential decoding algorithms examine their decision space (i.e. the set of all possible results) and choose a result from it, by gradually refining the space (i.e. eliminating some of the possible outcomes) until a predetermined number of outcomes remains (in SC the number is $1$, in SCL the number is $L$, from which the best outcome is chosen). In SC and SCL the decision space is described by the input vector to the encoder, ${\bf u}$. The decision space is refined by determining the components of ${\bf u}$ in a consecutive order.

It is quite common to describe the decision space of an algorithm by an edge-labeled directed tree dubbed a \textit{decoding tree}. Note, however,  that, strictly speaking, the graphical structure of the decoding tree is generally a forest, because we may have multiple nodes at the top of the tree (representing different input models) which are not connected to each other. These nodes are dubbed the \textit{roots} of the decoding tree.

\begin{figure}
\center
  \includegraphics[scale = 0.15]{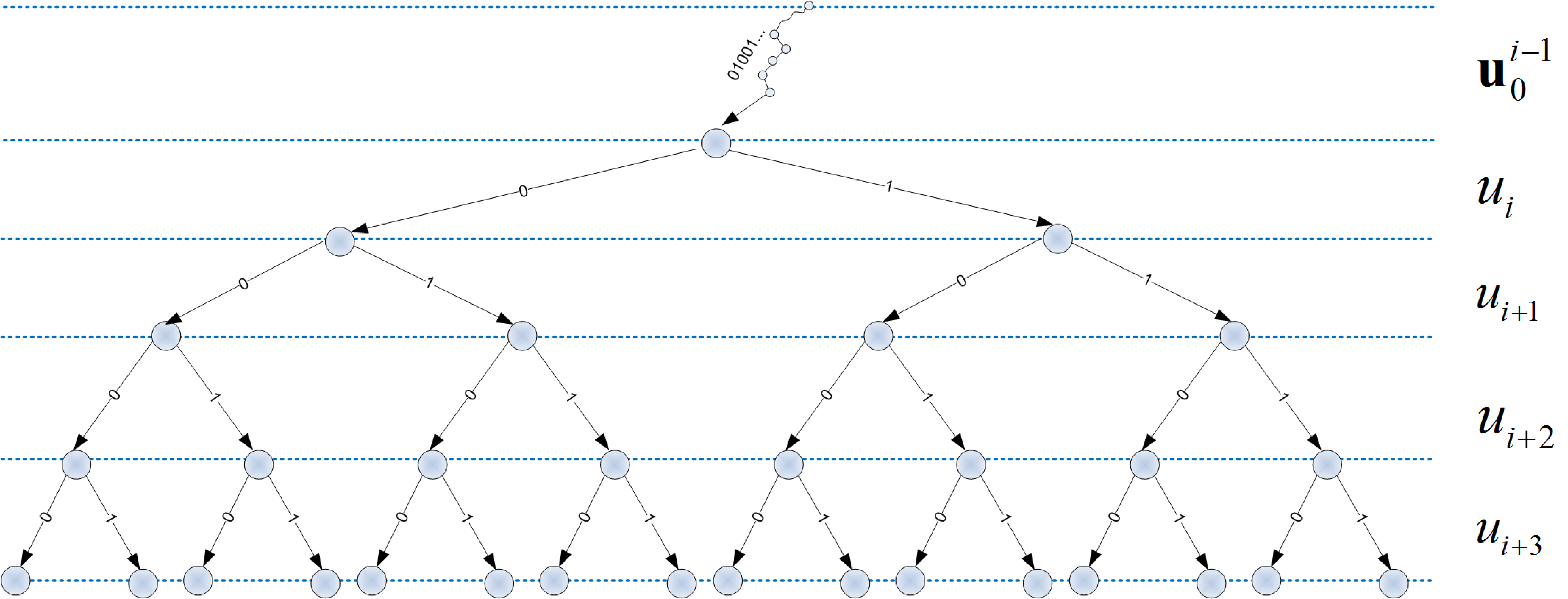}\\
  \caption{Decoding tree for $(u+v,v)$ polar code illustrating the decision space of the SC  and SCL algorithms  }\label{fig:decisionTree}
\end{figure}

Figure \ref{fig:decisionTree} illustrates such a decoding tree used in sequential decoding of Arikan's $(u+v,v)$ polar code. The decoding tree is a layered graph, such that the edges of each layer correspond to a single entry of the vector $\bf u$ (the layers boundaries are indicated by the dotted lines in Figure \ref{fig:decisionTree}). The nodes in the graph indicate  sequential decision junctions and the edges emanating from each node represent possible assignments to the variable of the layer.
The single path between the roots of the decoding tree and a node  appearing on the top of layer $u_i$ in the graph, indicates previous decisions (on variables ${\bf u}_{0}^{i-1}$) that preceded the decision on $u_i$. Consequently, the paths between the root of the tree and  the leaves of the tree correspond to all possible assignments to the vector $\bf u$. For example, in Figure   \ref{fig:decisionTree} the  paths of  the illustrated  tree correspond to  all the binary assignments to ${\bf u}_0^{i+3}$ given  that ${\bf u}_0^{i-1}$ is a fixed prefix (indicated in the figure by the string $01001...$, and further denoted by $\hat{\bf u}_0^{i-1}$) and ${\bf u}_{i}^{i+3}\in \{0,1\}^4$.

\begin{figure}
\center
  \includegraphics[scale = 0.15]{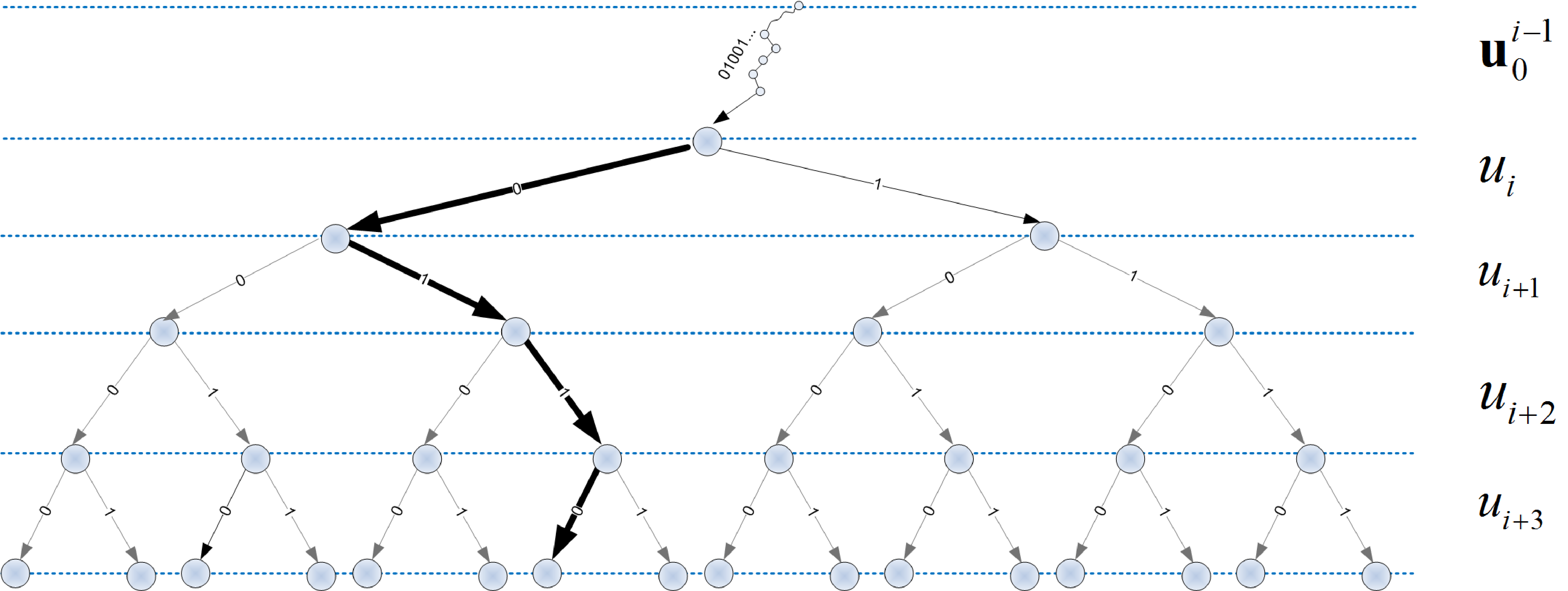}\\
  \caption{Representation of SC as a sequential walk on a decoding tree  }\label{fig:decisionTreeSC}
\end{figure}

In the SC algorithm the decoder always considers a single path (dubbed as a \textit{decoding-path}) among the possible tree paths. The decoding-path is gradually paved by sequentially joining to it edges emanating from nodes reached by the previous stages. On stage $\#i$ of the algorithm the edge selection  corresponds to the most preferable assignment to the variable $u_i$ (assuming that the path leading to $u_i$'s layer is fixed). Figure \ref{fig:decisionTreeSC} illustrates this abstraction. Given a certain prefix assignment to the sub-vector ${\bf u}_{0}^{i-1}=\hat{\bf u}_{0}^{i-1}$, we first turn to decide which edge emanating from the node corresponding to this prefix is better (i.e. we select the best assignment to  $u_i$ given the prefix). In SC this is done by calculating the likelihood of $u_i$ using  the channel observation vector $\bf y$ and the information prefix   ${\bf u}_{0}^{i-1}=\hat{\bf u}_{0}^{i-1}$, i.e. $W\left({\bf y},{\bf u}_0^{i-1}=\hat{\bf u}_0^{i-1}| u_{i}=b \right)$ where $b\in\{0,1\}$. We call  this likelihood function, the observed \textit{model} of the decision node and we view it as a function of only the variable $b$ (the value of $u_i$) while the other elements ($\bf y$ and ${\bf u}_{0}^{i-1}$) are considered as observations that define the statistical model.  Based on this model calculations the $'0'$ edge was chosen in Figure \ref{fig:decisionTreeSC} (indicated by the thick black edge). This decision, in turn, is used to update the model to  $W\left({\bf y},{\bf u}_0^{i-1}=\hat{\bf u}_0^{i-1},u_i = 0| u_{i+1}=b \right)$, which is further used to decide on  $u_{i+1}$ on the next step. Moving forward, the algorithm successively updates the model and choose to add to the existing path,  $\hat{\bf u}_{0}^{i+1}$, the edges corresponding to, $\hat{u}_{i+2}=1$ and $\hat{u}_{i+3}=0$. In case that a certain variable is frozen then the next edge is chosen as the one corresponding to its fixed value.

 When applied to polar codes, SC has an advantage in terms of its algorithmic complexity. Utilizing the recursive structure of the code, it is possible to efficiently compute the likelihoods needed for the decision on $u_i$ by reusing  previous computation results obtained when deciding on ${\bf u}_0^{i-1}$. In other words, in SC the channel observation model is easily updated given some results of the  calculations performed for determining the former observation model. The space needed to store these temporary calculations is linear in the code length (assuming the kernel size $\ell$ and the alphabet size $|F|$ are fixed).
  On the other hand, when SC algorithm decides on $u_i$ it does not take into account the existence of possible frozen values of its descendant nodes. In other words it assumes that all the assignments to ${\bf u}_i^{N-1}$ are possible when calculating the likelihoods, even though the code structure enforces certain variables to be fixed. This lack of "future awareness" and the inability of the algorithm to change its past decisions (i.e. the algorithm always advances in one direction in the tree, from "top" to "bottom") are fundamental reasons for its sub-optimality.

The SCL algorithm with list of size $L$  is a generalization of SC, in which the decoder  considers simultaneously at most $L$ possible decoding-paths. It can be seen  that the complexity of SCL both in time and in space will be approximately bounded from above by the complexity of SC times $L$.  The reason for this is that   operations associated with each tree junction in SCL are roughly same ones that would have been associated to the junction if it were  on a single SC path. We need however additional operations for choosing the $L$ edges with the maximal likelihoods for continuing the paths. This can be done in linear time (in $L$) per each decoding tree layer (information symbol, $u_i$). Secondly, tracking data structures need to be defined and utilized, in order to keep tabs on the existing decoding-paths while allowing an emulation of the SC algorithm for each separate decoding path. As we next see we can employ such structures and algorithms that will not exceed  $L$ times the asymptotic complexity of SC.

\begin{figure}
\center
  \includegraphics[scale = 0.09]{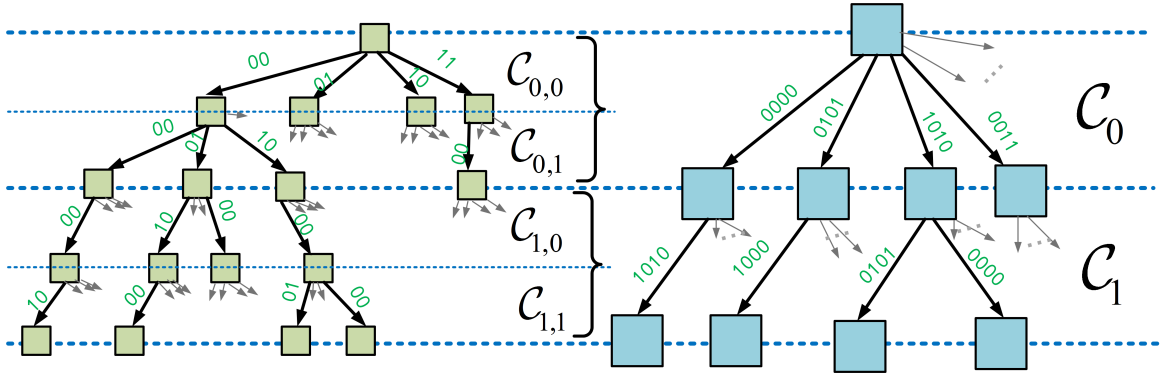}\\
  \caption{SCL ($L=4$) algorithm example of $(u+v,v)$ with $N=8$ bits (see Figure \ref{fig:decisionTree}) illustrated on the right a decoding tree on the outer-codes of the structure ($\mathcal{C}_{0},\mathcal{C}_{1}$). The left decoding tree expands each edge of the right tree into decoding-paths on the outer-codes of  $\mathcal{C}_{0}$ and $\mathcal{C}_{1}$ . The labels of the edges are the values of the outer-codes.  }\label{fig:decisionTreeAndGCC2}
\end{figure}

In Figure \ref{fig:decisionTreeAndGCC2} we described the SC and SCL algorithms as a sequence of decoding tree refinements. The models for these decisions are sequentially updated using past decisions (selected paths) and the current observation model. The connection between the input observation model and the input model for the next outer-code is defined by the inner-code layer. These updated input models are recursively provided to the smaller outer-codes, until the we reach codes of single symbols (corresponding to the elements of ${\bf u}_{0}^{7}$) in which decisions are  made. It is indeed a property of the the recursive description of SC and SCL that each recursion step utilizes a decoding tree of which the layers are the outer-codes. A fundamental property of the algorithms is that on each recursion step,  updating a model based on an edge selection is linear in the  outer-code length (assuming that the kernel is fixed). As a consequence, the number of operations of SC is $O\left(N\cdot \log N\right)$.

\subsubsection{Data Structures for Tracking Decoding Paths  in SCL}
 Tracking the employed observation-model is easy in SC because at any given point in time we assume only a single model (induced by previous SC decisions). On the other hand, in SCL, multiple models are considered simultaneously and it is therefore required to efficiently keeping track of them. Specifically, for each constituent code of the GCC we must store the tree structure connecting between its outer-codes.

We now propose  data structures for meeting this requirement.
 \begin{itemize}
 \item ${\bf S}^{(e)}$ - an $\ell\times L$ matrix describing the edges of the decoding tree. Specifically,  ${\bf S}^{(e)}_{0\rightarrow}$ contains indications for the edges in the $\mathcal{C}_{0}$ layer and   ${\bf S}^{(e)}_{1\rightarrow}$ corresponds to the $\mathcal{C}_{1}$ layer. The only interesting nodes in the tree  are the ones having a decoding path leading to them (we call them \textit{active nodes}). We use the arbitrary convention that active nodes are assigned numbers in  $[L]_{-}$ starting from the level's leftmost node to the rightmost node as appeared in the figure.  To represent this in our data structure we let $S^{(e)}_{i,j}$  contain the index of the single node at the top of layer $i$ that is connected to  node $j$ at the bottom of the layer. In case there are less than $L$ nodes at the bottom  of a layer, the matrix entries corresponding to the missing nodes are  assigned the null symbol, $\phi$.
     \item ${\bf S}^{(p)}$  - an $L\times \ell$ matrix, such that ${\bf S}^{(p)}_{i\rightarrow}$ defines the single path between the roots of the tree and the $i^{th}$ node at the bottom of the final layer. This path is specified in terms of the nodes indices, such that $S^{(p)}_{i,j}$ is the node located on the top of layer  $j$ in the path. Note that ${\bf S}^{(p)}$ is easily derived from ${\bf S}^{(e)}$.
         \item ${\bf s}$ - an $L$ length vector describing the origin model for each decoding path, i.e. ${\bf s} = \left({\bf S}^{(p)}_{\downarrow 0}\right)^{T}$.
         \item $\hat{\bf X}^{(i)}$ where $i\in [\ell]_{-}$ - $\ell$  matrices (of dimensions $L\times N/\ell$) used for keeping the labels of the selected edges in ${\bf S}^{(e)}$. Here $\hat{\bf X}^{(i)}_{r\rightarrow}$ contains the label of the edge pointing to node $r\in [L]_{-}$ at the bottom of layer $i\in [\ell]_{-}$ in the decoding tree. Note that this edge is represented by the $S^{(e)}_{i,r}$ entry.
             \item  $\hat{\bf U}^{(i)}$  where  $i\in [\ell]_{-}$ - $\ell$ matrices (of dimensions $L\times N/\ell$), such that $\hat{\bf U}^{(i)}_{r \rightarrow}$ is the information word (including the assignment of the frozen symbols) of the outer-code corresponding to $\hat{\bf X}^{(i)}_{r\rightarrow}$ codeword.
 \end{itemize}

 The decoding-paths data structures are generated throughout the  decoding process.  SCL sequential traversing the decoding tree from its first level to its leaves results in updating these matrices. After deciding on layer $i\in [\ell]_{-}$ we write row $i$ in     ${\bf S}^{(e)}$, and prepare matrices $ \hat{\bf X}^{(i)}$ and  $ \hat{\bf U}^{(i)}$. Following ${\bf S}^{(e)}$'s update, we prepare a new version of the paths matrix ${\bf S}^{(p)}$ and its corresponding source vector $\bf s$. Note that on stage $i$, we interpret  ${\bf S}^{(p)}_{r,0:(i-1)}$ as the single path leading from the roots to node $r$ at the bottom of layer $i$.

\subsubsection{SCL Recursive Definition}
Having defined the decoding paths tracking data structures, we are now ready to describe the SCL algorithms's inputs and outputs. Consider SCL for the $(u+v,v)$ polar code of length $N$ bits with  list size $L$. The inputs of the algorithm are listed below.
\begin{itemize}
\item Two likelihood matrices ${\bf \Pi}^{(0)}$ and ${\bf \Pi}^{(1)}$ of $L\times N$ dimensions. Each row of the matrices corresponds to a different input observation model  option, considered by the decoder. The plurality of  models exists, due to SCL's feature of constantly keeping a list of $L$ decoding-paths representing past decisions on the information word symbols.   Each  decoding-path  induces a different statistical model, in which it is assumed that the information sub-vector,  associated with it, is the one that was transmitted.  We have
    \begin{equation}\label{eq:defOfPi}
        { \Pi}^{(b)}_{i,j} = \Pr\left({  Y}_j^{(i)}={  y}_j^{(i)}|{  V }_j=b\right),
    \end{equation}
     where ${  Y}_j^{(i)}$ is the measurement of the $j^{\text{th}}$ channel ${ V }_j \rightarrow {  Y}_j $ of the  $i^{\text{th}}$ option in the list  and $b\in \{0,1\}$.
\item A scalar $\rho_{in}$ indicating how many rows in ${\bf \Pi}^{(0)}$ and ${\bf \Pi}^{(1)}$ are occupied. The algorithm supports tracking of $\rho_{in} \in [L]$ input models simultaneously.
\item A vector indicator ${\bf z} \in \{0,1\}^N$, in which $z_i=1$ if and only if the $i^{th}$ component of $\bf u$ is frozen.
\end{itemize}

The algorithm outputs the following structures.
\begin{itemize}
\item A matrix $ \hat{\bf U} $ of $L\times N$ dimensions, which represents $L$ arrays of information values (each array of length $N$) - this is the list of the possible information words that the decoder estimated.
\item A matrix $ \hat{\bf X} $ of $L\times N$ dimensions, which represents $L$ arrays of codewords  (each array of length $N$) - this is the list of codewords that correspond to the information words in   $\hat{\bf U}$.
\item A  vector ${\bf s}_{0}^{L-1}$, that indicates for each row in $\hat{ \bf U}$ and $\hat{\bf X}$ to which row in the input ${\bf \Pi}^{(0)}$ and ${\bf \Pi}^{(1)}$ it has originated from (i.e. it refers to the statical model that was assumed when estimating this row).
\item A scalar $\rho_{out}$ indicating how many rows in $  \hat{\bf U}$ or $ \hat{\bf X} $ are occupied.
\end{itemize}

The SCL function signature is defined as
\begin{equation}\label{eq:SCLUVVSign}
\left[\hat{\bf U},\,\,\, \hat{\bf X} ,\,\,\,\, {\bf s},\,\,\,\,\,\rho_{out}\right] = \text{SCLDecoder}\left(\left\{ {\bf \Pi}^{(b)} \right\}_{b\in\{0,1\}},\,\,\,\, \rho_{in},\,\,\,\,\,{\bf z}  \right).
\end{equation}

For  length $N=2$ bits code (i.e. the code induced by a single application of the kernel)  the procedure is described in Algorithm \ref{algo:SCLKernelUVV}.
\begin{algorithm}
\caption{SCL Decoding for the $(u+v,v)$ Kernel}          
\label{algo:SCLKernelUVV}
\begin{description}
\item[\ALGOIN Input: ] $\left\{ {\bf \Pi}^{(b)} \right\}_{b\in\{0,1\}}$; $\rho_{in}$; ${\bf z}$.
 \item[//Initialization:]  Initialize the decoding-paths data structures : ${\bf S}^{(e)}$, ${\bf S}^{(p)}$,   ${\bf s}$, $\hat{\bf U}^{(0)}$, $\hat{\bf X}^{(0)}$, $\hat{\bf U}^{(1)}$ and $\hat{\bf X}^{(1)}$.
   \item[//STEP 0:] \hfill \\
    \ALGOSTEP Generate two $\rho_{in}$ length vectors, ${ \bf p}^{(0)}$ and ${ \bf p}^{(1)}$.
    For each of the $\rho_{in}$ occupied rows of ${\bf \Pi}^{(0)}$ and ${ \bf \Pi}^{(1)}$ compute
    ${  p}^{(0)}_{r}=\frac{1}{2}\left({  \Pi}^{(0)}_{r,0}\cdot{ \Pi}^{(0)}_{r,1}+{ \Pi}^{(1)}_{r,0}\cdot{  \Pi}^{(1)}_{r,1}\right)$  and   ${  p}^{(1)}_{r}=\frac{1}{2}\left({  \Pi}^{(0)}_{r,0}\cdot{  \Pi}^{(1)}_{r,1}+{ \Pi}^{(1)}_{r,0}\cdot{  \Pi}^{(0)}_{r,1}\right)$, for $r \in \left[\rho_{in}\right]_{-}$.

   \item[//STEP 1:] \hfill \\
 \ALGOSTEP Concatenate the two vectors into one $2\cdot \rho_{in}$ length vector, ${\bf p} = [{\bf p}^{(0)} , {\bf p}^{(1)}]$.

\ALGOSTEP Let $\tilde {\bf p}$ be a vector that contains the  $\rho = \min\{2\cdot \rho_{in}, L \}$ largest values of $\bf p$.

 \ALGOSTEP For each $ r \in \left[\rho\right]_{-}$ have $ {   S}^{(e)}_{0,r} =\sigma$ and $ \hat{   U}^{(0)}_{r,0} =\beta$ if and only if the $r^{\text{th}}$ component of  $\tilde{\bf p}$ was originated from ${ p}^{(\beta)}_{\sigma}$. In other words, its source model is $\sigma$ and the decoding tree edge connecting between source model (represented by a node at the top level of the graph) and node $r$ at the bottom of the first layer has label $\beta$.

\underline{REMARK:} If $u$ is frozen (without loss of generality assume that it is set to the 0 value), then steps 0 and 1 can be skipped and $\rho=\rho_{in}$, ${\bf S}^{(e)}_{0,0:\rho_{in}-1} = \left[0,1,...,\rho_{in}-1\right]$ ,$\hat{\bf U}^{(0)} = {\bf 0}$.

 \ALGOSTEP Update ${\bf S}^{(p)}$ and ${\bf s}$ accordingly.
   \item[//STEP 2:] \hfill \\
   Generate two $\rho$ length vectors, ${ \bf p}^{(0)}$ and ${ \bf p}^{(1)}$. For each of the $\rho$ occupied rows of ${\bf S}^{(p)}$ compute ($\forall r\in [\rho]_{-}$).
\begin{equation}
{p}_{r}^{(0)} = \frac{1}{2}\cdot\left\{
                      \begin{array}{ll}
                        {  \Pi}^{(0)}_{{s}_r,0}\cdot{  \Pi}^{(0)}_{{  s}_r,1}, & \hbox{ $\hat{  U}^{(0)}_{r,0}=0$;} \\
                        {  \Pi}^{(1)}_{{  s}_r,0}\cdot{  \Pi}^{(0)}_{{  s}_r,1}, & \hbox{ $\hat{  U}^{(0)}_{r,0}=1$.}
                      \end{array}
                    \right.
\end{equation}

\begin{equation}
{  p}_{r}^{(1)} = \frac{1}{2}\cdot\left\{
                      \begin{array}{ll}
                        {  \Pi}_{{  s}_r,0}^{(1)}\cdot{  \Pi}_{{  s}_r,1}^{(1)}, & \hbox{ $\hat{   U}^{(0)}_{r,0}=0$;} \\
                        { \Pi}_{{ s}_r ,0}^{(0)}\cdot{  \Pi}^{(1)}_{{  s}_r,1}, & \hbox{ $\hat{   U}^{(0)}_{r,0}=1$.}
                      \end{array}
                    \right.
\end{equation}
   \item[//STEP 3:] \hfill \\
\ALGOSTEP Concatenate the two vectors into one $2\cdot \rho $ length vector, ${\bf p} = [{\bf p}^{(0)} , {\bf p}^{(1)}]$.

\ALGOSTEP Let $\tilde {\bf p}$ be a vector that contains the  $\rho_{out} = \min\{2\cdot \rho, L \}$ largest values of $\bf p$.

\ALGOSTEP  For each $r \in \left[\rho_{out}\right]_{-}$ have $ {  S}^{(e)}_{1,r} =\sigma$ and $ \hat{  U}^{(1)}_{r,0} =\beta$ if and only if the $r^{\text{th}}$ component of  $\tilde{\bf p}$ was originated from ${  p}^{(\beta)}_{\sigma}$.

\underline{REMARK:} If the second bit is frozen (without loss of generality assume that it is set to the 0 value), then steps 2 and 3 can be skipped and ${\bf S}^{(e)}_{1,0:\rho-1} = \left[0,1,\ldots, \rho-1\right],\,\,\hat{\bf U}^{(1)}= {\bf 0}, \rho_{out} = \rho$.

\ALGOSTEP Update ${\bf S}^{(p)}$ and ${\bf s}$ accordingly.

\item[\ALGOOUT Output: ]
\begin{itemize}
\item ${\hat{\bf U}_{r\rightarrow}} = [\hat{  U}^{(0)}_{S^{(p)}_{r,1},0}, \,\,\, \hat{ U}^{(1)}_{r,0}],\,\,\,\,\, \forall r\in \left[\rho_{out} \right]_{-}$;
\item $\hat{{\bf X}}= \left[ \hat{\bf U}_{\downarrow 0}+\hat{\bf U}_{\downarrow 1},\,\,\,\,\,  \hat{\bf U}_{\downarrow 1}\right];$
\item ${\bf s};$
\item $\rho_{out}.$
\end{itemize}
 \end{description}
\end{algorithm}
In order to specify the SCL decoder for length $N=2^{n}$ polar code, let us assume that we already developed an  SCL decoder for  length $N/2$  polar code. Using this assumption, a recursive decoder for length $N$ polar code is described in Algorithm \ref{algo:SCLDecoderUVV}.
\begin{algorithm}
\caption{SCL Decoder for Length $N=2^n$ Bits $(u+v,v)$ Polar Code }          
\label{algo:SCLDecoderUVV}                           
 \begin{description}
\item[\ALGOIN Input: ] $\left\{ {\bf \Pi}^{(b)} \right\}_{b\in\{0,1\}}$; $\rho_{in}$; ${\bf z}$.
 \item[//Initialization:]  \ALGOSTEP Initialize the decoding-paths data structures : ${\bf S}^{(e)}$, ${\bf S}^{(p)}$,   ${\bf s}$, $\hat{\bf U}^{(0)}$, $\hat{\bf X}^{(0)}$, $\hat{\bf U}^{(1)}$ and $\hat{\bf X}^{(1)}$.
   \item[//STEP 0:] \hfill \\
   \ALGOSTEP Prepare the probability transition matrices for the first  outer-code decoder.
   Specifically, generate two matrices ${\bf P}^{(b)}$ of dimensions $L\times N/2$, $b\in \{0,1\}$, such that
   \begin{equation}
   {  P}^{(0)}_{r,j}=\frac{1}{2}\left({  \Pi}^{(0)}_{r,2\cdot j}\cdot{  \Pi}^{(0)}_{r,2\cdot j+1}+{  \Pi}^{(1)}_{r,2\cdot j}\cdot{  \Pi}^{(1)}_{r,2\cdot j+1}\right)
   \end{equation}
    and
   \begin{equation}
   {  P}^{(1)}_{r,j}=\frac{1}{2}\left({ \Pi}^{(0)}_{r,2\cdot j}\cdot{ \Pi}^{(1)}_{r,2\cdot j+1}+{  \Pi}^{(1)}_{r,2\cdot j}\cdot{ \Pi}^{(0)}_{r,2\cdot j+1}\right), \,\,\, \forall r \in \left[\rho_{in}\right]_{-}, \forall j\in \left[N/2\right]_{-}
   \end{equation}

   \item[//STEP 1:] \hfill \\
   \ALGOSTEP Decode the first outer-code using the updated channel model matrix, i.e.
   \begin{equation} \label{eq:SCLUVVStep1}
   \left[\hat{\bf U}^{(0)},\,\,\, \hat{\bf X}^{(0)} ,\,\,\,\, {\bf S}^{(e)}_{0\rightarrow},\,\,\,\,\,\rho\right] = \text{SCLDecoder}\left(\left\{ {\bf P}^{(b)} \right\}_{b\in\{0,1\}},\,\,\,\, \rho_{in},\,\,\,\,\,{\bf z}_{0}^{N/2-1}  \right).
   \end{equation}
   \ALGOSTEP Update ${\bf S}^{(p)}$ and $\bf s$ following (\ref{eq:SCLUVVStep1}).

   \item[//STEP 2:] \hfill \\
  \ALGOSTEP Prepare the input matrices for the decoder of the second outer-code of length $N/2$.
     Specifically, generate two matrices ${\bf P}^{(b)}$ of dimensions $L\times N/2$, $b\in \{0,1\}$, such that
   \begin{equation}
   { P}^{(0)}_{r,j}=\frac{1}{2}\cdot\left\{
                               \begin{array}{ll}
                                 { \Pi}^{(0)}_{{ s}_r,2\cdot j}\cdot{ \Pi}^{(0)}_{{ s}_r,2\cdot j+1}, & \hbox{ $\hat{ X}_{r,j}^{(0)}=0$;} \\
                                 { \Pi}^{(1)}_{{ s}_r,2\cdot j}\cdot{ \Pi}^{(0)}_{{ s}_r,2\cdot j+1}, & \hbox{$\hat{ X}_{r,j}^{(0)}=1$,}
                               \end{array}
                             \right.
   \end{equation}
    and
      \begin{equation}
    {P}^{(1)}_{r,j}=\frac{1}{2}\cdot\left\{
                               \begin{array}{ll}
                                 { \Pi}^{(1)}_{ {s}_r,2\cdot j}\cdot{ \Pi}^{(1)}_{{ s}_r,2\cdot j+1}, & \hbox{ $\hat{ X}_{r,j}^{(0)}=0$;} \\
                                 { \Pi}^{(0)}_{{ s}_r,2\cdot j}\cdot{ \Pi}^{(1)}_{{ s}_r,2\cdot j+1}, & \hbox{$\hat{  X}_{r,j}^{(0)}=1$,}
                               \end{array}
                             \right. ,\,\,\,\,\, \forall r\in \left[\rho \right]_{-}, \forall j\in \left[N/2\right]_{-}.
   \end{equation}
   \item[//STEP 3:] \hfill \\
      \ALGOSTEP Decode the second outer-code using the updated channel model matrix, i.e.
   \begin{equation} \label{eq:SCLUVVStep2}
   \left[\hat{\bf U}^{(1)},\,\,\, \hat{\bf X}^{(1)} ,\,\,\,\, {\bf S}^{(e)}_{1\rightarrow},\,\,\,\,\,\rho_{out}\right] = \text{SCLDecoder}\left(\left\{ {\bf P}^{(b)} \right\}_{b\in\{0,1\}},\,\,\,\, \rho,\,\,\,\,\,{\bf z}_{N/2}^{N-1}  \right).
   \end{equation}
   \ALGOSTEP Update ${\bf S}^{(p)}$ and $\bf s$ following (\ref{eq:SCLUVVStep2}).

%

\item[\ALGOOUT Output: ]
\begin{itemize}
\item $\hat{\bf U}_{\rightarrow r}=\left[ \hat{\bf U}^{(0)}_{{ S}^{(p)}_{r,1}\rightarrow },\,\,\,\, \hat{\bf U}^{(1)}_{r}\right],\,\,\,\, \forall  r\in \left[\rho_{out}\right]_{-}$;
\item $\hat{\bf X}_{r,\text{even}}=\hat{\bf X}^{(0)}_{{ S}^{(p)}_{r,1}\rightarrow}+\hat{\bf X}^{(1)}_{r\rightarrow}$ and $\hat{\bf X}_{r,\text{odd}}=\hat{\bf X}^{(1)}_{ {r} \rightarrow}\,\,\,\,,  \forall r\in \left[\rho_{out}\right]_{-}$;
\item $\bf s$;
\item $\rho_{out}$.
\end{itemize}
Here  $\hat{\bf X}_{r,\text{even}}$ ($\hat{\bf X}_{r,\text{odd}}$) are the vectors of the  even (odd) indices columns of row number $r$ in  matrix $\hat{\bf X}$.
\end{description}
\end{algorithm}
Let $T(n)$ be the decoding time complexity, for length $N=2^n$ bits polar code. Then $T(n)= 2\cdot T(n-1) + O(L\cdot N)$, and $T(1)=O(L)$, which results in $T(n)=O(L\cdot N\cdot\log N)$. Similarly, the space complexity of the algorithm can be shown to be $O(L\cdot N)$.

The generalization of the decoding algorithm for a homogenous kernel of  $\ell$ dimensions with alphabet $F$ is quite straight-forward. Here we emphasize the principal changes, from the $(u+v,v)$ case. Firstly, the only change in the input is that we should have $|F|$  channel matrices, ${\bf \Pi}^{(b)}$, one for each alphabet symbol $b\in F$. With this change in alphabet the definition of each matrix in (\ref{eq:defOfPi}) remains. Consequently, the function signature is defined as follows.
\begin{equation}\label{eq:SCLGenSign}
\left[\hat{\bf U},\,\,\, \hat{\bf X} ,\,\,\,\, {\bf s},\,\,\,\,\,\rho_{out}\right] = \text{SCLDecoder}\left(\left\{ {\bf \Pi}^{(b)} \right\}_{b\in F},\,\,\,\, \rho_{in},\,\,\,\,\,{\bf z}  \right).
\end{equation}

In the decoding algorithm, we have $\ell$ pairs of steps, such that each one is dedicated to a different outer-code. Before reaching step $2\cdot i -1$, we already decoded outer-codes  $\left\{\mathcal{C}_{m}\right\}_{m=0}^{i-1}$. Using the decoding tree terminology, we can say that we have traversed $i$ layers of the tree (starting from the roots) generating at most $L$ decoding-paths. As a result we have the paths tracking data structures ${\bf S}^{(e)}$, ${\bf S}^{(p)}$,   ${\bf s}$, $\left\{\hat{\bf U}^{(m)}\right\}_{m=0}^{i-1}$ and $\left\{\hat{\bf X}^{(m)}\right\}_{m=0}^{i-1}$ updated and describing the possible paths, that reach  nodes at the top of the $i^{th}$ layer. Algorithm  \ref{algo:SCLDecodingOuterCodeCr} elaborates  on steps $2\cdot i $ and $2\cdot i+1$ which find the sequel to the $L$ paths in layer $i $ of the tree. The output generation of SCL is described in Algorithm \ref{algo:SCLDecodingOutput}.

\begin{algorithm}
\caption{SCL Decoding Steps Dedicated for Outer-Code $\mathcal{C}_i,\,\,\,\, i\in [\ell]_{-}$ }          
\label{algo:SCLDecodingOuterCodeCr}                           
//Let $\rho$ be set to the number of active nodes at the top of layer $i$. For $i=0$ set $\rho=\rho_{in}$.
\begin{description}
\item[STEP $2\cdot i$] \hfill \\
\ALGOSTEP Using the decoding results of the outer-codewords from the previous steps i.e. $\hat{\bf X}^{(m)}$, for $m\in \left[i-1\right]_{-}$, prepare the $N/{\ell}$ length likelihood lists,  $\left\{{\bf P}^{(b)}\right\}_{b\in F}$. Each item in the list is an $L\times N/{\ell}$ matrix, and all of them will serve as   inputs to the decoder of the $N/{\ell}$ length outer-code  $\#i$. For the computation of row $r$ of ${\bf P}^{(b)}$, use the input statistical model $s_r$, that is the likelihoods in rows $\left\{{\bf \Pi}^{(b)}_{s_r \rightarrow }\right\}_{b\in F}$.
\begin{equation}\label{eq:piCalcSCLGen}
P^{(b)}_{r,j}= \sum_{{\bf x}\in {\mathcal A} ^{(r,j,b)}}\Pi_{s_r,j\cdot \ell}^{(x_0)}\cdot \Pi_{s_r,j\cdot \ell+1}^{(x_1)}\cdot \Pi_{s_r,j\cdot \ell+2}^{(x_2)}\cdot \ldots \cdot \Pi_{s_r,(j+1)\cdot \ell-1}^{(x_{\ell-1})}, \,\,\,\, \forall r\in \left[\rho\right]_{-},\forall j\in \left[N/\ell\right]_{-},
\end{equation}
where ${\mathcal{ A}} ^{(r,j,b)}$ is defined to be the set of all possible codewords ${\bf c} = g({\bf v})$  of the inner-code (defined by the kernel $g(\cdot)$), having $v_i = b$, and  the prefix ${\bf v}_{0}^{i-1}$ defined by the $r^{\text{th}}$ decoding-path edge labels. Note that the $r^{\text{th}}$ decoding path nodes are ${\boldsymbol \sigma} = \left[{\bf S}^{(p)}_{r,0:(i-1)}\,\,\,,\,\,\,\,\, r\right]$ and their corresponding $j^{\text{th}}$ inner-code information prefix is ${\bf v} = \left[ \hat{X}_{\sigma_{m+1},j}^{(m)} \right]_{m=0}^{i-1}$. Consequently we have,
\begin{equation}\label{eq:cosetDefintion}
{\mathcal A} ^{(r,j,b)} \triangleq \left\{g({\bf v}) \left| {\bf v}_{i+1}^{\ell-1}\in F^{\ell-1-i} \bigwedge v_i = b \bigwedge v_{i-1}=X^{(i-1)}_{r,j }\bigwedge v_m = \hat{X}^{(m)}_{S^{(p)}_{r,m+1},j} \text{ where } m\in [i-2]_{-}   \right.\right\}.
\end{equation}

   \item[STEP $2\cdot i+1$] \hfill \\
     \ALGOSTEP SCL decode  the $i^{th}$ outer-code using the updated channel model matrix, i.e.

   \begin{equation} \label{eq:SCLGenStepr}
   \left[\hat{\bf U}^{(i)},\,\,\, \hat{\bf X}^{(i)} ,\,\,\,\, {\bf S}^{(e)}_{i\rightarrow},\,\,\,\,\,\rho\right] = \text{SCLDecoder}\left(\left\{ {\bf P}^{(b)} \right\}_{b\in F},\,\,\,\, \rho,\,\,\,\,\,{\bf z}_{i\cdot N/\ell}^{(i+1)\cdot N/\ell-1}  \right).
   \end{equation}

 \ALGOSTEP  Update ${\bf S}^{(p)}$ and $\bf s$ following (\ref{eq:SCLGenStepr}).

\end{description}
\end{algorithm}

\begin{algorithm}
\caption{SCL Decoding Algorithm Output Generation }          
\label{algo:SCLDecodingOutput}                           
\begin{description}
\item[\ALGOOUT Output: ] (occurs after applying Algorithm \ref{algo:SCLDecodingOuterCodeCr}  for all $i\in [\ell]_{-}$)\hfill\\
\begin{itemize}
\item $\hat{\bf U}_{\rightarrow r}=\left[ \hat{\bf U}^{(0)}_{{ S}^{(p)}_{r,1}\rightarrow },\,\,\hat{\bf U}^{(1)}_{{ S}^{(p)}_{r,2}\rightarrow },\ldots , \hat{\bf U}^{(\ell-2)}_{{ S}^{(p)}_{r,\ell-1}\rightarrow },\,\, \,\,\hat{\bf U}^{(\ell-1)}_{r}\right]$,\,\,\,\,  $ \forall r\in \left[\rho\right]_{-}$;
\item $\hat{\bf X}_{r, \left(j\cdot\ell\right):\left((j+1)\cdot\ell-1\right)}= g\left(\hat{\bf X}^{(0)}_{{ S}^{(p)}_{r,1}\rightarrow },\,\,\hat{\bf X}^{(1)}_{{ S}^{(p)}_{r,2}\rightarrow },\ldots, \hat{\bf X}^{(\ell-2)}_{{ S}^{(p)}_{r,\ell-1}\rightarrow },\,\, \,\,\hat{\bf X}^{(\ell-1)}_{r}\right)$,  $\,\,\,\, \forall r\in \left[\rho\right]_{-},\,\,\,\,\forall j\in \left[N/\ell\right]_{-}$;
\item $\bf s$;
\item $\rho_{out}=\rho$.
\end{itemize}
\end{description}
\end{algorithm}

 The decoder for the basic $N={\ell}$ length code  also contains $\ell$ pairs of steps. The procedure is similar to Algorithm \ref{algo:SCLDecodingOuterCodeCr}. However, instead of delivering the likelihood matrices $\left\{{\bf P}^{(b)}\right\}_{b\in F}$ (here these matrices are actually column vectors) to an outer-code decoder, we concatenate them to a vector $\tilde{{\bf p}}$ and choose the $\rho = \min\left\{L,|F|\cdot \rho\right\}$ maximal elements from it. Following this selection we update the decoding path tracking structures. This is a generalization of the case of  $N=2$  decoder in the $(u+v,v)$ construction.

 In case the kernel is mixed, the generalization is also quite easy. Let us consider the mixed-kernels example, from the end of Subsection \ref{sec:recSCDec}. The only changes we have in the decoding algorithm, are for the pair of steps in Algorithm \ref{algo:SCLDecodingOuterCodeCr} associated with the glued outer-code $\mathcal{C}_{0,1}$. In step $3$ (the preparation step for this outer-code), we prepare $|F|^{2}$ input matrices ${\bf P}^{(b_1,b_2)}$, for all $(b_1,b_2)\in F^2$. In order to do this, we modify  equations (\ref{eq:piCalcSCLGen}) and (\ref{eq:cosetDefintion})  replacing $b$ with the pair $(b_1,b_2)$, corresponding to $v_1$ and $v_2$ in (\ref{eq:cosetDefintion}). The decoder of $\mathcal{C}_{1,2}$ is supposed to return a list of estimations of the information words, their corresponding codewords and the model indicator vectors. These outputs and the temporary structures are re-organized, as is done in   step $2\cdot r$ for the  decoding algorithm of the homogenous kernel polar code. Note, however, that at the end of step $3$, there are three information words lists $\hat{\bf U}^{(0)}$, $\hat{\bf U}^{(1)}$ and $\hat{\bf U}^{(2)}$ along with their corresponding three outer-code codewords lists. This is because we have decoded $\mathcal{C}_{1,2}$'s glued symbols simultaneously, which resulted in retrieving  $\hat{\bf U}^{(1)}$, $\hat{\bf U}^{(2)}$, $\hat{\bf X}^{(1)}$ and $\hat{\bf X}^{(2)}$ in a single decoding step.

 \subsection{A Recursive Description of the BP Algorithm}\label{sec:BP}
BP is an iterative message-passing  decoding algorithm, which messages are sent over  Forney's normal factor graph \cite{Forney01}. Although being  an alternative to SC decoding \cite{Arikan} there is no evidence which algorithm has better performance over general channels, except for the BEC, in which BP is shown to outperform SC \cite{Hussami2009}. Simulations, however, suggest that in many cases BP outperforms SC. On the other hand, SCL with small list size $L$ outperforms BP in many cases.

The order of sending the messages on the graph is called the \textit{schedule} of the algorithm. Hussami \textit{et al.} suggested employing the "$Z$ shape schedule" for transferring the messages \cite[Section II.A]{Hussami2009}.  In this correspondence we introduce a serial schedule which is induced from the GCC structure of the code.

We begin by describing the types of messages that are computed throughout the algorithm for the $(u+v,v)$ polar code. Figure \ref{fig:GCCWithLayerUVVPart} depicts the normal factor graph representation of Arikan's kernel. We have four symbol half edges denoted by $u,v,x_0$ and $x_1$. These symbols have the following functional dependencies among them:
 $x_0 = u+v$ and $x_1=v$. The messages and the inputs that are sent on the graph are assumed to be LLRs, and their values are taken from $\mathbb{R}\bigcup\{\pm\infty\}$. The $\infty$ and $-\infty$ are special types of LLR values that indicate known assignment of $0$ and $1$, respectively. They are used to support the existence of the polar code's frozen symbols.

  We   associate four input LLR messages with the symbols half edges. These messages may be generated by the output of the channel, by known values associated with frozen bits   or by computations that were done in this iteration or previous ones. We represent these messages by $\mu^{(in)}_{u}$, $\mu^{(in)}_{v}$, $\mu^{(in)}_{x_0}$ and $\mu^{(in)}_{x_1}$. The algorithm computes four output LLR messages, $\mu^{(out)}_{u}$, $\mu^{(out)}_{v}$, $\mu^{(out)}_{x_0}$ and $ \mu^{(out)}_{x_1}$, indicating the estimations of $u,v,x_0$ and $x_1$, respectively,  by the decoding algorithm. The messages are computed according to the \textit{extrinsic information principle}, i.e. each message that is sent from a node on an adjacent edge is  a function of all the messages that were previously sent to the node, except the message that was received over the particular edge. The nodes of the graphs are denoted by $a_0$ (the adder functional) and $e_1$ (the equality functional). Using the ideas mentioned above we have the following computation rules.
\begin{equation}\label{eq:BPUV1}
\mu_{e_1  \rightarrow a_0 }=f_{(=)}(\mu^{(in)}_{x_1},\mu^{(in)}_{v}),
\end{equation}
\begin{equation}\label{eq:BPUV2}
\mu_{a_0 \rightarrow e_1}=f_{(+)}(\mu^{(in)}_{x_0},\mu^{(in)}_{u}),
\end{equation}
\begin{equation}\label{eq:BPUV3}
\mu^{(out)}_{u}=f_{(+)}(\mu^{(in)}_{x_0},\mu_{e_1  \rightarrow a_0 }),
\end{equation}
\begin{equation}\label{eq:BPUV4}
\mu^{(out)}_{v}=f_{(=)}(\mu^{(in)}_{x_1},\mu_{a_0 \rightarrow e_1}),
\end{equation}
\begin{equation}\label{eq:BPUV5}
\mu^{(out)}_{x_0}=f_{(+)}(\mu^{(in)}_{u},\mu_{e_1  \rightarrow a_0 }),
\end{equation}
\begin{equation}\label{eq:BPUV6}
\mu^{(out)}_{x_1}=f_{(=)}(\mu^{(in)}_{v},\mu_{a_0  \rightarrow e_1 }),
\end{equation}
where $f_{(=)}(z_0,z_1) \triangleq z_0+z_1$ and $f_{(+)}(z_0,z_1)  \triangleq 2\tanh^{-1}\left(\tanh(z_0/2)\cdot\tanh(z_1/2)\right)$. We denote by $\mu_{\alpha\rightarrow\beta}$ where $\alpha,\beta\in\{e_1,a_0\}$ the message sent from node $\alpha$ to node $\beta$. $\mu^{(out)}_{u}$ and $\mu^{(out)}_{x_0}$ are sent from $a_0$ over the half edges corresponding to symbols $u$ and $x_0$, respectively. $\mu^{(out)}_{v}$ and $\mu^{(out)}_{x_1}$ are sent from $e_1$ over the half edges corresponding to symbols  $v$ and $x_1$, respectively. Note that
\begin{equation}
f_{(=)}(\pm \infty,z_1)=f_{(=)}(z_0,\pm \infty)=\pm \infty
\end{equation}
\begin{equation}
f_{(+)}(\pm \infty,z_1)=\pm z_1,\,\,\,\,f_{(+)}(z_0,\pm \infty)=\pm z_0.
\end{equation}

We now turn to give a  recursive description of an iteration of the algorithm. As depicted in Figure \ref{fig:GCCWithLayer} the factor graph of the length $N$ bits code, has $\log_2N$ layers. In each layer, there exist $N/2$ copies of the kernel normal factor graph, depicted in Figure \ref{fig:GCCWithLayerUVVPart}.
As a consequence, for each layer, we have $N/2$  realizations of each type of input messages, output messages and inner messages (each one is corresponding to a different set of symbols and interconnect). To denote the $i^{\text{th}}$ realization of these messages, we use the notation $\mu_{\alpha\rightarrow\beta,i}$, $\mu_{\gamma,i}^{(in)}$ and $\mu_{\gamma,i}^{(out)}$, where $\alpha,\beta \in \{a_0,e_1\}$ and $\gamma\in \{x_0,x_1,u,v\}$.
As before, we denote the channel LLRs by the   $N$ length vector ${\boldsymbol \lambda}$. Each input message or inner message, unless given (by the channel output or by a prior knowledge on the frozen bits) is set to $0$ before the first iteration. It is assumed that the inner messages are preserved between the iterations (and see a further discussion in the sequel).

Let us  describe the BP decoder inputs and outputs for the $(u+v,v)$ code of length $N$ bits. The inputs of the algorithm are the following.
\begin{itemize}
\item  An $N$ length vector of input LLRs, $\boldsymbol \lambda$, containing the observation from the channel.
\item  A pointer to a matrix $ {\bf M}^{(u,in)}$ of $N \times \log_2(N)$ dimensions, which is used to hold the $\mu_{v}^{(in)}$ and  $\mu_{u}^{(in)}$ messages between iterations. We employ a pointer here, because we would like to be able to change the values of the matrix as the algorithm progresses.
\item  A vector indicator ${\bf z} \in \{0,1\}^N$, in which $z_i=1$ if and only if   the $i^{th}$ component  of the information vector $\bf u$ is frozen.
\end{itemize}
The algorithm outputs the following structures.

\begin{itemize}
\item An $N$ length binary vector $ \hat{\bf u} $ containing the information word that the decoder estimated (including its frozen symbols).
\item An $N$ length vector $\hat{\bf x}$  containing the LLRs for the estimated codeword symbols. This structure is used to store the $\mu_{x_0}^{(out)}$ and $\mu_{x_1}^{(out)}$ messages.
\end{itemize}
The BP function signature is defined as follows
\begin{equation}\label{eq:BPSignature}
\left[\hat{\bf u},\,\,\, \hat{\bf x} \right] = \text{BPDecoder}\left({\boldsymbol \lambda},\,\,\,\, {\bf M}^{(u,in)},\,\,\,\,  {\bf z}  \right).
\end{equation}

\begin{algorithm}
\caption{BP Decoder of Length $N=2^n$ Bits $(u+v,v)$ Polar Code }          
\label{algo:BPUVVLengthN}                           
\begin{description}
\item[\ALGOIN Input: ] ${\boldsymbol \lambda}$; ${\bf M}^{(u,in)}$; ${\bf z} $.
   \item[//Initializations:] \hfill \\
   We use the following aliasing for the inputs of the algorithm.
   $$ \mu_{x_0,r}^{(in)}:\equiv \lambda_{2r}\,\,\,\text{ and }\,\,\,\,\, \mu_{x_1,r}^{(in)}:\equiv\lambda_{2r+1} ,\,\,\,\,\,\, \forall r \in \left[N/2\right]_{-};$$
   $$ \mu_{u,r}^{(in)}:\equiv{ M}^{(u,in)}_{2r,0}\,\,\,\text{ and }\,\,\,\,\,   \mu_{v,r}^{(in)}:\equiv{ M}^{(u,in)}_{2r+1,0},\,\,\,\,\,\, \forall r \in \left[N/2\right]_{-} ;$$
   \item[//STEP 0:] \hfill \\
     \ALGOSTEP Compute   messages  $\left[\mu_{e_1 \rightarrow a_0,r}\right]_{r=0}^{N/2-1}$ using (\ref{eq:BPUV1}).

     \ALGOSTEP Compute   messages $\left[\mu_{u,r}^{(out)} \right]_{r=0}^{N/2-1}$ using (\ref{eq:BPUV3}).
   \item[//STEP 1:] \hfill \\
   \ALGOSTEP Perform an iteration on the first outer-code: give the vector $\left[\mu_{u,r}^{(out)} \right]_{r=0}^{N/2-1}$ as an input to the  polar code BP iterative decoder of length $N/2$ bits. Also provide the indices of the frozen bits from the first half of the codeword. The decoder outputs an estimation of the first outer-code codeword to  $\left[\mu_{u,r}^{(in)}\right]_{r=0}^{N/2-1}$ (manifested as LLRs) and an estimation of its information word to the binary vector $\hat{\bf u}^{(0)}$, i.e.
       \begin{equation}\label{eq:BPSignatureCall}
    \left[\hat{\bf u}^{(0)},\,\,\, \left[\mu_{u,r}^{(in)}\right]_{r=0}^{N/2-1} \right] = \text{BPDecoder}\left(\left[\mu_{u,r}^{(out)} \right]_{r=0}^{N/2-1},\,\,\,\, {\bf M}^{(u,in)}_{0:(N/2-1),1:\left(\log_2(N)-1\right)},\,\,\,\,  {\bf z}_0^{N/2-1}  \right).
    \end{equation}
   \item[//STEP 2:] \hfill \\
    \ALGOSTEP Compute the messages  $\left[\mu_{a_0 \rightarrow e_1,r}\right]_{r=0}^{N/2-1}$ using (\ref{eq:BPUV2}).

    \ALGOSTEP Compute the messages  $\left[\mu_{v,r}^{(out)} \right]_{r=0}^{N/2-1}$ using (\ref{eq:BPUV4}) (Note that these two steps can be combined into  one computation).
   \item[//STEP 3:] \hfill \\
   \ALGOSTEP  Perform an iteration on the second outer-code: give the vector $\left[\mu_{v,r}^{(out)} \right]_{r=0}^{N/2-1}$ as an input to the  polar code BP iterative decoder of length $N/2$. Also provide the indices of the frozen bits from the second half of the codeword. The decoder outputs an estimation of the second outer-code codeword to  $\left[\mu_{v,r}^{(in)}\right]_{r=0}^{N/2-1}$ (manifested as LLRs) and an estimation of its information word to the binary vector $\hat{\bf u}^{(1)}$, i.e.
       \begin{equation}\label{eq:BPSignatureCall}
    \left[\hat{\bf u}^{(1)},\,\,\, \left[\mu_{v,r}^{(in)}\right]_{r=0}^{N/2-1} \right] = \text{BPDecoder}\left(\left[\mu_{v,r}^{(out)} \right]_{r=0}^{N/2-1},\,\,\,\, {\bf M}^{(u,in)}_{N/2:(N-1),1:\left(\log_2(N)-1\right)},\,\,\,\,  {\bf z}_{N/2}^{N-1}  \right).
    \end{equation}
 \ALGOSTEP  Compute  messages  $\left[\mu_{e_1 \rightarrow a_0,r}\right]_{r=0}^{N/2-1}$ using (\ref{eq:BPUV1}).

 \ALGOSTEP Compute  messages  $\left[\mu_{x_0,r}^{(out)}\right]_{r=0}^{N/2-1}$ and $\left[\mu_{x_1,r}^{(out)}\right]_{r=0}^{N/2-1}$  using (\ref{eq:BPUV5}) and (\ref{eq:BPUV6}), respectively.

\item[\ALGOOUT Output: ]
\begin{itemize}
\item $\hat{\bf u} =\left[ \hat{\bf u}^{(0)}, \,\,\, \hat{\bf u}^{(1)} \right]$;
\item $\hat{ x}_{2\cdot r}= \mu_{x_0,r}^{(out)} ;\,\,\,\,   \hat{  x}_{2\cdot r+1}= \mu_{x_1,r}^{(out)}, \,\,\,\,\,\,\, \forall r\in \left[N/2\right]_{-}$.
\end{itemize}
 \end{description}
\end{algorithm}

Algorithm \ref{algo:BPUVVLengthN} outlines the BP iteration for length $N>2$ code. Algorithm \ref{algo:BPUVVLenN2} completes this recursive description by considering the case of length $N=2$ bits code.
Note that in the algorithms we use aliases for  several of our inputs in order to improve the procedure readability. We say that $s$ is an alias for a variable $w$ (and denote it by $s:\equiv w$), if $s$ is an alternative name for the memory space of $w$, and therefore any algorithmic operation  on $w$ has the   same results and side-effects as performing the operation on $s$.

\begin{algorithm}
\caption{BP Decoder for Length $N=2$ Bits $(u+v,v)$ Polar Code }          
\label{algo:BPUVVLenN2}                           
\begin{description}
\item[\ALGOIN Input: ] ${\boldsymbol \lambda}$; ${\bf M}^{(u,in)}$; ${\bf z} $.
\item[//Initializations:] \hfill \\
  \ALGOSTEP  We use the following aliasing to the inputs of the algorithm.
   $$ \mu_{x_0}^{(in)} :\equiv \lambda_{0} \,\,\,\text{ and }\,\,\,\,\,\mu_{x_1}^{(in)} :\equiv \lambda_{1} ;$$

\ALGOSTEP Initialize the $u,v$ input LLR messages (we assume that  frozen variables are fixed to the zero value)
\begin{equation}
\mu_{w}^{(in)}=\left\{
                 \begin{array}{ll}
                   0, & \hbox{$w$ is not frozen;} \\
                   \infty, & \hbox{$w$ is frozen.}
                 \end{array}
               \right.\,\,\,\,\,\,\,\,\,\,\,\,\,\, \forall w\in \{u,v\}.
\end{equation}
   \item[//STEP 0:] \hfill \\
 \ALGOSTEP Compute $\mu_{e_1  \rightarrow a_0 }$ according to (\ref{eq:BPUV1}).
 \item[//STEP 1:] \hfill \\
  \ALGOSTEP If $u$ is not frozen, compute $\mu_{u}^{(out)}$ according to (\ref{eq:BPUV3}), and make a hard decision  on this bit, based on its sign (denote it by $\hat{u}$). Otherwise, $\hat{u} = 0$.
 \item[//STEP 2:] \hfill \\
 \ALGOSTEP Compute $\mu_{a_0  \rightarrow e_1 }$ according to (\ref{eq:BPUV2}).
 \item[//STEP 3:] \hfill \\
 \ALGOSTEP If $v$ is not frozen, compute $\mu_{v}^{(out)}$ according to (\ref{eq:BPUV4}), and make a hard decision on it,  based on its sign (denote it by $\hat{v}$). Otherwise, $\hat{v} = 0$.

\ALGOSTEP Compute $\mu_{x_0}^{(out)}$ and $\mu_{x_1}^{(out)}$ according to (\ref{eq:BPUV5}) and (\ref{eq:BPUV6}), respectively.

\item[\ALGOOUT Output: ]
\begin{itemize}
\item $\hat{\bf u} =\left[ \hat{u}, \,\,\, \hat{v} \right]$;
\item $\hat{\bf x} = \left[\mu_{x_0}^{(out)},\,\,\, \mu_{x_1}^{(out)}\right] $.
\end{itemize}
\end{description}
\end{algorithm}

General schedules of BP may require to hold   a dedicated memory for storing
 $\mu_{u}^{(in)}$, $\mu_{v}^{(in)}$, $\mu_{x_0}^{(in)}$, $\mu_{x_1}^{(in)}$ and $\mu_{a_0\rightarrow e_1}$ type of messages that were previously computed. This memory may be needed for each realization of such messages, specifically, for each layer of the graph and for each $(u+v,v)$ normal subgraph, as in Figure \ref{fig:GCCWithLayerUVVPart}. However, for our GCC schedule, excluding $\mu_{v}^{(in)}$,  we do not need to save any message beyond the iteration boundary. This is because that in each  iteration, all the messages except $\mu_{v}^{(in)}$ are re-computed before their first usage (in the iteration). The implication of this observation is that the required memory consumption can be reduced (see Subsection \ref{sec:UVLineDecoderBP}). Furthermore, the memory used for the other messages is temporary and needed only for the same iteration. It can be seen that the memory for these temporary messages is linear in the block length. The requirement to keep all the $\mu_{v}^{(in)}$ type of messages beyond the iteration boundary of the algorithm results in memory consumption of $\Theta\left(N\cdot \log (N)\right)$.

In each iteration, we send one instance for each of the possible messages and for each $(u+v,v)$ block realization in the code, except for the $\mu_{e_1\rightarrow a_0}$ type of message for which we send two messages (for all the layers, besides the last one). Consequently the iteration time complexity is $\Theta\left(N\cdot\log(N)\right)$.

A complete BP implementation may require several iterations. The number of iterations may be fixed  or set adaptively, which means that the algorithm continues until some consistency constraints are satisfied. An example for such a constraint, is that the signs of the LLR estimations for all the frozen bits agree with their know values (i.e. if all the frozen bits are set to zero, then  $\mu^{(out)}_{w}>0$ for all the frozen bits, $w$). In this case, it is possible to stop an iteration in the middle by keeping a counter in a similar way to the method that is usually used in BP decoding of LDPC codes using the check-node based serial schedules (see e.g. \cite{Sharon07}). We note, however, that in the LDPC case,  the consistency is manifested by the fact that all the parity check equations are satisfied.

\begin{figure}
\centering
\begin{subfigure}{.5\textwidth}
  \centering
\includegraphics[scale = 0.12]{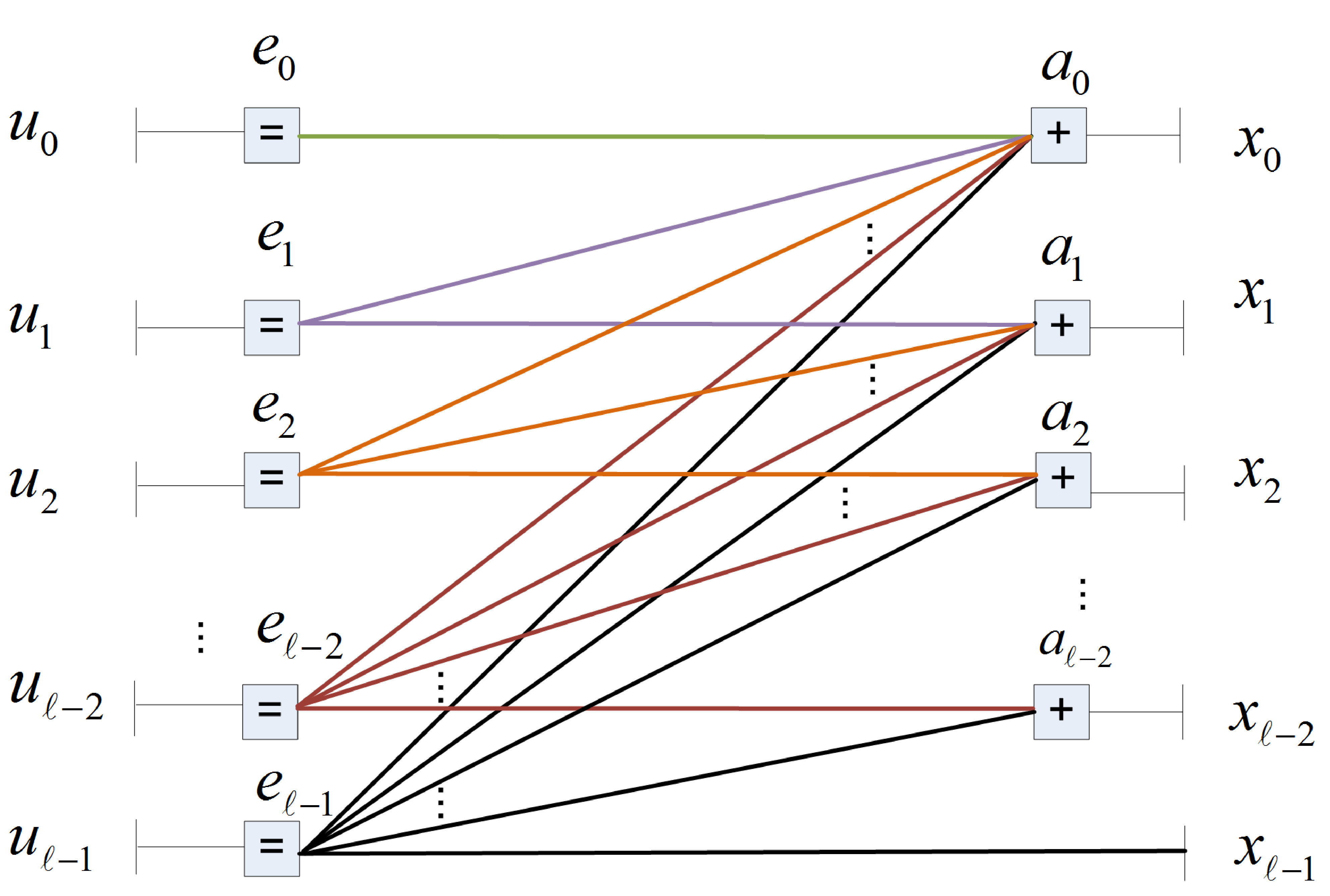}\\
  \caption{A binary linear kernel with lower triangular generating matrix }\label{fig:normFactorLDimension}
\end{subfigure}%
\begin{subfigure}{.5\textwidth}
  \centering
  \includegraphics[scale = 0.14]{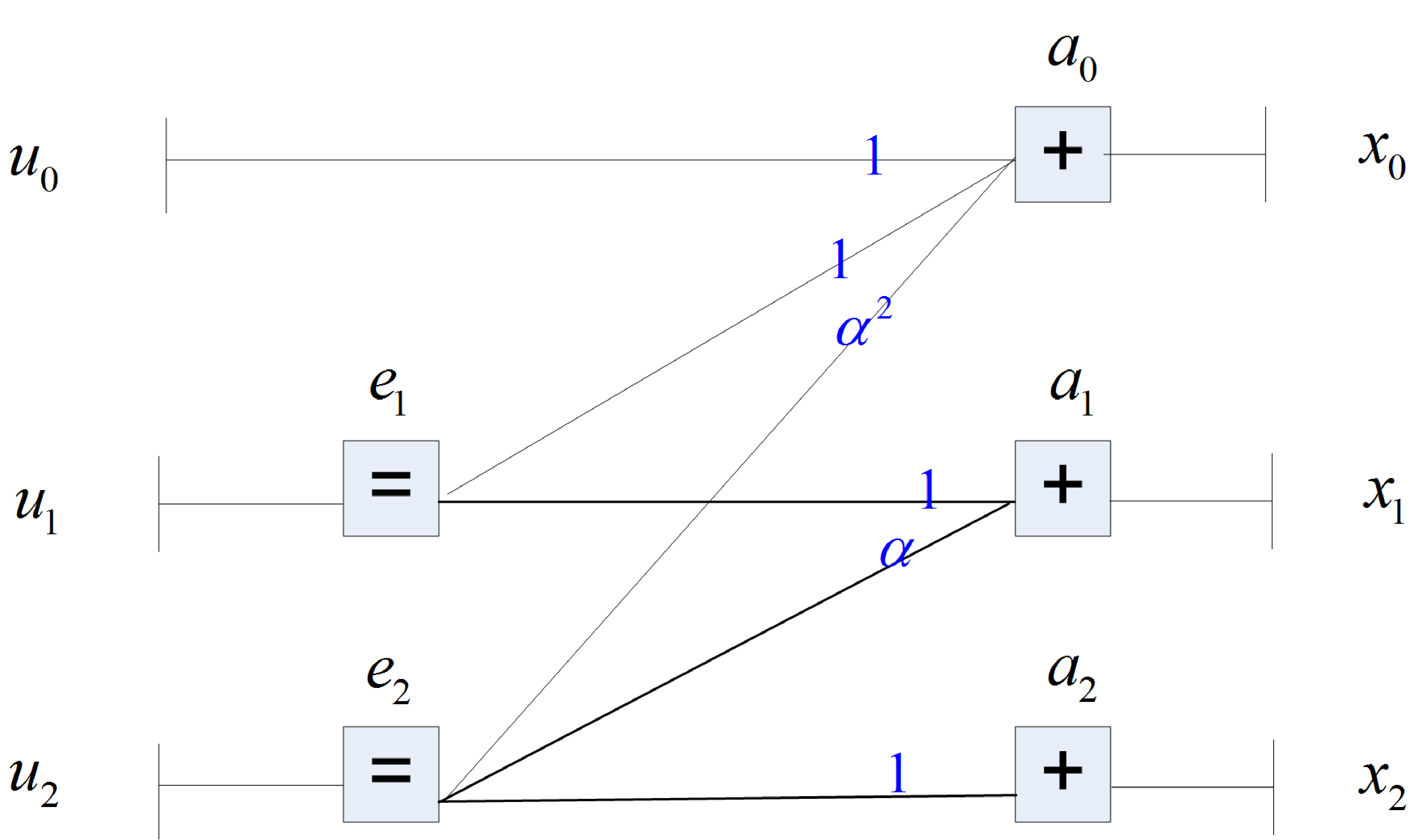}\\
  \caption{The $RS3$ kernel defined in (\ref{eq:RS3GenMat}) }\label{fig:normFactorGraphRS4}
\end{subfigure}
\caption{Normal factor graphs representations of polar codes kernels}\label{fig:normFactGraphRepGen}
\end{figure}

Let us now consider BP for polar codes with general kernels. For this description we require the kernels to be linear and be represented by a lower triangular generating matrix.  The input to the kernel and the output of the kernel are  $\ell$ length vectors $\bf u$ and $\bf x\in F^{\ell}$, respectively, satisfying ${\bf x } = g\left({\bf u}\right) =  {\bf u}\cdot {\bf G}$, where $\bf G$ is an $\ell\times\ell$ lower triangular generating matrix. Figure \ref{fig:normFactorLDimension} depicts a normal factor graph for such an $\ell$ dimensions binary kernel. We have that an edge $e_{i}\rightarrow a_j$ exists in the graph if and only if $G_{i,j}\neq 0$. Being a lower triangular matrix means that in the factor graph there are no edges $e_{i}\rightarrow a_j$ , such that $j>i$. In case the kernel is non-binary, each edge $e_{i}\rightarrow a_j$ also has a label equal to its $G_{i,j}$ value. For example, Figure \ref{fig:normFactorGraphRS4} depicts the normal factor graph corresponding to the $RS3$ kernel, with generating matrix
\begin{equation}\label{eq:RS3GenMat}
{\bf G}_{RS3}=\left[
          \begin{array}{cccc}
            1 & 0 & 0  \\
            1 & 1 & 0  \\
            \alpha^2 & \alpha & 1  \\
          \end{array}
        \right].
\end{equation}

Similarly to the discussion on the $(u+v,v)$ code, we define input messages to the factor graph denoted by $\mu_{u_i}^{(in)(t)}$ and  $\mu_{x_j}^{(in)(t)}$ and their corresponding output messages $\mu_{u_i}^{(out)(t)}$ and  $\mu_{x_j}^{(out)(t)}$ where $i,j\in \left[\ell\right]_{-}$ and $t\in F\backslash\{0\}$. Moreover, we have messages  $\mu_{e_i\rightarrow a_j}^{(t)}$ and $\mu_{a_j \rightarrow e_i }^{(t)}$ for every edge $e_i \rightarrow a_j$. All of these messages are LLRs  such that for a message $\mu^{(t)}$ we have $\mu^{(t)} = \ln\left(\frac{\Pr\left({\bf y}\left| \omega = 0 \right.\right)}{\Pr\left({\bf y}\left| \omega = t \right.\right)}\right)$, where $\bf y$ is a vector of observations, and $\omega$ is the variable associated with the edge on which the message $\mu^{(t)}$ is transmitted. In case the code is binary the  letter indication $t$ may be omitted.  All the messages are calculated using the (typically false) assumption that the factor graph is  cycle free, and consequently for each node the messages sent to it are  statistically independent.   Let $a_j$ be an adder node corresponding to column $j$ of the generating matrix $G_{\downarrow j}$, such that $\sum_{i=j}^{\ell-1} G_{i,j} \cdot  u_j = x_j$. We have for $i,j\in [\ell]_{-}$ and $i\geq j$
\begin{equation}\label{eq:BPGenEiAj}
\mu_{e_i \rightarrow  a_j}^{(t)} = f_{(=)}\left(t,\left[\left[\mu_{a_r \rightarrow  e_i}^{(\tau)} \right]_{\tau\in F\backslash \{0\}}\right]_{r=0,r\neq i}^{j},\left[ \mu_{u_i}^{(in)(\tau)} \right]_{\tau\in F\backslash \{0\}}\right);
\end{equation}
\begin{equation}\label{eq:BPGenAjEi}
\mu_{a_j \rightarrow e_i }^{(t)} = f_{(+)}\left(t,\left[\left[\mu_{e_r \rightarrow a_j }^{(\tau)} \right]_{\tau \in F\backslash \{0\}}\right]_{r=j,r\neq i}^{\ell-1},\left[ \mu_{x_j}^{(in)(\tau)} \right]_{\tau\in F\backslash \{0\}},G_{i,j}^{-1}\cdot \left[\left[G_{r,j} \right]_{r=j,r\neq i}^{\ell-1},\,\,\,1 \right]\right);
\end{equation}
\begin{equation}\label{eq:BPGenUOut}
\mu_{u_i}^{(out)(t)} = f_{(=)}\left(t,\left[\left[\mu_{a_r \rightarrow  e_i}^{(\tau)} \right]_{\tau\in F\backslash \{0\}}\right]_{r=0}^{j}\right);
\end{equation}
\begin{equation}\label{eq:BPGenXOut}
\mu_{x_j}^{(out)(t)} = f_{(+)}\left(t,\left[\left[\mu_{e_r \rightarrow a_j }^{(\tau)} \right]_{\tau \in F\backslash \{0\}}\right]_{r=j,}^{\ell-1}, \left[G_{r,j} \right]_{r=j}^{\ell-1} \right).
\end{equation}

The functions $f_{(=)}(\cdot)$ and $f_{(+)}(\cdot)$ are  generalizations of the functions that were presented before for the $(u+v,v)$ case. Note that for these functions, the number of input arguments that follow $t$   (the alphabet symbol)  may vary. This number is equal to the degree of the sending node  minus one. Consequently, we denoted them in (\ref{eq:BPGenEiAj})-(\ref{eq:BPGenXOut}) as vector of vectors (i.e. using the $\left[\left[\mu^{(\tau)}_r\right]_{\tau \in F\backslash \{0\}}\right]_{r\in \mathcal{B}}$ notation), however it should be understood that each element of this vector of vectors is a different argument to the functions.

\begin{figure}
\centering
\begin{subfigure}{.5\textwidth}
  \centering
  \includegraphics[scale = 0.075]{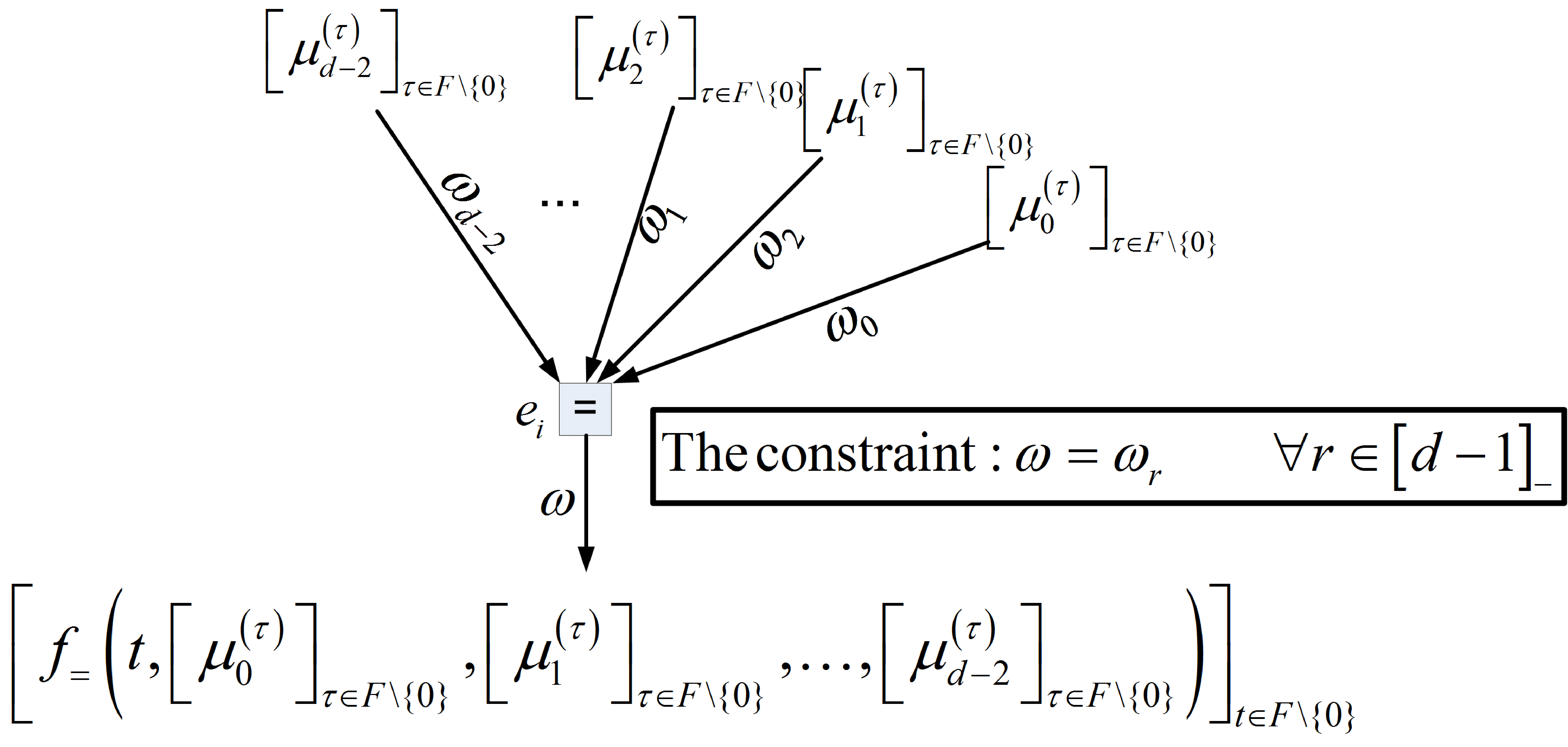}
  \caption{Equality node messages}
  \label{fig:messagesInBPEqualityConstraint}
\end{subfigure}%
\begin{subfigure}{.5\textwidth}
  \centering
  \includegraphics[scale = 0.075]{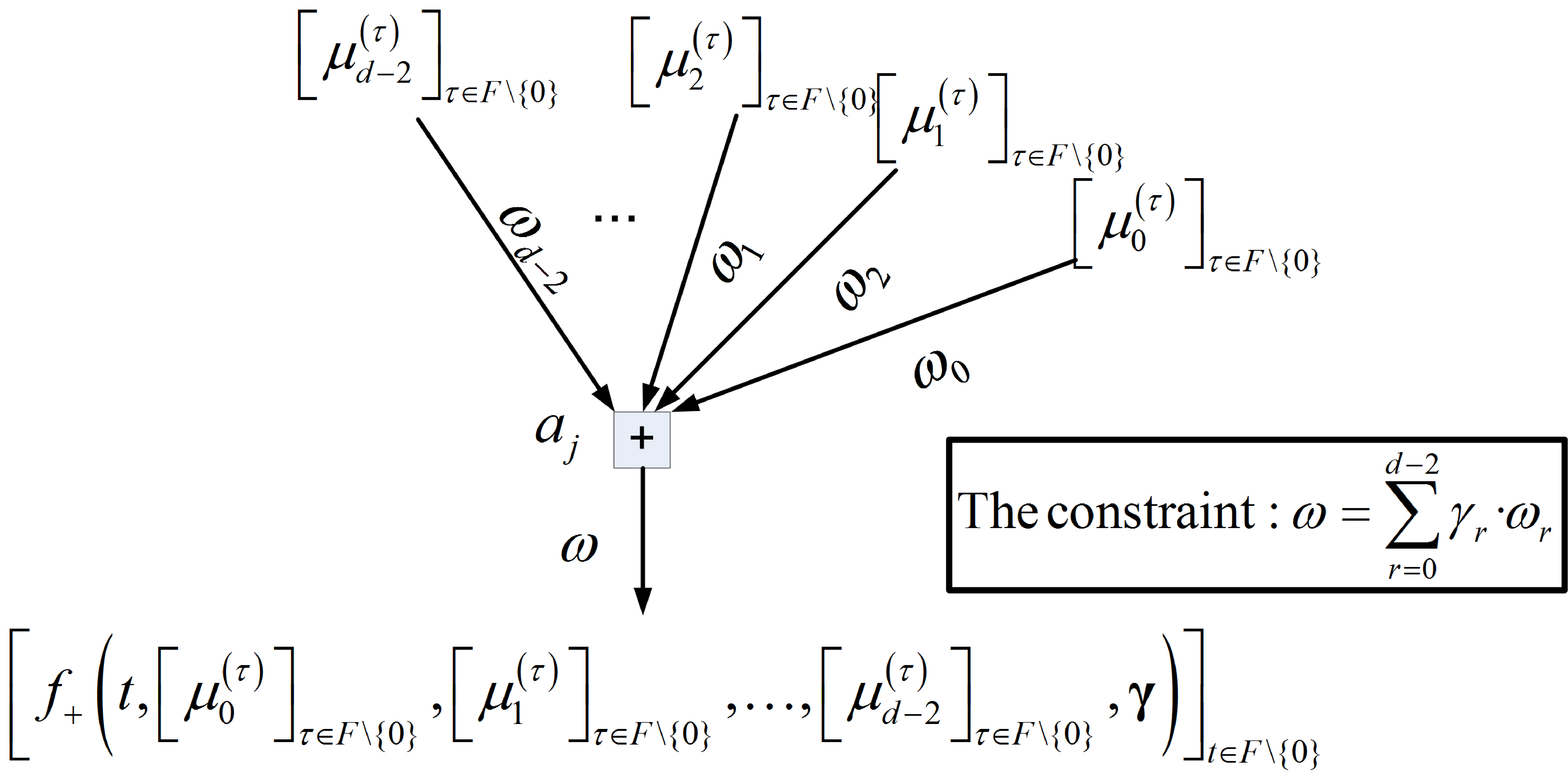}
  \caption{Adder node messages}
  \label{fig:messagesInBPAdderConstraint}
\end{subfigure}
\caption{Messages of BP algorithm}\label{fig:messagesInBP}
\label{fig:test}
\end{figure}

Given an equality node $e_i$ with degree $d$, the function $f_{(=)}(\cdot)$ is defined as follows
\begin{equation}\label{eq:fEqDefinedGen}
f_{(=)}\left(t, \left[\mu_{0 }^{(\tau)} \right]_{\tau\in F\backslash \{0\}},\left[\mu_{1 }^{(\tau)} \right]_{\tau\in F\backslash \{0\}},\ldots,\left[\mu_{d-2 }^{(\tau)} \right]_{\tau\in F\backslash \{0\}} \right) \triangleq \sum_{r=0}^{d-2} \mu_{r}^{(t)},
\end{equation}
where $t\in F\backslash\{0\}$ and $\left[\mu_{r }^{(\tau)}\right]_{\tau\in F\backslash \{0\}}$  are LLR messages received at the node from  $d-1$ edges adjacent to it. Denote the variable associated with these edges by $\omega_r$ for $r\in[d-1]_{-}$ and let $\omega$ denote the variable associated with the  edge which messages were not given as input (we  refer to this edge as the "missing edge"). The output of this function is the LLR message sent from $e_i$ on the missing edge. Being a repetition constraint ($\omega = \omega_r$ for all $r\in[d-1]_{-}$) the LLR calculated in (\ref{eq:fEqDefinedGen}) appears as a summation of the LLRs corresponding to the same alphabet letter $t$. See Figure \ref{fig:messagesInBPEqualityConstraint} for an illustration of this case.

Given an adder node $a_j$ with degree $d$, the function $f_{(+)}(\cdot)$ is defined as follows
\begin{equation}\label{eq:fSumDefinedGen}
f_{(+)}\left(t, \left[\mu_{0 }^{(\tau)} \right]_{\tau\in F\backslash \{0\}},\left[\mu_{1 }^{(\tau)} \right]_{\tau\in F\backslash \{0\}},\ldots,\left[\mu_{d-2 }^{(\tau)} \right]_{\tau\in F\backslash \{0\}}, {\boldsymbol \gamma} \right)  \triangleq \ln\left(\frac{\sum_{{\boldsymbol \omega}_0^{d-2} \in \mathcal{A}^{\left({\boldsymbol \gamma},0\right)}} \exp \left( -\sum_{r=0}^{d-2}\mu_r^{(\omega_r)} \right)}{\sum_{{\boldsymbol \omega}_0^{d-2} \in \mathcal{A}^{\left({\boldsymbol \gamma},t\right)}} \exp \left(- \sum_{r=0}^{d-2}\mu_r^{(\omega_r)} \right)}\right)
\end{equation}
where $t\in F\backslash\{0\}$, $\mu_r^{(0)}\triangleq 0$ and $\left[\mu_{r }^{(\tau)}\right]_{\tau\in F\backslash \{0\}}$  are LLR messages received at the node from   $d-1$ edges adjacent to it,  $r\in [d-1]_{-}$. Let us denote the variables corresponding to these edges by $\left\{\omega_r\right\}_{r=0}^{d-2}$ and the variable corresponding to the missing edge by $\omega$. It is assumed that the generating matrix equation corresponding to node $a_j$ is
 \begin{equation}\label{eq:summationNode}
 \omega = \sum_{r=0}^{d-2}\gamma_r\cdot \omega_r.
 \end{equation}
Consequently, the set $\mathcal{A}^{\left({\boldsymbol \gamma},t\right)}$ is defined as the set of all assignments to ${\boldsymbol \omega}_{0}^{d-2}$, such that $\omega=t$ in (\ref{eq:summationNode}), i.e. $\mathcal{A}^{\left({\boldsymbol \gamma},t\right)} = \left\{ {\boldsymbol \omega}_0^{d-2} | \sum_{r=0}^{d-2}\gamma_r\cdot \omega_r = t \right\}$. See Figure \ref{fig:messagesInBPAdderConstraint} for an illustration of this case. Note that naive calculation of (\ref{eq:fSumDefinedGen}) for all $t\in F\backslash\{0\}$ has time complexity of  $O( |F|^d)$  while computation  that uses trellis has complexity of $O(d \cdot |F|^2)$.

We are now ready to describe the BP algorithm for general lower triangular kernels. The inputs of the algorithm are as follows.
\begin{itemize}
\item   $N$ length vectors of input LLRs, $\left\{\boldsymbol \lambda ^{(t) }\right\}_{t\in F\backslash \{0\}}$, containing the observations of the channel ($t$ indicates the code alphabet letters).
\item  Pointers to matrices $\left\{\left\{{\bf M}^{(u_i,in)(t)}\right\}_{t\in F\backslash \{0\}}\right\}_{i\in [\ell]_{-}}$ of $N/{\ell} \times \log_{\ell}(N)$ dimensions, which are used to hold the $\mu_{u_i}^{(in)(t)}$  messages between iterations. Pointers are employed here, because we would like to be able to change the values of the matrix as the algorithm progresses.
\item  Pointers to matrices $\left\{\left\{{\bf M}^{(a_j \rightarrow e_i)(t)}\right\}_{t\in F\backslash \{0\}}\right\}_{0 \leq j\leq i\leq \ell-1 }$ of $N/{\ell} \times \log_{\ell}(N)$ dimensions, which are used to hold the $\mu_{a_j \rightarrow e_i}^{(t)}$  messages between iterations.
\item  Vector indicator ${\bf z} \in \{0,1\}^N$, in which $z_i=1$ if and only if the $i^{th}$  component of the information vector $\bf u$ is frozen.
\end{itemize}
The algorithm outputs the following structures.
\begin{itemize}
\item An $N$ length vector $ \hat{\bf u} \in F^N $, containing the information vector that the decoder estimated (including its frozen symbols).
\item  $N$ length vectors $\left\{\hat{\bf x}^{(t)}\right\}_{t\in F\backslash\{0\}}$ which are the LLRs for the estimated codeword. This structure is used for delivering the $\mu_{x_j}^{(out)(t)}$ messages.
\end{itemize}
\normalsize
The BP function signature is defined as follows (note that the $i,j$ in the third argument are limited such that $0 \leq j\leq i\leq \ell-1$)
   \begin{equation} \label{eq:BPSignatureGeneral}
   \left[\hat{\bf u},\,\,\, \left\{\hat{\bf x}^{(t)}\right\}_{t\in F\backslash\{0\}} \right] = \text{BPDecoder}\left(\left\{ {\boldsymbol \lambda }^{(t)} \right\}_{t\in F\backslash \{0\}},\,\,\,\, \left\{{\bf M}^{(u,in)(t)}\right\}_{t\in F\backslash \{0\}},\,\,\,\left\{\left\{{\bf M}^{(a_j \rightarrow e_i)(t)}\right\}_{t\in F\backslash \{0\}}\right\}_{i,j}, \,\,{\bf z}  \right).
   \end{equation}

We start with the $N=\ell$ symbols case. Algorithm \ref{algo:GenBPSchemeForNl} gives the description for this case. Algorithm \ref{algo:BPGenLengthN} consider the $N>\ell$ symbols case.

\begin{algorithm}
\caption{BP Decoder for Length $N=\ell$ $F$-Symbols Polar Code }          
\label{algo:GenBPSchemeForNl}                           
\begin{description}
\item[\ALGOIN Input: ] $\left\{ {\boldsymbol \lambda }^{(t)} \right\}_{t\in F\backslash \{0\}}$; $\left\{{\bf M}^{(u,in)(t)}\right\}_{t\in F\backslash \{0\}}$; $\left\{\left\{{\bf M}^{(a_j \rightarrow e_i)(t)}\right\}_{t\in F\backslash \{0\}}\right\}_{i,j}$; ${\bf z}$.
\item[//Initializations:] \hfill \\
   \ALGOSTEP  Use the following aliases to the inputs of the algorithm.
   $$ \left[ \mu_{x_j}^{(in)(t)}\right]_{t\in F\backslash\{0\}}:\equiv\left[ \lambda^{(t)}_j\right]_{t\in F\backslash\{0\}}, \,\,\,\,\,\,\,\,\,\forall j\in [\ell]_{-};$$
   $$
        \mu_{a_j\rightarrow e_i}^{(t)} :\equiv M_{0,0}^{(a_j\rightarrow e_i)(t)},\,\,\,\,\,\,\,\,\,\, \forall t\in F\backslash\{0\},\,\,\,\,\,\,\,\, \forall 0 \leq j\leq i\leq \ell-1.
   $$

\ALGOSTEP Initialize the vector $\left[\mu_{u_i}^{(in)(t)}\right]_{t\in F\backslash\{0\}}$
    $$ \mu_{u_i}^{(in)(t)}= \left\{
                             \begin{array}{ll}
                               0, & \hbox{$z_i = 0$;} \\
                               \infty, & \hbox{$z_i \neq  0$.}
                             \end{array}
                           \right. \,\,\,\,\,\, \forall i\in [\ell]_{-},\,\,\,\,\,\, \forall t\in F\backslash \{0\}.
   $$

 \item[//Iteration:]\hfill \\
  \ALGOSTEP \textbf{For} $j = \ell-1$ to $0$ \textbf{Do}
        \begin{itemize}
                \item Compute $\left[ \mu_{e_i  \rightarrow a_j }^{(t)}\right]_{t\in F\backslash\{0\}}, \,\,\,\, \forall i,\,\,\,\, \text{s.t.} \,\,\,\, j < i \leq \ell-1$ using (\ref{eq:BPGenEiAj});
                \item Compute $\left[ \mu_{a_j\rightarrow e_j}^{(t)}\right]_{t\in F\backslash\{0\}}$ using (\ref{eq:BPGenAjEi}).
        \end{itemize}
    \ALGOSTEP \textbf{For} $i = 0 $ to $\ell-1$ \textbf{Do}
        \begin{itemize}
                \item Compute $\left[ \mu_{a_j   \rightarrow  e_i}^{(t)}\right]_{t\in F\backslash\{0\}}, \,\,\,\, \forall j, \,\,\,\, \text{s.t.} \,\,\,\, 0 \leq j <i$  using (\ref{eq:BPGenAjEi});
                \item If $u_i$ is not frozen, compute $\left[\mu_{u_i}^{(out)(t)}\right]_{t\in F\backslash \{0\}}$ according to (\ref{eq:BPGenUOut}), and make a hard decision  on this symbol, based on the LLR vector (denote the hard decision  by $\hat{u}_i$). If $u_i$ is frozen, set $\hat{u}_i=0$;
                \item Compute $\left[ \mu_{e_i\rightarrow a_j}^{(t)}\right]_{t\in F\backslash\{0\}} , \,\,\,\, \forall j,\,\,\,\, \text{s.t.} \,\,\,\, 0 \leq j <i$  using (\ref{eq:BPGenEiAj}).
        \end{itemize}
    \ALGOSTEP Compute $\left[ \mu_{x_j}^{(t)}\right]_{t\in F\backslash\{0\}}, \,\,\,\, \forall j\in [\ell]_{-}$ using (\ref{eq:BPGenXOut}).
    \item[ \ALGOOUT Output:]
\begin{itemize}
\item $\hat{\bf u} =\left[ \hat{u}_0, \,\,\, \hat{u}_1,  \ldots ,\hat{u}_{\ell -1} \right]$;
\item $\hat{\bf x} = \left[\left[\mu_{x_0}^{(out)(t)}\right]_{t\in F\backslash\{0\}},\,\,\,  \left[\mu_{x_1}^{(out)(t)}\right]_{t\in F\backslash\{0\}},\ldots, \left[\mu_{x_{\ell-1}}^{(out)(t)}\right]_{t\in F\backslash\{0\}} \right] $.
\end{itemize}
\end{description}
\end{algorithm}

\begin{algorithm}
\caption{BP Decoder of Length $N=\ell^n$ $F$-Symbols   Polar Code }          
\label{algo:BPGenLengthN}                           
\begin{description}
\item[\ALGOIN Input: ] $\left\{ {\boldsymbol \lambda }^{(t)} \right\}_{t\in F\backslash \{0\}}$; $\left\{{\bf M}^{(u,in)(t)}\right\}_{t\in F\backslash \{0\}}$; $\left\{\left\{{\bf M}^{(a_j \rightarrow e_i)(t)}\right\}_{t\in F\backslash \{0\}}\right\}_{i,j}$; ${\bf z}$.
   \item[//Initializations:] \hfill \\
   \ALGOSTEP  Use the following aliases to the inputs of the algorithm.
   $$ \left[ \mu_{x_j,r}^{(in)(t)}\right]_{t\in F\backslash\{0\}}:\equiv\left[ \lambda^{(t)}_{r\cdot\ell+j}\right]_{t\in F\backslash\{0\}} \,\,\,\forall j\in [\ell]_{-} \text{ and } \forall r\in [N/\ell]_{-} ;$$
    $$ \left[ \mu_{u_i,r}^{(in)(t)}\right]_{t\in F\backslash\{0\}}:\equiv\left[ M_{r,0}^{(u_i)(t)}\right]_{t\in F\backslash\{0\}} \,\,\,\forall i\in [\ell]_{-} \text{ and } \forall r\in [N/\ell]_{-} ;$$
   $$
        \mu_{a_j\rightarrow e_i,r}^{(t)} :\equiv M_{r,0}^{(a_j\rightarrow e_i)(t)}\,\,\,\,\,\,\,\,\,\, \forall t\in F\backslash\{0\},\,\,\,\, \forall 0 \leq j\leq i\leq \ell-1 \text{ and } \forall r\in [N/\ell]_{-} .
   $$

 \item[//Iteration:]\hfill \\
  \ALGOSTEP \textbf{For} $j = \ell-1$ to $0$ \textbf{Do}
        \begin{itemize}
                \item Compute $\left[ \mu_{e_i  \rightarrow a_j,r }^{(t)}\right]_{t\in F\backslash\{0\}}, \,\,\,\, \forall i,\,\,\,\, \text{s.t.} \,\,\,\, j < i \leq \ell-1  \text{ and } \forall r\in [N/\ell]_{-} $ using (\ref{eq:BPGenEiAj});
                \item Compute $\left[ \mu_{a_j\rightarrow e_j,r}^{(t)}\right]_{t\in F\backslash\{0\}} \text{ and } \forall r\in [N/\ell]_{-} $ using (\ref{eq:BPGenAjEi}).
        \end{itemize}

    \ALGOSTEP \textbf{For} $i = 0 $ to $\ell-1$ \textbf{Do}
        \begin{itemize}
            \item Run steps $2\cdot i$  and $2\cdot i+1$ of Algorithm \ref{algo:BPIterationStepsForCi}.
        \end{itemize}
    \ALGOSTEP Compute $\left[ \mu_{x_j}^{(out)(t)}\right]_{t\in F\backslash\{0\}}, \,\,\,\, \forall j\in [\ell]_{-}$ using (\ref{eq:BPGenXOut}).

    \item[\ALGOOUT Output:]
\begin{itemize}
\item $\hat{\bf u} =\left[ \hat{\bf u}^{(0)}, \,\,\, \hat{\bf u}^{(1)},  \ldots, \hat{\bf u}^{(\ell-1)}\right]$;
\item $\hat{ x}_{r\cdot \ell+j}^{(t)}= \mu_{x_j,r}^{(out)(t)},\,\,\,\,\, \forall j\in \left[\ell\right]_{-},\,\,\,\,\,\,\forall t\in F\backslash\{0\} \text{ and } \forall r\in \left[N/\ell\right]_{-}$.
\end{itemize}
 \end{description}
\end{algorithm}

\begin{algorithm}
\caption{BP Iterations Steps Dedicated for Decoding of Outer-Code $\mathcal{C}_i,\,\,\,\, i\in [\ell]_{-}$ }          
\label{algo:BPIterationStepsForCi}                           
\begin{description}
\item[//STEP $2\cdot i$:] \hfill\\
                \ALGOSTEP Compute $\left[ \mu_{a_j   \rightarrow  e_i,r}^{(t)}\right]_{t\in F\backslash\{0\}}, \,\,\,\, \forall j, \,\,\,\, \text{s.t.} \,\,\,\, 0 \leq j <i \text{ and } \forall r\in [N/\ell]_{-}$  using (\ref{eq:BPGenAjEi}).

\ALGOSTEP Compute $\left[\mu_{u_i,r}^{(out)(t)}\right]_{t\in F\backslash \{0\}},\,\,\,\,\,\forall r\in [N/\ell]_{-}$ according to (\ref{eq:BPGenUOut}).
\item[//STEP $2\cdot i+1$:] \hfill\\
\ALGOSTEP Give the vector $\left\{\left[\mu_{u_i,r}^{(out)(t)} \right]_{r=0}^{N/\ell-1}  \right\}_{t\in F\backslash\{0\}}$ as an input to the  polar code decoder of length $N/\ell$ symbols. Also provide to this decoder the indices of the  frozen symbols corresponding to $\mathcal{C}_i$ and pointers to the matrices containing the messages of this outer-code. Assume that the decoder outputs $\left[\left[\mu_{u_i,r}^{(in)(t)} \right]_{t\in F\backslash\{0\}}\right]_{r=0}^{N/\ell-1}$ and the estimation of the information word of $\mathcal{C}_i$, i.e.

$$
 \left[\hat{\bf u}^{(i)},\,\,\,\left[\left[\mu_{u_i,r}^{(in)(t)} \right]_{t\in F\backslash\{0\}}\right]_{r=0}^{N/\ell-1} \right]  =\text{BPDecoder}\left(\left\{\left[\mu_{u_i,r}^{(out)(t)} \right]_{r=0}^{N/\ell-1}  \right\}_{t\in F\backslash\{0\}}\right.,\,\,\,\,
$$
   \begin{equation} \label{eq:BPGeneralCall}
 \left. \left\{{\bf M}^{(u,in)(t)}_{i\cdot N/{\ell}:\left((i+1)\cdot N/{\ell}-1\right) ,1:\left(\log_{\ell}N-1\right)}\right\}_{t\in F\backslash \{0\}},\,\,\,\left\{\left\{{\bf M}^{(a_{j'} \rightarrow e_{i'})(t)}_{i\cdot N/{\ell}:\left((i+1)\cdot N/{\ell}-1\right) ,1:\left(\log_{\ell}N-1\right)}\right\}_{t\in F\backslash \{0\}}\right\}_{i',j'}, \,\,{\bf z}_{i \cdot N/{\ell}}^{(i+1)\cdot N/{\ell}-1}  \right).
   \end{equation}
// Note that  $i',j'$ in the third argument of (\ref{eq:BPGeneralCall}) are limited such that $0 \leq j'\leq i'\leq \ell-1$.

\ALGOSTEP Compute $\left[ \mu_{e_i\rightarrow a_j,r}^{(t)}\right]_{t\in F\backslash\{0\}} , \,\,\,\, \forall j,\,\,\,\, \text{s.t.} \,\,\,\, 0 \leq j <i \text{ and } \forall r\in [N/\ell]_{-}$  using (\ref{eq:BPGenAjEi}).
\end{description}
\end{algorithm}

Thus far we discussed homogenous kernels. BP on mixed-kernels polar codes can be defined in a similar manner. In mixed-kernels structures we have at least two types of constituent kernels, each one with different alphabet. In order to connect these kernels, we combine several input symbols of the first kernel and consider them as a single entity for decoding purposes.   We say that these symbols are "glued" together, thereby creating a symbol of the larger-alphabet kernel. The output symbols of the larger alphabet size kernel  are given as input to the glued input entry of the  inner mapping defined by the first kernel. In order to support this gluing operation we introduce an additional node to the normal factor graph, and label it by the '$\&$' symbol.   This node serves as a "bridge" between the two alphabets.

\begin{example}[BP on Mixed-Kernels]\label{ex:BPMixed}
Let us consider the mixed-kernels code  discussed in Example \ref{ex:MxdKernelsExample}. In this example we use the ${\bf G}=\left[\begin{array}{cc}
                                                                1 & 0 \\
                                                                1 & 1
                                                              \end{array}\right]^{\bigotimes 2}$ binary matrix as our first kernel and glue its input components $u_1$ and $u_2\in GF(2)$ into one entity called $u_{(1,2)} \in GF(4)$. The second kernel is the $RS4$ kernel described by the generating matrix (\ref{eq:RS4GenMat}). All the BP messages sent  over the edges of this  kernel and the $RS4$ kernel were already discussed above, except the ones sent and received by the  $\&_{(1,2)}$ node. Note that the correspondance between the binary representation of $u_{(1,2)}$ and its representation over $GF(4)$ is as follows:  $[u_2,u_1] =[0,0] \equiv 0$; $[0,1] \equiv 1$; $[1,0] \equiv \alpha$ and $[1,1] \equiv \alpha^2$.

\begin{equation}\label{eq:MxdKernelsBP1}
{\mu_{{\&_{(1,2)}} \rightarrow {e_1}}} = \ln \left( {\frac{{\exp \left\{ { - {\mu_
{{e_2} \rightarrow }}_{{\&_{(1,2)}}} - \mu_{{u_{(1,2)}}}^{\left( {in} \right)\left( \alpha  \right)}}
\right\}}+1}{{\exp \left\{ { - {\mu_{{e_2} \rightarrow }}_{{\&_{(1,2)}}} - \mu_{{u_{(1,2)}}}^{\left( {in}
\right)\left( {{\alpha^2}} \right)}} \right\} + \exp \left\{ { - \mu_{{u_{(1,2)}}}^{\left( {in}
\right)\left( 1 \right)}} \right\}}}} \right)
\end{equation}
\begin{equation}\label{eq:MxdKernelsBP2}
{\mu_{{\&_{(1,2)}}
\rightarrow {e_2}}} = \ln \left( {\frac{{\exp \left\{ { - {\mu_{{e_1} \rightarrow }}_{{\&_{(1,2)}}} - \mu_{{u_
{(1,2)}}}^{\left( {in} \right)\left( 1 \right)}} \right\}}+1}{{\exp \left\{ { - {\mu_{{e_1} \rightarrow }}
_{{\&_{(1,2)}}} - \mu_{{u_{(1,2)}}}^{\left( {in} \right)\left( {{\alpha^2}} \right)}} \right\} +
\exp \left\{ { - \mu_{{u_{(1,2)}}}^{\left( {in} \right)\left( \alpha  \right)}} \right\}}}}
\right)
\end{equation}

\begin{equation}\label{eq:MxdKernelsBP3}
\mu_{{u_{(1,2)}}}^{\left( {out} \right)\left( \alpha\right)} = {\mu_{{e_2} \rightarrow {\&_{(1,2)}}}}
\end{equation}
\begin{equation}\label{eq:MxdKernelsBP4}
\mu_{{u_{(1,2)}}}^{\left({out} \right)\left( 1 \right)} = {\mu_{{e_1} \rightarrow {\&_{(1,2)}}}}
\end{equation}
\begin{equation}\label{eq:MxdKernelsBP5}
\mu_{{u_{(1,2)}}}^{\left( {out} \right)\left( {{\alpha^2}} \right)} = {\mu_{{e_1} \rightarrow {\&_{(1,2)}}}} + {\mu _{{e_2} \rightarrow {\&_{(1,2)}}}}
\end{equation}

We use the following aliases between the messages mentioned in (\ref{eq:MxdKernelsBP1})- (\ref{eq:MxdKernelsBP5}) and the messages of the standard homogenous kernel defined in (\ref{eq:BPGenEiAj})- (\ref{eq:BPGenXOut}): $ \mu_{u_i}^{(in)}:\equiv {\mu_{{\&_{(1,2)}} \rightarrow {e_i}}} $; $ \mu_{u_i}^{(out)}:\equiv {\mu_{ {e_i}  \rightarrow {\&_{(1,2)}}}} $; for $i\in\{1,2\}$.
The BP schedule suggested in Algorithm \ref{algo:BPGenLengthN}  is preserved, i.e. each iteration starts in an initialization step and then moves to  BP decoding of its outer-codes. Messages (\ref{eq:MxdKernelsBP3})-(\ref{eq:MxdKernelsBP5}) are computed before calling the BP decoder of the $RS4$ outer-code, in order to convert binary LLRs into quaternary ones. Moreover, messages  (\ref{eq:MxdKernelsBP1}) and  (\ref{eq:MxdKernelsBP2}) are employed after the BP iteration on the $RS4$ outer-code has finished, in order to convert the quaternary LLRs into binary ones.
\end{example}

 \begin{figure}
\center
  \includegraphics[scale = 0.15]{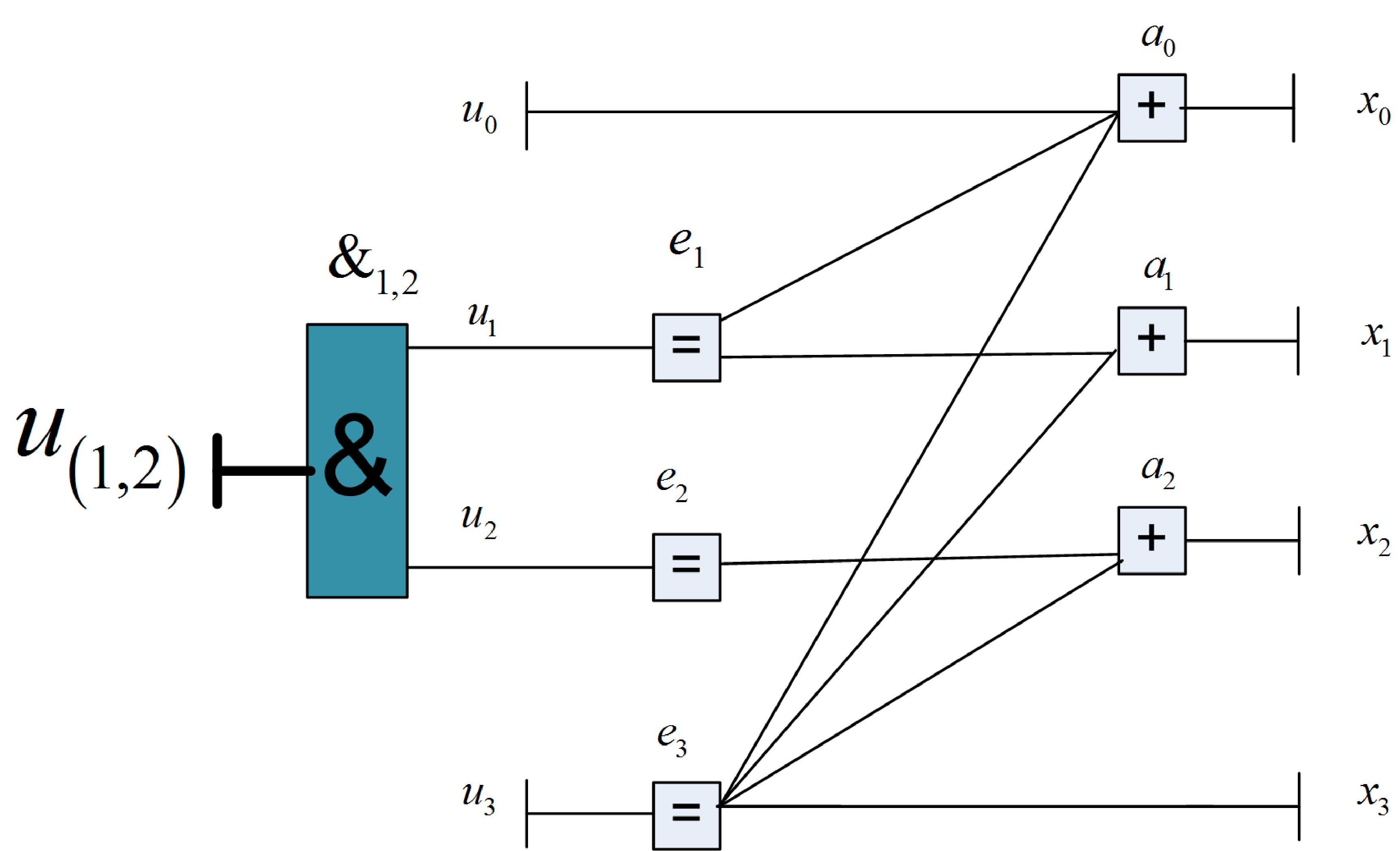}\\
  \caption{Normal factor graph representation for the first kernel of Example \ref{ex:BPMixed}. This kernel is constructed by gluing inputs $u_{1},u_2$ of the mapping defined by the generating matrix $\bf G$. }\label{fig:normFactorGraphGen}
\end{figure}


In the next section we describe  architectures implementing  the decoding algorithms we covered so far.

 \section{Recursive Descriptions of Polar Code Decoders Hardware Architectures}\label{sec:HrdwreArchi}

In this section  we study schematic architectures that are induced from the recursive decoding algorithms presented in Section \ref{sec:RecDescOfDecAlgor}. Indeed most of the algorithmic details were given in that section, therefore the purpose of our discussion here is to consider aspects of hardware algorithms, such as possible parallelism, scheduling and memory resources managements. Note, however, that throughout the  discussion, our presentation is relatively abstract, emphasizing the important concepts and features of the recursive designs without dwelling into all the specifics. Consequently, the figures representing the block diagrams should not be considered as  full detailed specifications of the implementation, but rather as illustrations that aim to guide the reader in the task of designing the decoder.


Throughout this section we use the same notations for signals array and registers arrays. Let $u(0:N-1)$ be an $N$ length   signals array. We denote its $i^{\text{th}}$ component by $u(i)$. If $v$ is a two dimensional array (i.e. a matrix) of $L$ rows and $N$ columns, we denote it by  $v(0:M-1,0:N-1)$. Naturally, the $i^{\text{th}}$ row of this array is denoted by $v(i,0:N-1)$, and it is a one dimensional array of $N$ elements, of which the $j^{\text{th}}$ element is denoted by $v(i,j)$.

\subsection{Arikan's Construction Decoders}\label{sec:HrdwreArikConstr}
 This subsection covers  architectures for Arikan's $(u+v,v)$ construction. Generalizations of this discussion for other polar code types are presented in Subsection \ref{sec:HardArchiForOthKer}. We begin by the simple SC pipeline decoder (Subsection \ref{sec:SCPipeUV}), and then proceed to the more efficient SC line decoder (Subsection \ref{sec:UVLineDecoder}). Both of these designs were previously presented by Leroux \textit{et al.} \cite{Leroux10,Leroux2012} in a non-recursive fashion. We conclude by introducing a  BP line decoder (Subsection \ref{sec:UVLineDecoderBP}).

\subsubsection{The Processing Element}\label{sec:PE}
\begin{figure}
\center
  \includegraphics[scale = 0.17]{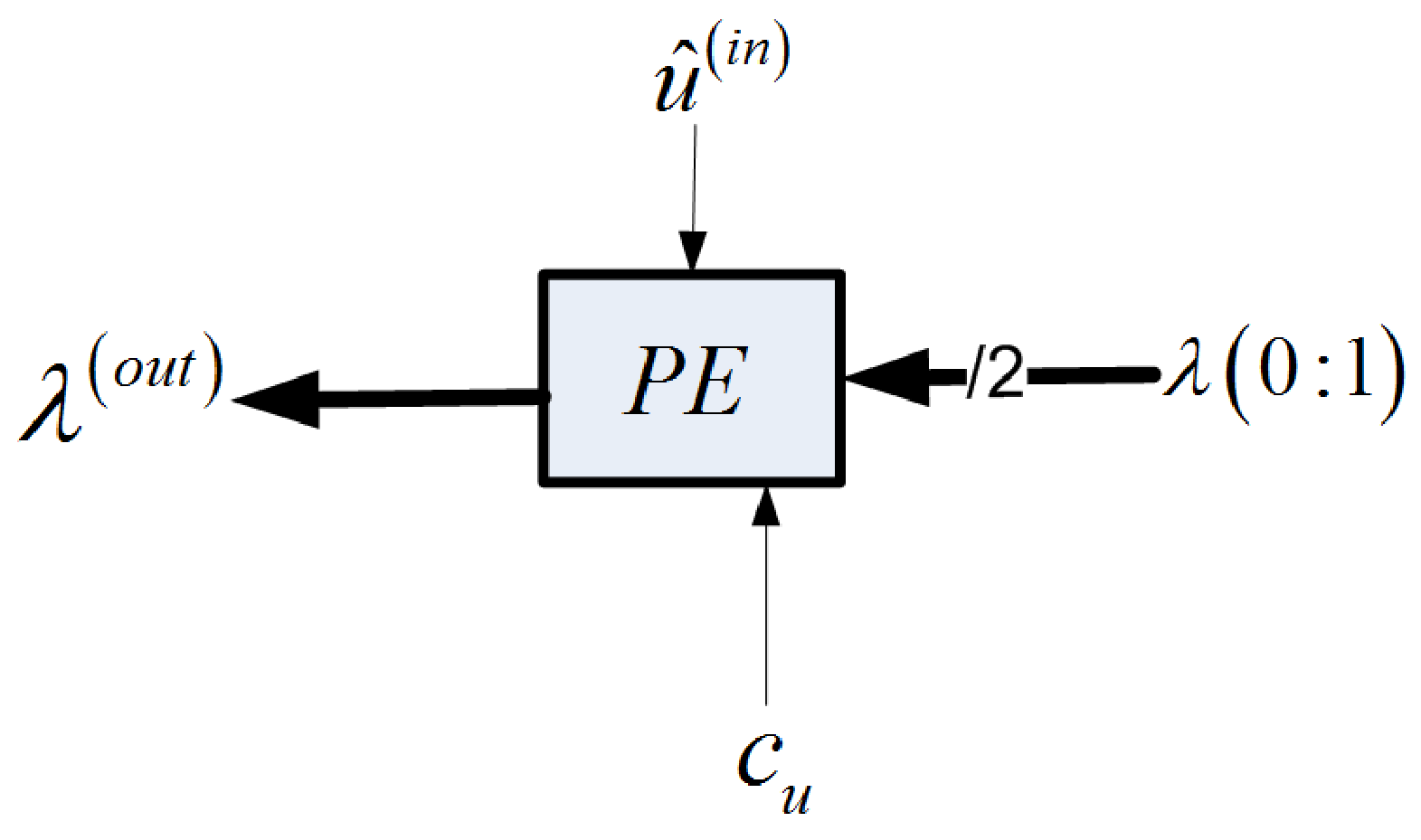}\\
  \caption{ $(u+v,v)$ polar code PE block}\label{fig:PEblkDiag}
\end{figure}

The basic computation element of the decoding circuits, described in Subsections \ref{sec:SCPipeUV} and  \ref{sec:UVLineDecoder}, is the processing element (PE). Figure \ref{fig:PEblkDiag} depicts the PE block. Note that throughout Subsection \ref{sec:HrdwreArikConstr} we use thick arrows to designate signals corresponding to real numbers (to be represented by some quantization method)  and thin arrows to designate binary signals.
The PE block has three inputs:
\begin{itemize}
\item $\lambda(0:1)$ - an array of two input LLRs.
\item $\hat{u}^{(in)}$ - an estimation of the "$u$" bit from the coded pair $(u+v,v)$.
\item $c_u$ - a binary control signal determining the type of LLR that the circuit gives as output in $\lambda^{(out)}$; $c_u=0$ means that we calculate the LLR of $u$ and $c_u=1$ means that we calculate the LLR of $v$ given the estimation of $u$ (the input signal $\hat{u}^{(in)}$).
\end{itemize}
The circuit  outputs the LLR of $u$ or $v$ depending on the control signal $c_u$
\begin{equation}\label{eq:PE}
\lambda^{(out)} = \left\{
                    \begin{array}{ll}
                      2\tanh^{-1}\left(\tanh\left(\lambda(0)/2\right)\tanh\left(\lambda(1)/2\right)\right), & \hbox{$c_u=0$;} \\
                      (-1)^{\hat{u}^{(in)}}\cdot \lambda(0)+ \lambda(1), & \hbox{$c_u=1$.}
                    \end{array}
                  \right.
\end{equation}

 \subsubsection{The SC Pipeline Decoder}\label{sec:SCPipeUV}

\begin{figure}
\centering
\begin{subfigure}{.5\textwidth}
  \centering
  \includegraphics[scale = 0.18]{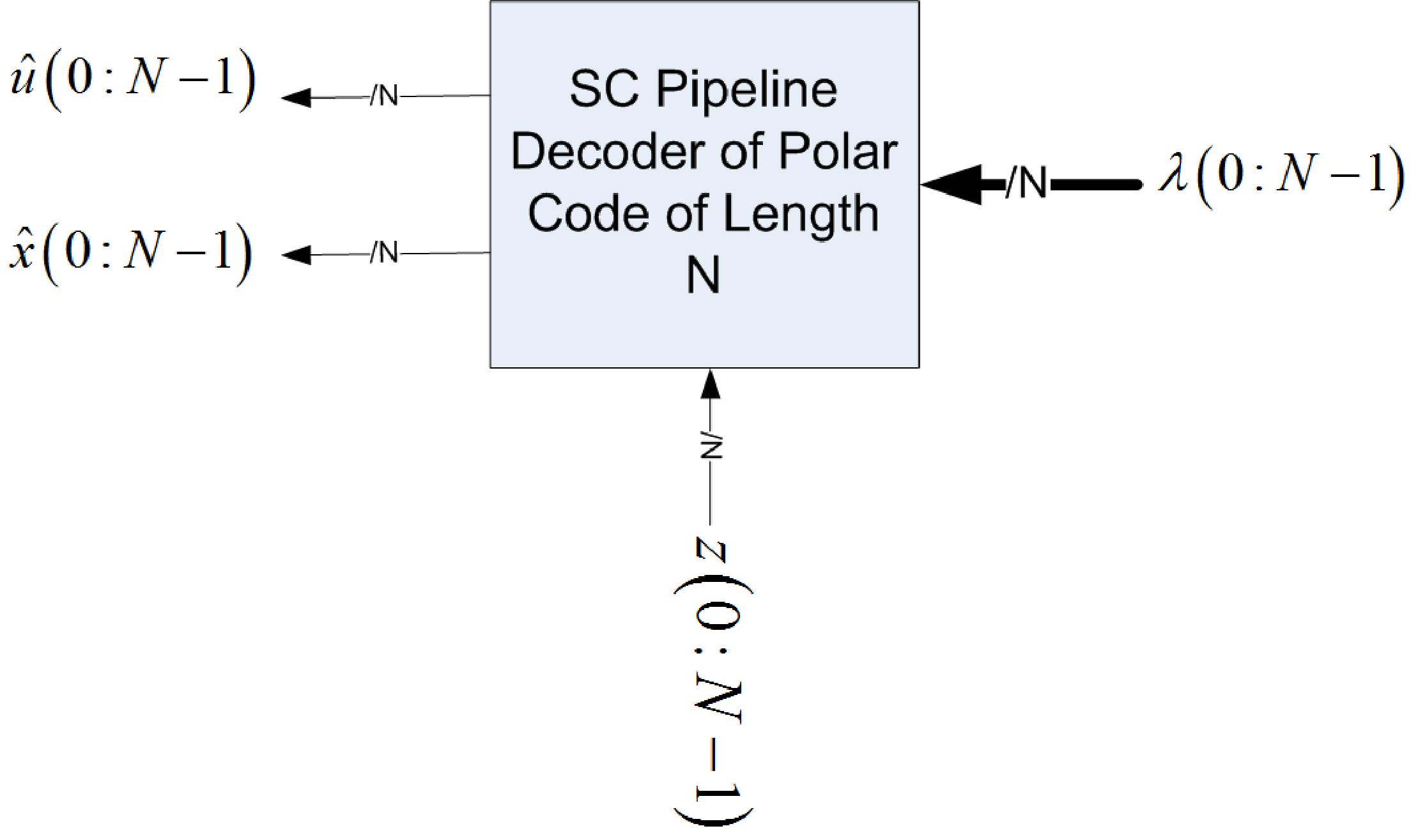}
  \caption{Pipeline decoder}
  \label{fig:arikanPPlnBlk}
\end{subfigure}%
\begin{subfigure}{.5\textwidth}
  \centering
  \includegraphics[scale = 0.13]{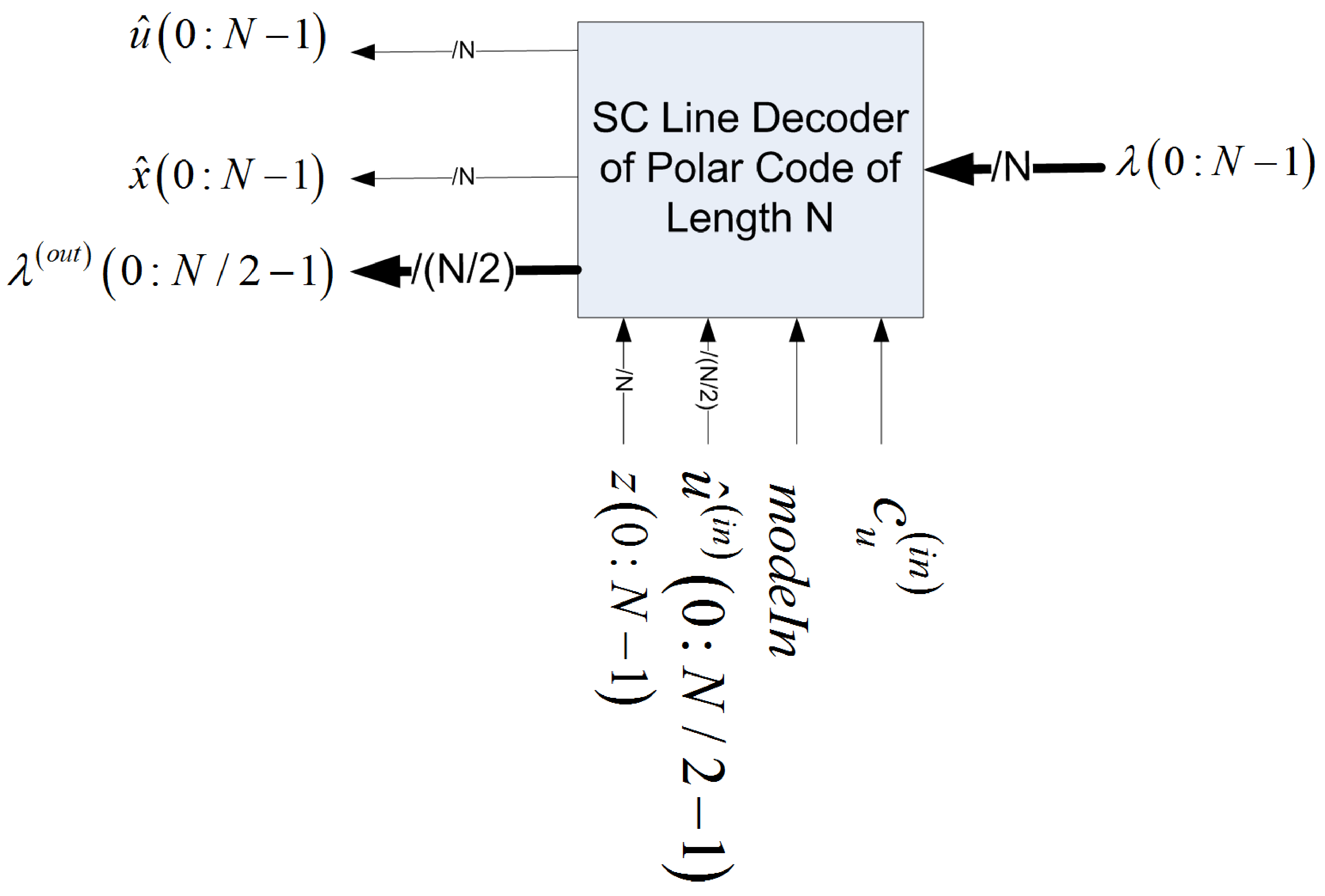}
  \caption{Line decoder}
  \label{fig:arikanLineBlock}
\end{subfigure}
\caption{Blocks of the $(u+v,v)$ polar code decoders of length $N$ bits}\label{fig:arikansBlocks}
\end{figure}

\begin{figure}
\center
  \includegraphics[scale = 0.12]{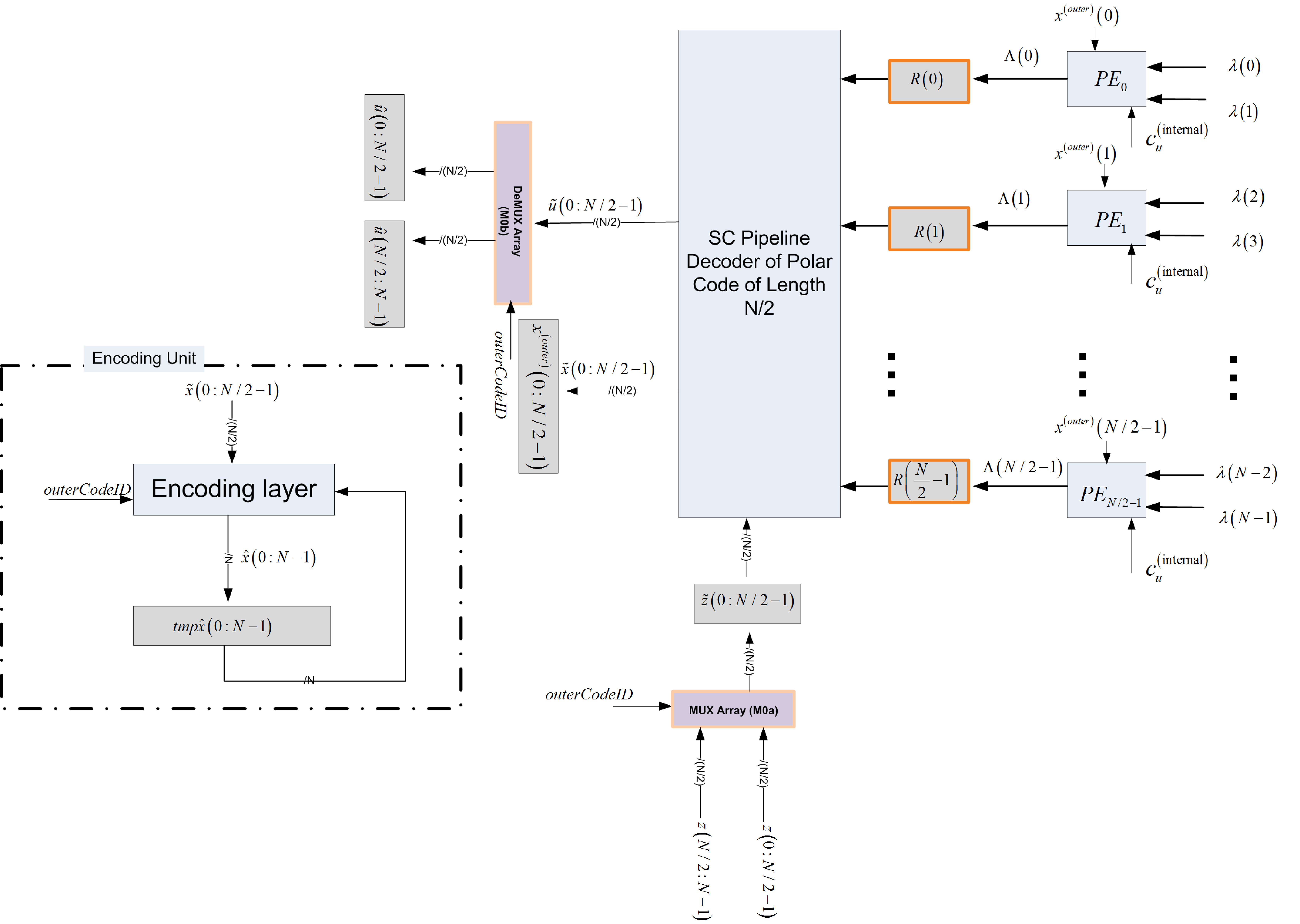}\\
  \caption{Block diagram for the SC pipeline decoder}\label{fig: pipArikan}
\end{figure}

Figure \ref{fig: pipArikan} contains a block description of the SC pipeline decoder. The  decoder's signals $\lambda(0:N-1)$, $z(0:N-1)$, $\hat{u}(0:N-1)$ and $\hat{x}(0:N-1)$ correspond to the inputs and outputs of the SCDecoder function (\ref{eq:SCSignature}) $\boldsymbol \lambda$, $\boldsymbol z$, $\hat{\boldsymbol u}$ and $\hat{\boldsymbol x}$, respectively.  For code length $N=2$ bits, the SC decoder includes a single PE and a slicer. It operates according to Algorithm \ref{algo:SCRecsDescUVV}.

 A block diagram of the implementation of this decoder for $N>2$ is depicted in Figure \ref{fig: pipArikan}. Scanning the diagram from right to left we can observe the following ingredients. The $\lambda(0:N-1)$ LLR input to the circuit is given as input  to an array of $N/2$ PEs, $\left\{PE_j \right\}_{j=0}^{N/2-1}$, which all of them are controlled by the same control signal, $c_u^{(\text{internal})}$. The output of these PEs is denoted by the array of signals $\Lambda(0:N-1)$ and stored in an array of $N/2$ registers $R(0:N/2-1)$ (depicted as rectangle blocks with the register names, $R(i)$, written in them).  These registers are given as the LLR input to a SC pipeline decoder of length $N/2$ bits. This decoder is referred to as the \textit{embedded $N/2$ length decoder} within the $N$ length decoder.

 The embedded decoder is also given as input the frozen bits indicator signals $\tilde{z}(0:N/2-1)$ (binary array), which is generated by splitting the $z(0:N-1)$ binary array into two halves using the MUX array (M0a). The multiplexers in (M0a) are controlled by the internal binary signal $outerCodeID$ that indicates the ordinal of the outer-code that the embedded decoder decodes. For instance, if $outerCodeID =0$ then the embedded decoder handles the first outer-code and therefore it should be given as input the first half of the $z$ array. The two outputs of the embedded decoder are denoted by signals arrays $\tilde{u}(0:N/2-1)$ and $\tilde{x}(0:N/2-1)$. The array $\tilde{u}(0:N/2-1)$ is given as input to the two halves of the output decoded information bits array $\hat{u}(0:N-1)$. The DeMUX array (M0b) determines to which part of the $\hat{u}$ array   $\tilde{u}$ is written.

 The Encoding Unit performs the encoding  of the outer-code's estimated codewords into the estimated codeword of the $N$ length code. The binary register $tmp\hat{x}(0:N-1)$ stores the temporary value of the estimated codeword $\hat{x}$ which is the signals array at its input. The encoding layer is given as input the $outerCodeID$ signal and the two signals arrays $\tilde{x}(0:N/2-1)$ and $tmp\hat{x}(0:N-1)$. Its output is derived as follows
\begin{equation}\label{eq:encodlayer}
\hat{x}\left(2j:(2j+1)\right) = \left\{
                                 \begin{array}{ll}
                                    \left[\tilde{x}(j), \,\,\, 0 \right], & \hbox{$outerCodeID = 0$;} \\
                                      \left[\tilde{x}(j)+tmp\hat{x}(2j), \,\,\, \tilde{x}(j)  \right], & \hbox{$outerCodeID =1$}
                                 \end{array}
                               \right.  , \,\,\,\, \forall j\in \left[\frac{N}{2}\right]_{-}.
\end{equation}
Note  that in order to avoid  delays due to sampling by  registers, it is important that the codeword estimation (which is one of the outputs of the decoder) will be the output of the encoding layer and not the register following it. This issue and further timing concerns are considered in the next subsection.

\begin{algorithm}
\caption{SC Pipeline Decoder of Length $N$ $(u+v,v)$ Polar Code  }          
\label{algo:HWUVVSC}                           
\begin{description}
   \item[//STEP 0:] \hfill \\
  \ALGOSTEP  Set $c_u^{(\text{internal})} = outerCodeID = 0$.

Using the PEs array $\left\{PE_j \right\}_{j=0}^{N/2-1}$, prepare the LLRs  input array for the embedded decoder of the first $N/2$ length outer-code and output it on the signals array $\Lambda(0:N/2-1)$,
such that $$\Lambda(j)= 2\tanh^{-1}\left(\tanh(\lambda({2j})/2)\tanh(\lambda({2j+1})/2)\right),\,\,\,\,\,\, j\in [N/2]_{-}.$$
Sample the $\Lambda(0:N/2-1)$ array by the registers array $R(0:N/2-1)$. Sample the first half of the frozen bits indicator $z$ by the $\tilde{z}$ register, i.e. $\tilde{z}(0:N/2-1) = z(0:N/2-1)$.
   \item[//STEP 1:] \hfill \\
\ALGOSTEP   Execute the embedded decoder on $R\left(0:N/2-1\right)$ and $\tilde{z}(0:N/2-1)$.

 \ALGOSTEP  Sample the $\tilde{ u}(0:{N/2-1})$ output array  by the first half of $\hat{u}$, i.e.  $\hat{u}(0:{N/2-1})=\tilde{ u}(0:{N/2-1})$. Sample the $\tilde{ x}(0:{N/2-1})$ output array by the $x^{(outer)}(0:N/2-1)$ register, i.e. ${x}^{(outer)}(0:{N/2-1})=\tilde{ x}(0:{N/2-1})$.   Let the Encoding Unit process $\tilde{ x}(0:{N/2-1})$ according to (\ref{eq:encodlayer}).
   \item[//STEP 2:] \hfill \\
  \ALGOSTEP     Set $c_u^{(\text{internal})} = outerCodeID = 1$.

Using the PEs array $\left\{PE_j \right\}_{j=0}^{N/2-1}$, prepare the LLRs  input  array for the embedded decoder of the second $N/2$ length outer-code and output it on the signals array $\Lambda(0:N/2-1)$, such that $$\Lambda(j) = (-1)^{{x}^{(outer)}(j)}\lambda({2j})+\lambda({2j+1}),\,\,\,\,\,\,\, j\in [N/2]_{-}.$$
Sample the $\Lambda(0:N/2-1)$ array by the registers array $R(0:N/2-1)$. Sample the second half of the frozen bits indicator $z$ by the $\tilde{z}$ register, i.e. $\tilde{z}(0:N/2-1) = z(N/2:N-1)$.
   \item[//STEP 3:] \hfill \\
 \ALGOSTEP   Execute the embedded decoder on $R\left(0:N/2-1\right)$ and $\tilde{z}(0:N/2-1)$.

 \ALGOSTEP  Sample the $\tilde{ u}(0:{N/2-1})$ output array  by the second half of $\hat{u}$, i.e.  $\hat{u}(N/2:{N-1})=\tilde{ u}(0:{N/2-1})$. Let the Encoding Unit process $\tilde{ x}(0:{N/2-1})$ according to (\ref{eq:encodlayer}).

 \end{description}
\end{algorithm}

We describe the recursive schematic decoding procedure for $N>2$ in Algorithm \ref{algo:HWUVVSC}. Let us consider the complexity of this circuit. We assume that a PE finishes its operation in one clock cycle. Denote by $T(n)$ the  time (in terms of the number of clock cycles) that is required to complete the decoding of $N=2^n$ length polar code. Then, $T(n)=2+2\cdot T(n-1) \,\,\,\,\,\, n> 1$ and $T(1) = 2$. This recursion yields $T(n) = 2N-2$.
Denote by $P(n)$ the number of PEs for a decoder of length $N=2^n$ bits polar code, we have $P(n) = 2^{n-1} + P(n-1) \,\,\,\,\, n > 1$ and $P(1) = 1$, resulting in $P(n) = 2^n - 1 = N-1$. The cost of the encoding unit is of $2\cdot \sum_{i=1}^n 2^i = 4\cdot(N-1)$ bits registers, and $\sum_{i=0}^{n-1}2^i=N-1$ xor circuits. We should have $\rho(n)$ registers for holding LLR values, so $\rho(n) = 2^{n-1} +\rho(n-1) \,\,\,\,\, n>1$ and $\rho(1) = 0$, so  $\rho(n) = N-2$. Note, that in this design, we assume that the encoding layer is a combinatorial circuit.

\subsubsection{ The SC Line Decoder}\label{sec:UVLineDecoder}
In the decoder pipeline design of length $N$ polar code, the $N/2$ processing elements $\left\{ PE_j \right\}_{j=0}^{N/2-1}$, are only employed during steps $0$ and $2$ of the algorithm. During the other steps (that ideally consume $2\cdot T(n-1) = 2N-4$ clock cycles of the total $2N-2$ clock cycles) these processors are idle, resulting in an inefficient design. In order to increase the processors utilization we observe that the maximum number of operations that can be done  in parallel by the PEs in the SC decoding algorithm is $N/2$.  As a consequence, in order to support the maximum level of parallelism, the design has to include at least $N/2$ PEs. The line decoder\footnote{ Note that strictly speaking, the original line decoder, presented by Leroux \textit{et al.} \cite[Section 3.3]{Leroux2012}, is not precisely the  same design, discussed here. The differences, however,  appear to be minor (existing mostly in the routing between the LLR registers and the PEs). As a consequence we preferred not to distinguish it from Leroux's design.}, that we describe in this subsection, achieves this lower-bound.

\begin{figure}
\center
  \includegraphics[angle=90,scale = 0.06]{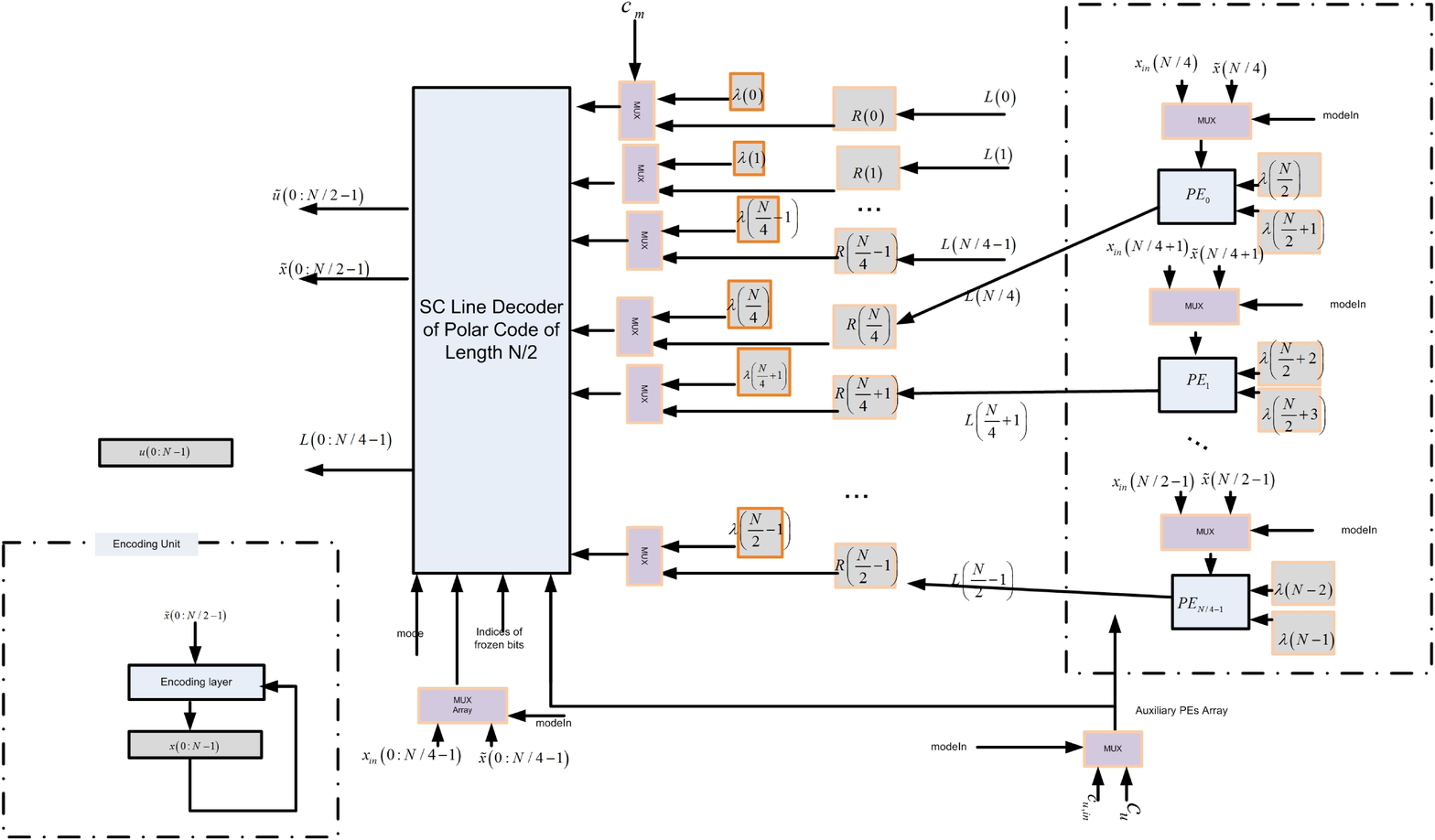}\\
  \caption{Block diagram for the SC line decoder}\label{fig: linArikan}
\end{figure}

Figure \ref{fig:arikanLineBlock} depicts the line decoder block for length $N$ bits code. The line decoder has two operation modes.
\begin{description}
   \item[Standard Mode (S-Mode): $modeIn = 0$] \hfill \\
 The decoder gets as input LLRs  array, $\lambda(0:N-1)$, and the frozen bits indicator vector, $z(0:N-1)$. Upon completion of its operation the decoder outputs the hard decision on the information word $\hat{u}(0:N-1)$ and its corresponding codeword $\hat{x}(0:N-1)$ (this is the operation mode we supported thus far in the pipeline decoder).
   \item[PE-Array Mode (P-Mode): $modeIn = 1$] \hfill \\
   The decoder gets as input a  signals array of LLRs ${ \lambda}({0}:{N -1})$, a control signal $c_u^{(in)}$ and a binary vector ${\hat{u}}^{(in)}(0:N/2-1)$. The output is a signals array ${ \lambda^{(out)}}(0:N-1)$ of LLRs, where
 \begin{equation}
    {\lambda^{(out)}}(j)=\left\{
     \begin{array}{ll}
       2\cdot \tanh^{-1}\left(\tanh\left(\lambda({2j})/2 \right)\cdot\tanh\left(\lambda({2j+1})/2 \right)\right), & \hbox{$c_u^{(in)}=0$;} \\
       (-1)^{{\hat{u}}^{(in)}(j)}\cdot \lambda({2j})+\lambda({2j+1}), & \hbox{$c_u^{(in)}=1$,}
     \end{array}
   \right. \,\,\,\,\,\,\,\,\,\,\,\,\, \forall j\in \left[\frac{N}{2}\right]_{-}.
\end{equation}
    \end{description}
In Figure \ref{fig: linArikan}, we provide a block diagram for this decoder. Note, that in order to maintain  the maximum level of parallelism,  the length $N$  polar code decoder ought to have $N/2$ processors. Thus, in order to build the length $N$  polar code decoder using an embedded $N/2$ length polar code decoder (already having $N/4$ processors), we use an additional array of $N/4$ PEs, which is referred to as the \textit{auxiliary array}. The input signal \textit{modeIn} indicates wether the decoder is used in  S-Mode  or in  P-Mode. The \textit{mode} signal is an internal signal that controls  whether the $N/2$ length embedded decoder is in P-Mode.

Let us scan Figure \ref{fig: linArikan} from right to left and observe its important ingredients. The auxiliary PEs array contains $N/4$ processors $\left\{PE_j \right\}_{j=0}^{N/4-1}$ to which the second half of input array $\lambda(N/2:N-1)$ is connected. The first half of the input LLRs array $\lambda(0:N/2-1)$ is connected to the embedded line decoder via the MUX array (M2), in which all the multiplexers are controlled by the binary signal $c_m$. The other input alternative of the (M2) array  is the registers array $R(0:N/2-1)$. The $c_u$ input of  $\left\{PE_j \right\}_{j=0}^{N/4-1}$ is determined by the (M3) multiplexer, such that in S-Mode ($modeIn = 0$) the input is $c_u^{(\text{internal})}$ (an internal signal) and otherwise $c_u = c_u^{(in)}$ (one of the inputs to the $N$ length decoder). The output of (M3) also serves as the $c_u^{(in)}$ input to the embedded decoder.  The $modeIn$   signal further controls the (M4) MUX array, such that in S-Mode the $\hat{u}^{(in)}$ input to $\left\{PE_j \right\}_{j=0}^{N/4-1}$ is the input sub-vector ${\hat{u}}^{(in)}\left(N/4:N/2-1\right)$ and otherwise the input is  $x^{(outer)}\left(N/4:N/2-1\right)$ (the second half of the estimated codeword output of the embedded decoder). Furthermore, $modeIn$ also controls the (M1) MUX array that selects between ${\hat{u}}^{(in)}\left(0:N/4-1\right)$ and $x^{(outer)}\left(0:N/4-1\right)$ for $modeIn=1$ and $modeIn=0$ respectively. The output of (M1) serves as the ${\hat{u}}^{(in)}$ input of the embedded decoder. The internal binary signal,  $mode$,  is given to the embedded decoder as its $modeIn$ input.

The S-Mode and the P-Mode procedures of the line decoder are described in Algorithms \ref{algo:SModeLineMode} and  \ref{algo:PModeLineMode}, respectively. Let us discuss the complexity of the decoder. Let $P(n)$ be the number of processors of the $N=2^n$ decoder. Then, $P(n)=2^{n-2}+P(n-1)\,\,\,\,\,\, P(1)=1$, so $P(n)=2^{n-1}=N/2$. The number of LLR registers is $\rho(n) = 2^{n-1} +\rho(n-1),\,\,\,\,\,\, \rho(1) = 1$, so we have $\rho(n) = 2^n-1 = N-1$. Note that $\rho(n)$ doesn't account for the binary registers for $\tilde{z}, tmp\hat{x}$ and $\hat{u}$.
\begin{algorithm}
\caption{S-Mode of SC Line-Decoder of Length $N$ $(u+v,v)$ Polar Code ($modeIn = 0$)  }          
\label{algo:SModeLineMode}                           
\begin{description}
   \item[//STEP 0:] \hfill \\
   \ALGOSTEP Set $c_m =c_u^{(internal)} = outerCodeID= 0$, $mode = 1$.

 Operate the embedded decoder in P-Mode, such that at the output of the decoder we have
     $$  {\Lambda}(j) = 2\cdot \tanh^{-1}\left(\tanh\left(\lambda({2j})/2 \right)\cdot\tanh\left(\lambda({2j+1})/2 \right)\right)\,\,\,\,\,\,\, \forall j\in[N/4]_{-}.$$

Use the auxiliary PEs array and compute
$${\Lambda}(j) = 2\cdot \tanh^{-1}\left(\tanh\left(\lambda({2j})/2 \right)\cdot\tanh\left(\lambda({2j+1})/2 \right)\right) \,\,\,\,\,\, \forall N/4 \leq j \leq N/2-1$$

 Sample the $\Lambda(0:N/2-1)$ array by the registers array $R(0:N/2-1)$. Sample the first half of the frozen bits indicator $z$ by the $\tilde{z}$ register, i.e. $\tilde{z}(0:N/2-1) = z(0:N/2-1)$.
   \item[//STEP 1:] \hfill\\
\ALGOSTEP Set $mode = 0$ and $c_m=1$.

 Execute the embedded decoder in S-Mode on $R\left(0:N/2-1\right)$ and $\tilde{z}(0:N/2-1)$.

 \ALGOSTEP  Sample the $\tilde{ u}(0:{N/2-1})$ output array  by the first half of $\hat{u}$, i.e.  $\hat{u}(0:{N/2-1})=\tilde{ u}(0:{N/2-1})$.  Sample the $\tilde{ x}(0:{N/2-1})$ output array by the $x^{(outer)}(0:N/2-1)$ register, i.e. $ {x}^{(outer)}(0:{N/2-1})=\tilde{ x}(0:{N/2-1})$. Let the Encoding Unit process $\tilde{ x}(0:{N/2-1})$ according to (\ref{eq:encodlayer}).

   \item[//STEP 2:] \hfill \\
  \ALGOSTEP Set $c_m =0 ,c_u^{(internal)}= mode = outerCodeID= 1 $.

Operate the embedded decoder in P-Mode, such that at the output of the decoder we have
     $$ {\Lambda}(j)   = (-1)^{{x}^{(outer)}(j)}\cdot\lambda({2j})+\lambda({2j+1})\,\,\,\,\,\,\, \forall j\in[N/4]_{-}.$$

 Use the auxiliary PEs array and compute
$${\Lambda}(j) =  (-1)^{{x^{(outer)}}(j)}\cdot\lambda({2j})+\lambda({2j+1}) \,\,\,\,\,\, \forall N/4 \leq j \leq N/2-1$$

Sample the $\Lambda(0:N/2-1)$ array by the registers array $R(0:N/2-1)$.
Sample the second half of the frozen bits indicator $z$ by the $\tilde{z}$ register, i.e. $\tilde{z}(0:N/2-1) = z(N/2:N-1)$.

   \item[//STEP 3:] \hfill \\
   \ALGOSTEP Set $mode = 0 $ and $c_m= 1$.

Execute the embedded decoder in S-Mode on $R\left(0:N/2-1\right)$ and $\tilde{z}(0:N/2-1)$.

  \ALGOSTEP  Sample the $\tilde{ u}(0:{N/2-1})$ output array  by the second half of $\hat{u}$, i.e.  $\hat{u}(N/2:{N-1})=\tilde{ u}(0:{N/2-1})$. Let the Encoding Unit process $\tilde{ x}(0:{N/2-1})$ according to (\ref{eq:encodlayer}).

 \end{description}
\end{algorithm}

\begin{algorithm}
\caption{P-Mode of SC Line-Decoder of Length $N$ $(u+v,v)$ Polar Code ($modeIn = 1$)  }          
\label{algo:PModeLineMode}                           
\begin{description}
   \item
   \ALGOSTEP Set $c_m = 0 , mode = 1$.

 Operate the embedded decoder in P-Mode, such that at the output of the decoder we have

     $$  {\Lambda}(j) = \left\{
                          \begin{array}{ll}
                            2\cdot \tanh^{-1}\left(\tanh\left(\lambda({2j})/2 \right)\cdot\tanh\left(\lambda({2j+1})/2 \right)\right) & \hbox{$c^{(in)}_u = 0$;} \\
                              (-1)^{{{\hat{u}}^{(in)}}(j)}\cdot\lambda({2j})+\lambda({2j+1}) & \hbox{$c^{(in)}_u = 1$;}
                          \end{array}
                        \right.\,\,\,\,\,\,\, \forall j\in \left[N/4-1\right]_{-}.$$

Use the auxiliary PEs array and compute
     $$  {\Lambda}\left(j\right) = \left\{
                          \begin{array}{ll}
                            2\cdot \tanh^{-1}\left(\tanh\left(\lambda({2j})/2 \right)\cdot\tanh\left(\lambda({2j+1})/2 \right)\right), & \hbox{$c^{(in)}_u = 0$;} \\
                              (-1)^{{{\hat{u}}^{(in)}}(j)}\cdot\lambda({2j})+\lambda({2j+1}) & \hbox{$c^{(in)}_u = 1$;}
                          \end{array}
                        \right.\,\,\,\,\,\,\,\forall N/4 \leq j \leq N/2-1.$$

\item //Note that the signals array $\Lambda(0:N/2-1)$ is wired to the output signals array $\lambda^{(out)}(0:N/2-1)$.
 \end{description}
\end{algorithm}

At this point, we would like to make a remark regarding the efficiency of the proposed design. The recursive design has  the benefit of being a comprehensible reflection of the implemented algorithm. It also has the advantage of emphasizing the parts of the system that may be reused. However, it might be argued that it has a disadvantage considering the routing of signals in the circuit. This is because we use the embedded decoder  as a black box and consequently we route all the signals from it and to it, using its interface.  As a result, some of the signals traverse  lengthy paths before reaching their target processor. These paths may be too long for the decoder circuit to have an adequate  clock frequency, thereby resulting in degradation of the achievable throughput. We therefore recommend that after constructing the circuit in a recursive manner, it should be optimized by unfolding the recursive units and  contracting the paths. Furthermore, we advise that for building a decoder of length $2N$ bits  code, the designer will use the already optimized design of the $N$ length decoder (for the embedded unit), thereby taking advantage of the recursion.

We give below two examples of   long paths hazards, that are likely to pose a problem. Workarounds for these challenges are further provided.
\begin{enumerate}
\item The (M2) MUX array at the input of the embedded line decoder of the length $N/2$  code was included because of the introduction of  P-Mode. A closer examination of our design, reveals that some of (M2) input signals traverse long paths before reaching their destination PE. For example, the inputs $\lambda(0)$ and $\lambda(1)$ need to traverse $\log_2(N)-1$ multiplexer layers before reaching their processor. Since   P-Mode needs to be accomplished in a single clock cycle, this long path  might be prohibitive. By unfolding  the $N/2$ length embedded decoder block, the designer is able to control the lengths by carefully routing the signals.
\item The  encoding layer also suffers from long routing. In our analysis, we assumed that the encoding procedure is combinatorial, and therefore has to be completed within one clock cycle. This may be a problem when several encoding circuits are operated one after the other. For instance, this is the case of step $3$ of the decoder of length $N/2^{i}$  code, that occurs within  step $3$ of the decoder of length  $N/2^{i-1}$ code  for all $i \in \left[2N-2 \right] $. In this case, $O(\log N)$ operations need to occur in a sequential manner in one clock cycle. For large $N$  and high clock frequency circuit, this might not be feasible. The idea of Leroux  \textit{et al.} \cite{Leroux2012} was to use flip-flops for saving the partial encoding for each code bit in the different layers of the decoding circuit. Each such flip-flop, is connected using a xor circuit to the signal line of the estimated information bit. As such, whenever the SC decoder decides on an information bit, the flip-flops corresponding to the code bits that are dependent on this information bit are updated accordingly. These flip-flops need to be reset whenever we start decoding their corresponding outer-code. For example, when we start using the embedded $N/2$ length decoder (on step $1$ and step $3$) its flip-flops of partial encoding need to be erased (because they correspond to a new instance of outer-code).

    The above notion may also be described recursively, by changing the specification of the length $N$  polar code decoder in S-mode, and requiring it to output the estimated information bits as soon as they're ready.  The decoder should also have an $N$ length binary indicator vector, that indicates which code bits are dependent on the currently estimated information bit. It is easy to see that using the indicator vector of the length $N/2$ decoder, it is possible to calculate the  $N$  length indicator vector, by using the $(u+v,v)$ mapping. This, however, generates again a computation path of length $\Theta(\log N)$. This problem, can be addressed,  by having a fixed indicator circuit for each partially encoded-bit flip-flop. This circuit will indicate which information bit should be accumulated depending on the ordinal number of this bit.  For example, for the decoder of the code of length $N$, we should have an array of $N/2$ flip-flops, each one corresponds to a bit of the codeword of the $N/2$ length first outer-code. Each one of these flip-flops, should have an indicator circuit, that gets as input a value of a counter signaling the ordinal number of the information bit that has been estimated, and returns $1$ iff its corresponding codeword bit is influenced by this information bit.  For example, the indicator circuit, corresponding to the first code bit, is a constant $1$, because ${  x}_0 =\sum_{i=0}^{N/2-1}{ u}_i$, i.e. it is dependent on all the information bits. On the other hand, the last bit's indicator (i.e. of $x_{N/2-1}$)  returns $1$ iff its input equals to $N/2-1$, because $x_{N/2-1}=u_{N/2-1}$. Using the global counter (that is advanced whenever an information bit is estimated) and the indicator circuits, each code bit that is influenced by this information bit change its flip-flop state accordingly.

    Using the Kronecker power form of the generating matrix of the $(u+v,v)$ polar code, it can be seen that each of such indicator circuits can be designed by using no more than  $O(\log n) = O(\log\log N)$ AND and NOT circuits, therefore the total cost of these circuits will be of $O(N\log \log N)$ in terms of space complexity. Further improvements to the efficiency of the circuit can be achieved by employing Fan and Tsui's high performance partial sum network \cite{Fan2014}. This network implements the indicator circuits with constant space complexity and delay (per circuit).
\end{enumerate}

In summary, the recursive architecture may be developed and modified to achieve the timing requirements of the circuit. This may be done by "opening the box" of the embedded decoders, and altering them to support more efficient designs.

A careful examination of the line-decoder reveals that the \textit{auxiliary PEs array} is only used on steps $0$ and $2$, and is idle on the other steps. This fact motivates us to consider two variations on this design. The first one    adds hardware and use these arrays to increase the throughput, while the second one   decreases the throughput and thereby reduces the required hardware.

\subsubsection{Parallel Decoding of Multiple Codewords}\label{sec:ParDecLine}
High throughput communication systems may require support of simultaneous decoding of multiple codewords. A naive approach to meet this challenge is implementing $\mathrm{p}$ instances of the decoder  when there is a need for decoding $\mathrm{p}$ codewords simultaneously. However, because the \textit{PEs auxiliary array}   is idle most of the time, it seems like a good idea to "share" this array among several decoders. By   appropriately scheduling the commands to the processors, it is possible to have a decoder implementation for $\mathrm{p}$ parallel  codewords which is less expensive than just duplicating the decoders.

Since the array is idle during steps $1$ and $3$, in which the embedded length $N/2$  decoder is  active, it is possible to have $\mathrm{p}\leq T(n-1)+1=N-1$ decoders sharing the same \textit{auxiliary array}. The decoding of each one of them is issued in a delay of one clock cycle from each other. Assuming  that $\mathrm{p}=N-1$, we have a decoding time $T(n)+N-2=3N-4$ for $N-1$ codewords while having $\mathrm{p}\cdot P(n-1)+N/4 = (N-1)\cdot N/4+N/4 = N^2\cdot 4 $ processors, which is about half of the number of processors of the naive solution.

This notion can be further developed. For the embedded $N/2$ length decoder, there is a an auxiliary array of $N/8$ processors. This auxiliary array is used on steps $0$ and $2$ of the decoders of length $N$ and length $N/2$. Therefore, it is idle  most of the time, and we can share it among the $\mathrm{p}$ decoders of length $N/2$. Assuming that $\mathrm{p} = N-1$, we may allocate three auxiliary arrays that will be shared among  the decoders, each one is dedicated to one of these different steps: one array for step $0$ (and $2$) of the $N$ length decoder, one array for step $0$ of the $N/2$ length decoder and one array for step $2$ of the $N/2$ length  decoder. For each of the decoded codewords the number of clock cycles between these steps is at least $\mathrm{p}$, therefore there will be no contention on these resources and the throughput will not suffer because of this hardware reduction.

In general, for $\mathrm{p} = N-1$, the \textit{auxiliary array} within the embedded decoder of length $\frac{N}{2^i}$ polar decoder ($i \in \left[\log_2(N)-2\right]$), can be shared among the $\mathrm{p}$ decoders, provided that we allocate an instance of the array for each of the decoding steps it is used in,  during the first half of the decoding algorithm for the length $N$ code  (i.e. during the $N$ length decoder's steps $0$ and $1$). As a consequence, for this specific array, we have one call in step $0$ of the $N$ length  decoder, one call for step $0$ and one call for step $2$ of the embedded $\frac{N}{2}$ length  decoder, two calls for step $0$ and two calls for step $2$ of the $\frac{N}{2^{2}}$ length embedded decoder, ..., $2^i$ calls for step $0$ and $2^i$ calls for step $2$ for the length $\frac{N}{2^i}$ embedded decoder. In summary, we require $\sum_{t=0}^{i}2^t = 2^{i+1}-1$ \textit{auxiliary arrays} of processors, each one contains $\frac{N}{2^{i+2}}$ PEs. In particular, we need $N-1$ PEs for the length $2$ decoder (each PE is allocated to a specific decoder), and  $\frac{N}{2}\cdot \sum_{i=0}^{\log_2(N)-2}\frac{2^{i+1}-1}{2^{i+1}}\approx \frac{N}{2}\left(\log_2(N)-1\right)$ PEs for the other decoders lengths. This adds up to approximately $\frac{N}{2}\left(1+\log_2(N)\right)$ PEs. We conclude that this solution allows an increase of the throughput in a multiplicative factor of $N$, while the PEs hardware is only increased by  approximately $\log_2(N)$ factor. Note, that the number of registers should be increased by a multiplicative factor of $O(\mathrm{p})=O(N)$ as well.

A closer look at the above  design, reveals that we actually allocated for each sub-step of steps $0$ and $1$ of the $N$ length decoder a different array of processors. The decoding operations of the $\mathrm{p}$ codewords will go through these units in a sequential order. However, each decoder should have its own set of registers saving the state of the decoding algorithm. Another observation is that when we finish decoding the first codeword (i.e. the one we started decoding in time $0$), we can start decoding codeword number $N$ in the next time slot (and then codeword number $N+1$, etc.), in a pipelined fashion. Note that Leroux \textit{et al.} considered a similar idea, and referred to it as the \textit{vector-overlapping} architecture \cite{Leroux10}.

\subsubsection{Limited Parallelism Decoding}\label{sec:LimitedParDecLine}
An alternative approach for addressing the problem of low utilization of the \textit{auxiliary PEs arrays} is to limit the number of processing elements that may be allowed to operate simultaneously. This is a  practical consideration, since typically, a system design has a parallelism limitation which is due to power consumption and silicon area constraints. The limited parallelism, inevitably results in an increase of the decoding time, and thereby a decrease of the throughput.

The length $N$  line decoder has  PE parallelism of $N/2$, because it may simultaneously compute  at most $N/2$ LLRs using the $N/2$ PEs. Let us consider a line decoder of length $N$ code with limited parallelism of $N/{2^i}$, where $i\in \left[\log_2N\right]$. This means, that the decoder has exactly $\frac{N}{2^i}$ PEs. If $i=1$ then the decoder is actually  the standard line decoder. Figure \ref{fig:lineArikanLimPar} depicts the block diagram of the decoder for  $i>1$.  We highlight the changes that were applied to the standard line decoder (Figure \ref{fig: linArikan}) in creating Figure  \ref{fig:lineArikanLimPar}.
\begin{itemize}
\item The \textit{auxiliary PEs array} was omitted.
\item The embedded line decoder of the $N/2$ length  code was replaced by a limited parallelism line decoder, with parallelism  of $N/{2^i}$.
\item The input  to the registers array $R({N/4}:{N/2-1})$ is the  signals array $\Lambda(0:N/4-1)$.
\item A MUX array (M2a) was added providing the "channel" inputs to the (M2) MUX array. The control signal of the (M2a) array is an internal binary signal called \textit{subStep}, such that the output of the array is $\lambda(0:N/2-1)$ if $subStep = 0$ and otherwise it equals to $\lambda(N/2:N-1)$. Similarly, $subStep$ is the control signal of two additional MUX arrays (M1a) and (M1b) providing inputs to the (M1) MUX array. We have the outputs of these arrays equal ${\hat{u}}^{(in)}\left(0:N/4-1\right)$ and ${x^{(outer)}}\left(0:N/4-1\right)$ for  $subStep=0$ and ${\hat{u}}^{(in)}\left(N/4:N/2-1\right)$ and ${x^{(outer)}}\left(N/4:N/2-1\right)$ otherwise.
\item The output LLR signals array $\lambda^{(out)}\left(0:N/2-1\right)$ is routed such that
\begin{equation}\label{eq:RoutingLimParOut}
 \lambda^{(out)}\left(0:N/4-1\right) = R\left(0:N/4-1\right)  \text{ and } \lambda^{(out)}\left(N/4:N/2-1\right) = \Lambda\left(0:N/4-1\right).
\end{equation}
\end{itemize}

 \begin{figure}
\center
  \includegraphics[angle=90,scale = 0.08]{PolarLineDecoderUV2LimParallelism.eps}\\
  \caption{Block diagram for the limited parallelism line decoder}\label{fig:lineArikanLimPar}
\end{figure}

The limited parallelism S-mode decoding algorithm   has four steps as before, however steps $0$ and $2$ are modified including now two sub-steps. On each sub-step we calculate half of the LLRs because we don't have an auxiliary array. Note that depending on the parallelism of the embedded decoder, those sub-steps may require more than one clock cycle. In a similar manner the P-mode operation   is also amended, and now contains two sub-steps.

Let us analyze the time complexity of this algorithm. We denote by $T(n,n-i)$ the S-Mode running time (in terms of clock cycles) for length $N=2^n$ bits polar code with limited parallelism of $N/{2^i}=2^{n-i}$. We note that $T(n,n-1) = T(n)$, where $T(n)=2N-2$ is the time complexity of the standard line decoder. The following recursion formula is derived
\begin{equation}
T(n,n-i)=2\cdot T(n-1,n-i)+ 4\cdot T_p(n-1,n-i),
\end{equation}
where $T_p(n,m)$ is the running time of the $N=2^{n}$ bits length decoder with $2^{m}$ limited parallelism in P-Mode.
\begin{equation}
T_p(n,m) = \left\{
                 \begin{array}{ll}
                   1, & \hbox{$n-m\leq 1$;} \\
                   2\cdot T_p(n-1,m), & \hbox{otherwise.}
                 \end{array}
               \right.
\end{equation}
Therefore,
\begin{equation}
T_p(n,m)=\left\{
                 \begin{array}{ll}
                   1, & \hbox{$n-m\leq 1$;} \\
                   2^{n-m-1}, & \hbox{otherwise.}
                 \end{array}
               \right.
\end{equation}
It can be shown that
\begin{equation}\label{eq:timeTradeoffLimPara}
T(n,n-i)= 2\cdot N +(i-2)\cdot 2^i\,\,\,\,\,\,\,\,, i\geq 1.
\end{equation}
Equation (\ref{eq:timeTradeoffLimPara}) reveals the tradeoff between the number of PEs and the running time of the algorithm. For example,  decreasing the number of processors by a multiplicative factor of $8$, compared to the standard case (i.e. $i=4$), results in an increase of only $34$ clock cycles in the decoding time. We note however, that in order to implement such a decoder, additional routing circuitry (e.g. multiplexers layers) should be included.

\begin{remark}[SCL Implementation]\label{rem:SCLImplementation}
For a limited  list size, the SCL decoder may also be implemented by a line decoder. This requires to duplicate the hardware by the list size, $L$, and to introduce the appropriate logic (i.e. comparators and multiplexer layers).  It  is possible to provide an implementation with $O(f(L)\cdot N)$ time complexity, where $f(\cdot)$ is a polynomially bounded function, that is dependent on the efficiency of algorithms for selection of $L$ most likely decoding paths from a list of $2L$ paths (which is done by the $N=2$ length decoder). Furthermore, the normalization of the likelihoods should be carefully considered, and also should have its impact on the precise (i.e. non asymptotic) time complexity. As was mentioned in Subsection \ref{sec:LimitedParDecLine} by limiting the parallelism of the decoder, it is possible to reduce the number of processors with reasonable hit to the throughout.
\end{remark}

\subsubsection{ The BP Line Decoder}\label{sec:UVLineDecoderBP}

\begin{figure}
\begin{subfigure}{.5\textwidth}
  \centering
  \includegraphics[scale = 0.13]{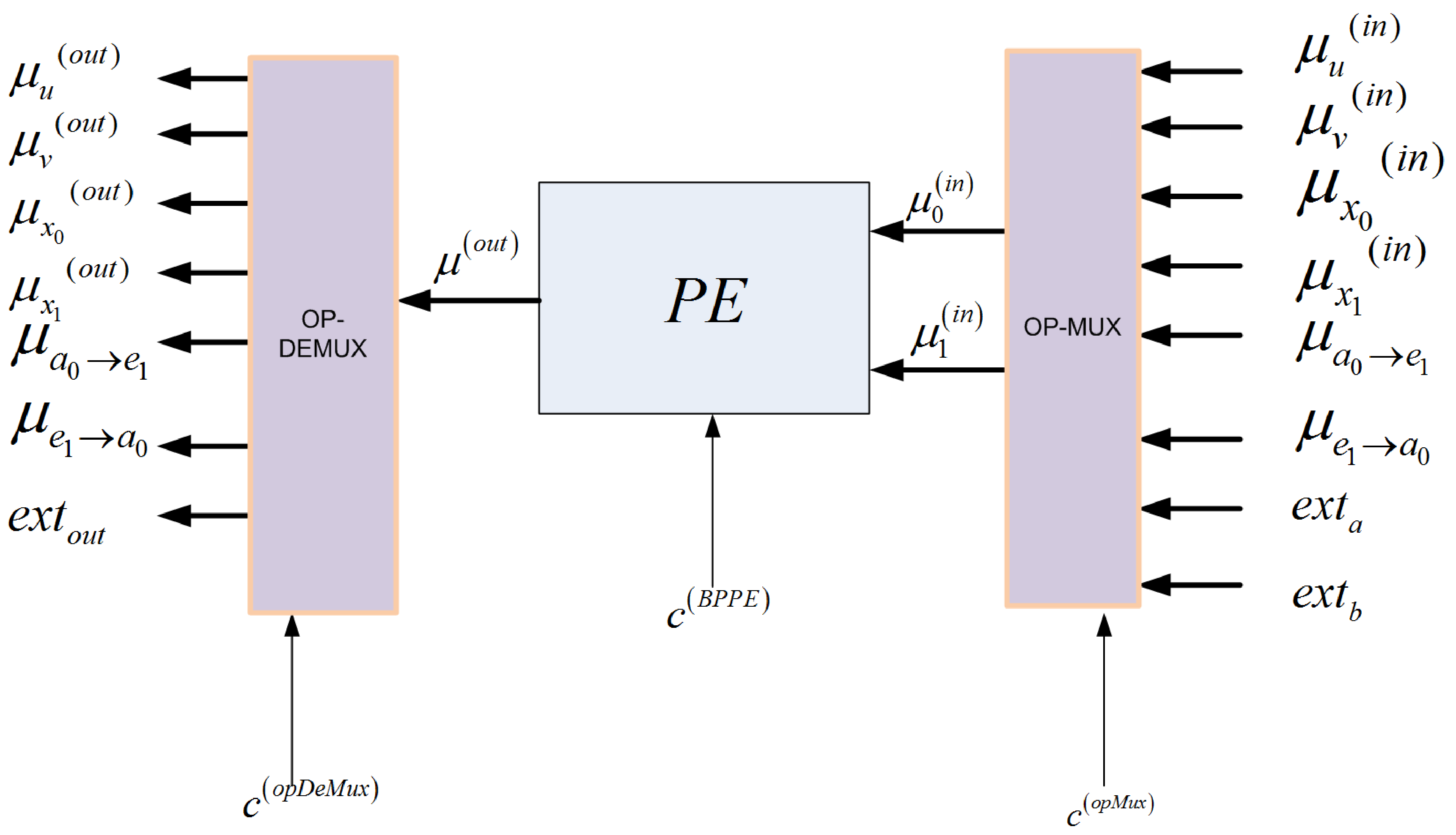}
  \caption{Processing element and routing layers}
  \label{fig: BPPEPRocessor}
\end{subfigure}
\begin{subfigure}{.5\textwidth}
  \centering
  \includegraphics[scale = 0.13]{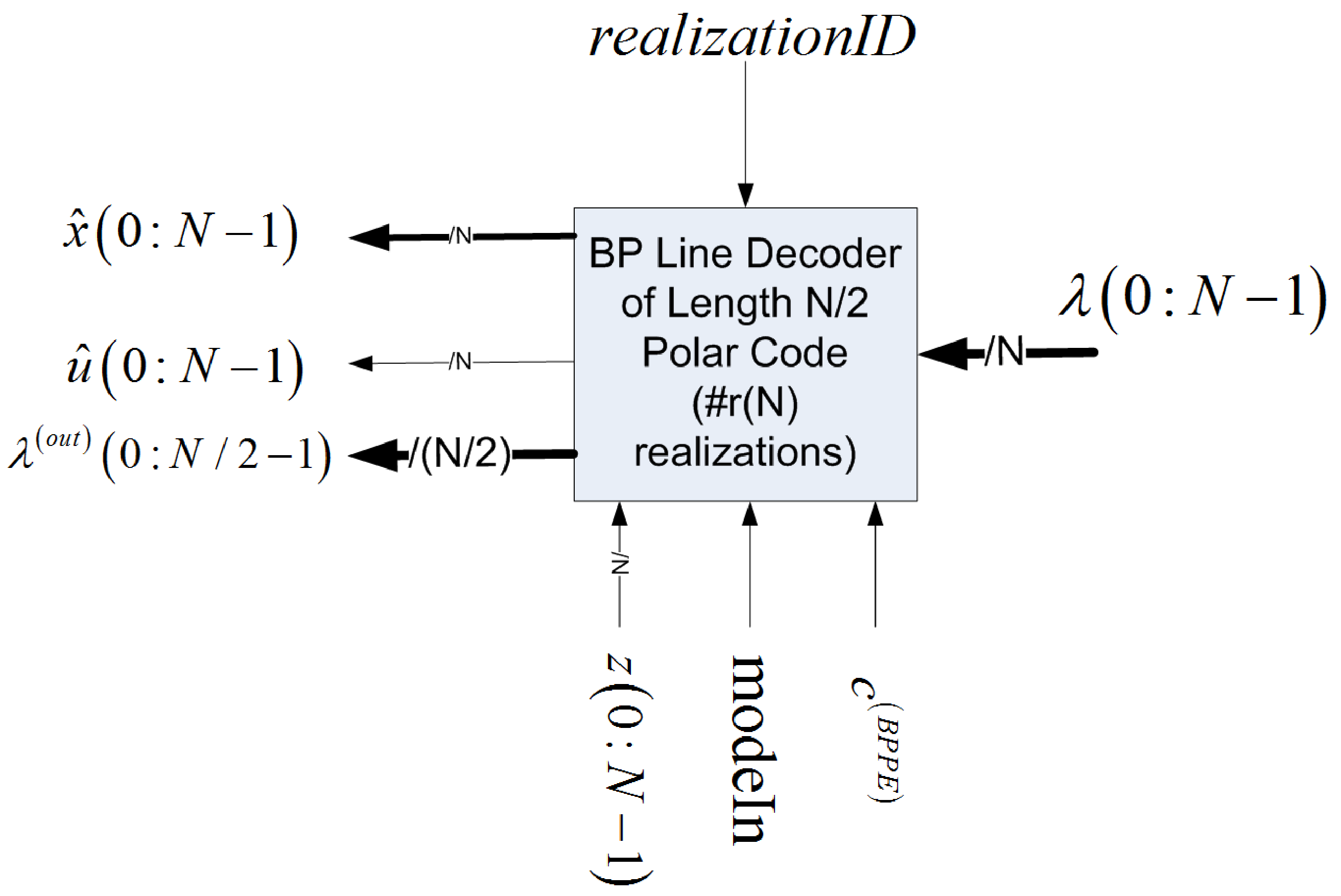}
  \caption{BP line decoder block}
  \label{fig:arikanPPlnBlk}
\end{subfigure}%
\caption{BP line decoder  components definitions}
\label{fig:BPBlockDefinitions}
\end{figure}


 \begin{figure}
\center
  \includegraphics[angle =90,scale = 0.07]{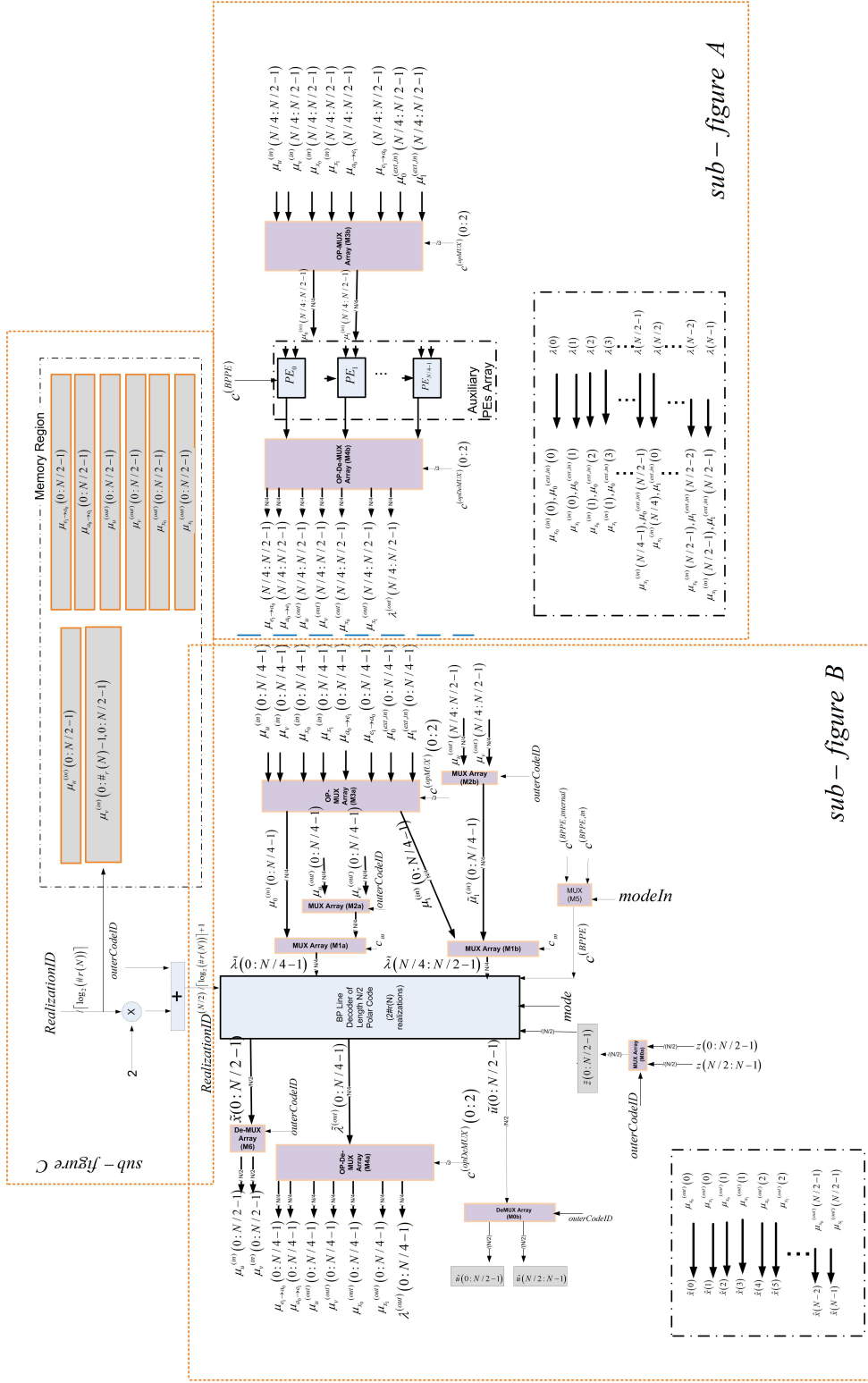}\\
  \caption{Block diagram for the BP line decoder. Details of figure appear in Figures \ref{fig: BPLineDecodeUV_A}, \ref{fig: BPLineDecodeUV_B} and \ref{fig: BPLineDecodeUV_C} corresponding to sub-figures A, B and C respectively.}\label{fig: BPLineDecodeUV}
\end{figure}
 \begin{figure}
\center
  \includegraphics[angle =0,scale = 0.09]{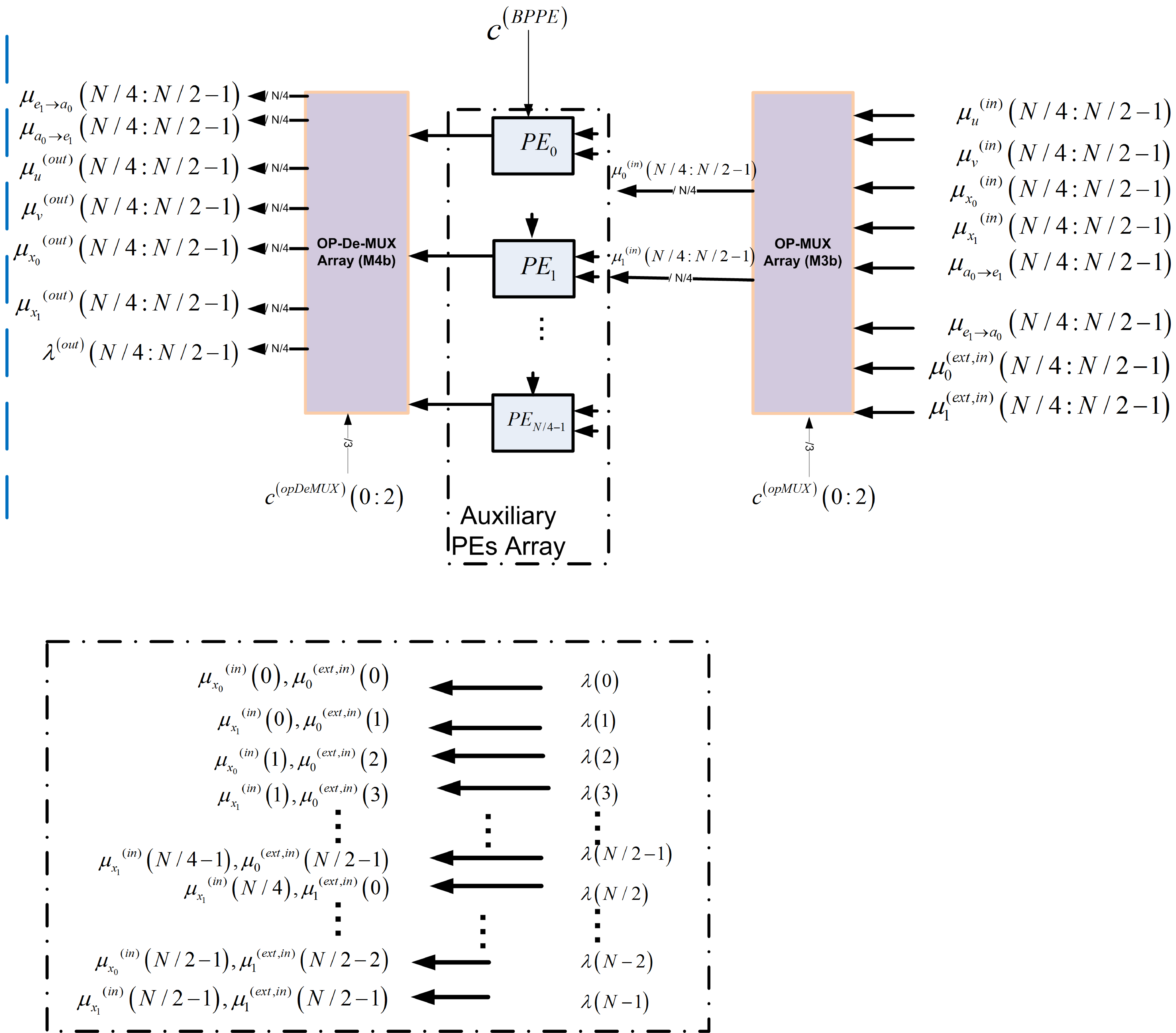}\\
  \caption{Block diagram for the BP line decoder (Figure \ref{fig: BPLineDecodeUV}) - zoom-in: Sub-figure A }\label{fig: BPLineDecodeUV_A}
\end{figure}
 \begin{figure}
\center
  \includegraphics[angle =0,scale = 0.09]{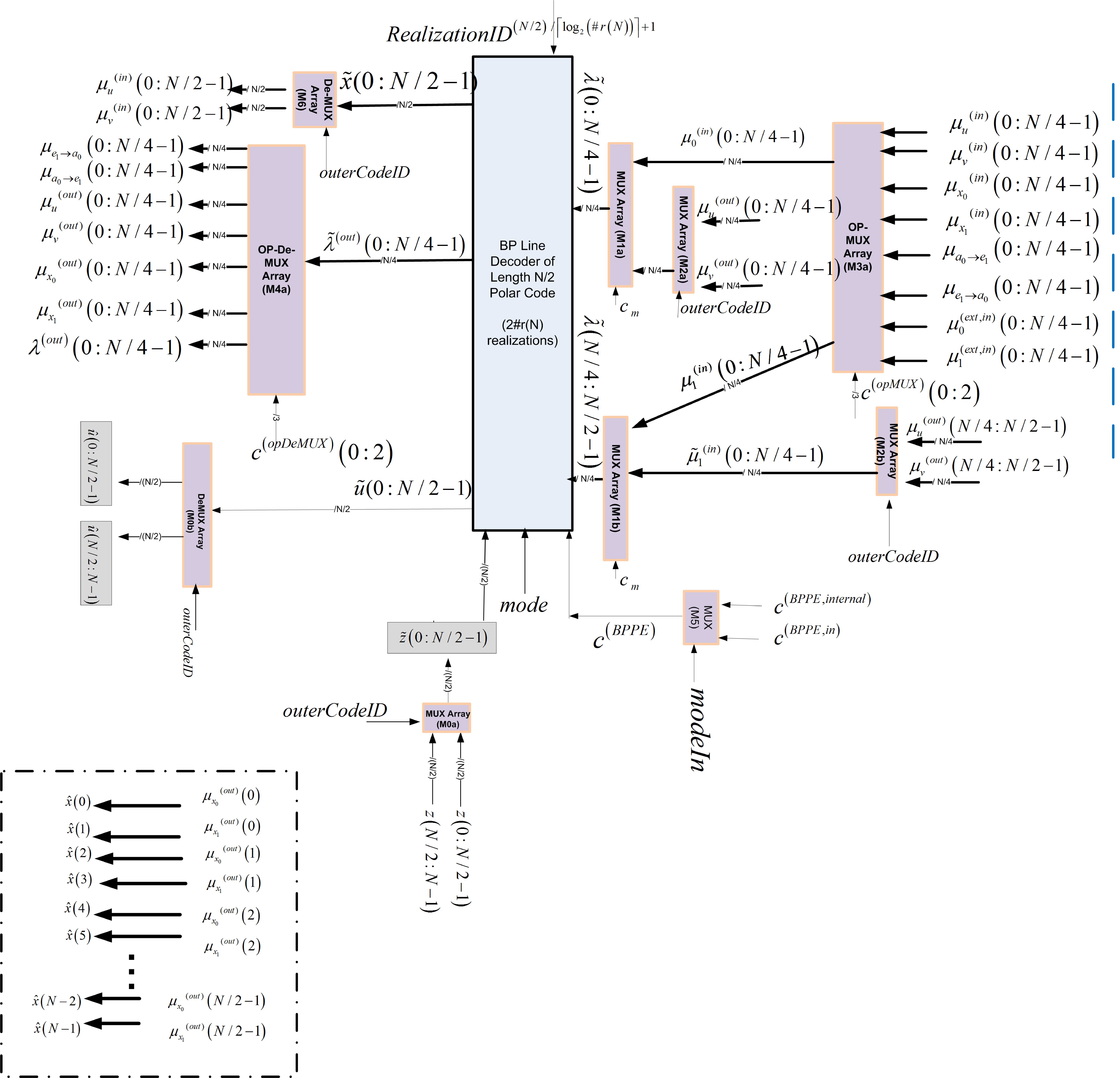}\\
  \caption{Block diagram for the BP line decoder (Figure \ref{fig: BPLineDecodeUV}) - zoom-in: Sub-figure B }\label{fig: BPLineDecodeUV_B}
\end{figure}
 \begin{figure}
\center
  \includegraphics[angle =0,scale = 0.09]{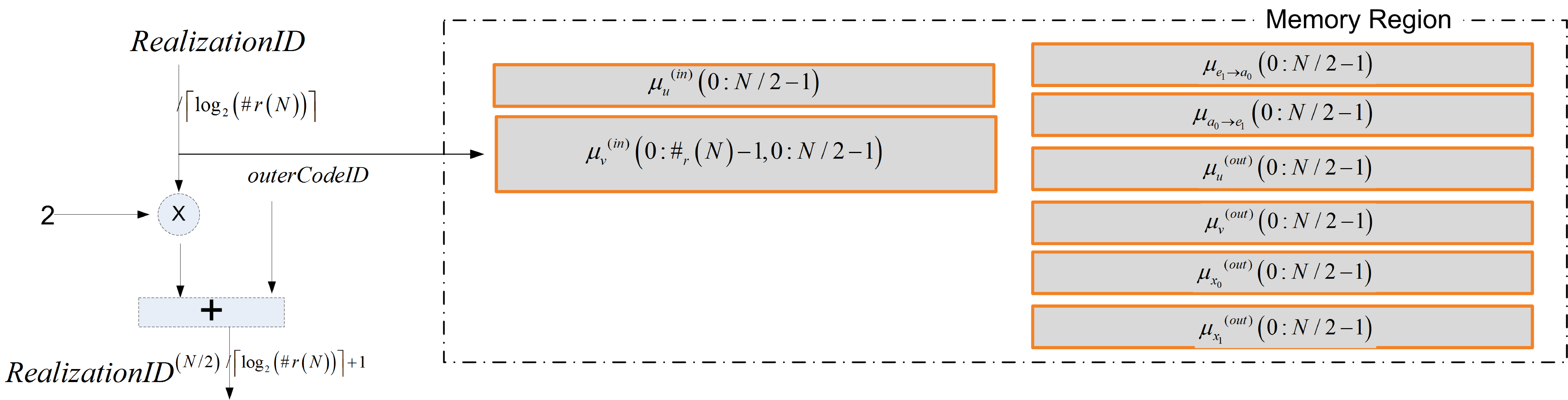}\\
  \caption{Block diagram for the BP line decoder (Figure \ref{fig: BPLineDecodeUV}) - zoom-in: Sub-figure C }\label{fig: BPLineDecodeUV_C}
\end{figure}
As we already noticed in Subsection \ref{sec:BP}, BP is an iterative algorithm, in which messages are sent on the normal factor graph representing the code. In this subsection, we consider an implementation of the BP decoder that employs the  GCC  serial schedule. Figure \ref{fig: BPPEPRocessor}, depicts the proposed design  processing element (PE). This unit has two inputs for message LLRs ($\mu_0^{(in)}$ and $\mu_1^{(in)}$), and depending on the control signal $c^{(BPPE)}$ it performs either the $f_{(+)}(\cdot,\cdot)$ function or the $f_{(=)}(\cdot,\cdot)$, i.e.
 \begin{equation}\label{eq:PEBlockDef}
 \mu^{(out)} = \left\{
                 \begin{array}{ll}
                   f_{(=)}\left(\mu_0^{(in)},\mu_1^{(in)}\right), & \hbox{$c^{(BPPE)}=0$;} \\
                   f_{(+)}\left(\mu_0^{(in)},\mu_1^{(in)}\right), & \hbox{$c^{(BPPE)}=1$.}
                 \end{array}
               \right.
 \end{equation}
 Since the PE has to support the implementation of  equations (\ref{eq:BPUV1})-(\ref{eq:BPUV6}), we introduce two routing layers for the inputs (OP-MUX) and the outputs (OP-De-MUX) that ensure that the proper inputs are given to the processor and that its output is dispatched to the appropriate destination. These routing units are controlled by two control signals $c^{(opMux)}$ and $c^{(opDeMux)}$ which have seven possible values, and is thereby represented by three bits. Table \ref{tbl:BPPeRoutingTbl} specifies the valid assignments   of $c^{(BPPE)}$, $c^{(opMux)}$ and $c^{(opDeMux)}$ for implementing different operations. The last option ($c^{opMux}=c^{opDeMux}=6$) is used in the decoder's P-Mode, that is defined in the sequel.

\begin{table}\center
\begin{tabular}{||c|c|c|c|c|c||}
  \hline
  \hline
  $c^{(opMux)},c^{(opDeMux)}$ & $c^{(BPPE)}$ & $\mu^{(in)}_0$ & $\mu^{(in)}_1$ & $\mu^{(out)}$ & Equation\\
\hline
\hline
  $0$ & $0$ & $\mu^{(in)}_{x_1}$  &  $\mu^{(in)}_v$ & $\mu_{e_1\rightarrow a_0}$ & (\ref{eq:BPUV1})\\
  $1$ & $1$& $\mu^{(in)}_{x_0}$ &  $\mu^{(in)}_u$& $\mu_{a_0 \rightarrow e_1}$ &(\ref{eq:BPUV2})\\
  $2$ & $1$ & $\mu^{(in)}_{x_0}$ & $\mu_{e_1\rightarrow a_0}$ & $\mu_{u}^{(out)}$  &(\ref{eq:BPUV3})\\
  $3$ & $0$ & $\mu^{(in)}_{x_1}$ &  $\mu_{a_0\rightarrow e_1}$ & $\mu_v^{(out)}$  &(\ref{eq:BPUV4})\\
  $4$ & $1$ & $\mu^{(in)}_u$ & $\mu_{e_1\rightarrow a_0}$ & $\mu_{x_0}^{(out)}$  &(\ref{eq:BPUV5})\\
  $5$ & $0$ & $\mu^{(in)}_v$ & $\mu_{a_0\rightarrow e_1}$ & $\mu_{x_1}^{(out)}$ &(\ref{eq:BPUV6})\\
  $6$ & $0$ or $1$ & $\mu^{(ext,in)}_0$ & $\mu^{(ext,in)}_1$ & $\mu^{(ext,out)}$  &  (\ref{eq:PEBlockDef})\\
  \hline
\end{tabular}
\caption{ Routing tables for OP-MUX and OP-DEMUX in Figure \ref{fig:BPBlockDefinitions}  }
\label{tbl:BPPeRoutingTbl}
\end{table}

The proposed decoder structure is inspired by  the recursive structure of the SC line decoder.  Figure \ref{fig:arikanPPlnBlk} depicts the BP line decoder block. Similarly to the SC line decoder we specify two  operation modes:
\begin{itemize}
\item S-Mode ($modeIn = 0$): the decoder completes a single iteration of the BP decoder, given the inputs $\lambda(0:N-1), z(0:N-1)$ and outputs $\hat{u}(0:N-1)$ and $\hat{x}(0:N-1)$ (defined in the BP signature, (\ref{eq:BPSignature})).
\item P-Mode ($modeIn = 1$): the decoder serves as an array of $N/2$ processors and performs simultaneously a parallel computation  on the input array $\lambda(0:N-1)$ such that $\forall i\in [N/2]_{-}$ the output $\lambda^{(out)}(i)$ is the outcome of applying the BP PE on inputs $\lambda( i)$ and $\lambda( i+N/2)$ with $C^{(BPPE, in)}$ as the control signal.
\end{itemize}

Figure  \ref{fig: BPLineDecodeUV} contains a block diagram for this design. Due to the vast number of details in this figure, we chose to enlarge three parts of this figures, named sub-figures A, B and C, in Figures \ref{fig: BPLineDecodeUV_A}, \ref{fig: BPLineDecodeUV_B} and \ref{fig: BPLineDecodeUV_C}, respectively.
The memory plays a fundamental role in the design, as it enables storing messages within the iteration boundary and beyond it. The basic requirement is that each "butterfly" realization of the $(u+v,v)$ factor graph, should have memory resources to store its messages. To allow messages to be kept within the iteration boundary, it is only required to have one registers array for each length of outer-code and for each message type. However, the need for keeping a message beyond the iteration boundary requires a dedicated memory array for each  outer-code instance. Note that messages  which their values are calculated before being used for the first time  in each iteration are not required to be kept beyond the iteration boundary.     In the case of the $(u+v,v)$ code and the GCC schedule, only messages of type $\mu_{v}^{(in)}$ need to be kept beyond the iteration boundary. We suggest to satisfy this requirement  in the following way. In the length $N$ decoder, we associate a registers matrix  $\mu_{v}^{(in)}(0:\#_r(N)-1,0:N/2-1)$. Here, $\#_r(N)$ is the \textit{number of realizations} of factor graphs corresponding to outer-codes of size $N$ that exist  in our code.

For the $N$ bits length code, there is only one factor graph of this size (i.e. the entire graph), and therefore for this decoder $\#_r(N)=1$.
Consider  now the $N/2$ bits length decoder that is embedded within the $N$ length decoder. We see in Figure \ref{fig: BPLineDecodeUV}, that this decoder has its number of realizations as $2\cdot \#_r(N)$, i.e. for the $N$ bits  length decoder we have $\#_r(N/2)=2$. This is because we have two outer-codes of length $N/2$ bits in the $N$ length code. Therefore, the memory matrix associated with it has two rows and $N/4$ columns. The first row is dedicated to the first realization of the outer-code and the second row is dedicated to the second realization.  Within this $N/2$ bits length decoder, there is an embedded $N/4$ length decoder with $2\cdot\#_r(N/2)$ realizations, so in this case $\#_r(N/4)= 4$. As a result, it has a registers matrix with $4$ rows and $N/8$ columns (each row is dedicated to one of the $4$ outer-codes of length $N/4$ in this GCC scheme). This development continues, until we reach the embedded  decoder of length $2$, which, by induction, has $\#_r(2)=N/2$ realizations for the $N$ length decoder, so it requires a registers matrix with $N/2$ rows and one column.

For a correct operation of the decoder, it is required to inform the embedded decoders to which realization of the outer-code's factor graph they are currently referring. This is the role of the \textit{realizationID} input signal in Figure \ref{fig:arikanPPlnBlk}, that takes decimal values in $\left[\#_r\left(N\right)\right]_{-}$, and therefore requires $\lceil\log_2
\left(\#_r\left(N\right) \right)\rceil$ bits for their representation. Moving to the implementation in Figure \ref{fig: BPLineDecodeUV} we can observe that indeed \textit{RealizationID} is used to select the row of $\mu_v^{(in)}$ corresponding to the outer-code realization that is currently processed.  Furthermore,  an internal signal $RealizationID^{(N/2)}$ is defined as the \textit{RealizationID} input of the embedded $N/2$ length decoder, such that
\begin{equation}\label{eq:BPRealizationID}
RealizationID^{(N/2)} = 2\cdot RealizationID + outerCodeID,
\end{equation}
where $outerCodeID \in \{0,1\}$ indicates the ordinal of the  $N/2$ bits length outer-code (of the current decoded length $N$ code) that is currently processed.

We also need to have  registers arrays for the messages of type $\mu_{e_1\rightarrow a_0},\mu_{a_0 \rightarrow e_1},\mu_{u}^{(in)},\mu_{u}^{(out)}$ and $\mu_{v}^{(out)}$, each one of them of length $N/2$. We denote them by $\mu_{e_1\rightarrow a_0}(0:N/2-1), \mu_{a_0\rightarrow e_1}(0:N/2-1), \mu_{u}^{(in)}(0:N/2-1), \mu_{u}^{(out)}(0:N/2-1)$ and $\mu_{v}^{(out)}(0:N/2-1)$, respectively. Note, that as opposed to the memory structure for the $\mu_{v}^{(in)}$ messages, these arrays do not need to be available beyond the iteration boundary, therefore it  is sufficient to have them as arrays and not matrices. Furthermore, the arrays for messages $\mu_{e_1\rightarrow a_0}$, $\mu_{u}^{(out)}$ and $\mu_{v}^{(out)}$, can be replaced by a single temporary array of length $N/2$. However, in the description of the hardware structure, we chose not to do this, in order to keep the discussion more comprehensible.

The routing units OP-MUX and OP-De-MUX that appeared in Figure \ref{fig: BPPEPRocessor} were grouped together in Figure \ref{fig: BPLineDecodeUV} into routing arrays (M3a), (M3b), (M4a) and (M4b).
The inputs and  outputs to these routing arrays are arrays of inputs and outputs corresponding to the types of inputs and outputs that appear in Figure \ref{fig: BPPEPRocessor}. The convention is that in these routing arrays, the $i^{\text{th}}$ output corresponds to the $i^{\text{th}}$ input from each signals array (the signals array is selected by the control signal of the routing array). Moreover, the $i^{\text{th}}$ output of the OP-MUX array corresponds to the consecutive $i^{\text{th}}$ processor from the array of processors it serves. Similarly, the $i^{\text{th}}$ input of the OP-De-MUX array corresponds to the $i^{\text{th}}$ consecutive processor from the array  of processors it serves.

MUX arrays (M1a), (M1b), (M2a) and (M2b) are used to select the LLR inputs to the embedded decoder, $\tilde{\lambda}\left(0:N/2-1\right)$. The select signal $c_m$ determines if the inputs to the embedded decoder comes from the outputs of the OP-MUX arrays (M3a) and (M3b) if $c_m=0$, or from the MUX-Arrays (M2a) and (M2b) if $c_m=1$.  We shall see that $c_m = 0$ is used when the embedded decoder is employed in S-Mode, while $c_m=1$ is used when it is employed in P-Mode. The multiplexer (M5) selects the appropriate source for the $c^{(BPPE)}$ control signal, such that in S-Mode ($modeIn = 0$), $c^{(BPPE)}$ takes the internal $c^{(BPPE,internal)}$ signal, and in  P-Mode  it takes the input signal $c^{(BPPE,in)}$. Finally, note that the $\lambda(0:N-1)$ inputs signals array is wired both to $\mu_{x_0}^{(in)}(0:N/2-1), \mu_{x_1}^{(in)}(0:N/2-1)$ signals arrays (used in S-Mode) and to $\mu_{0}^{(ext,in)}(0:N/2-1)$ and $\mu_{1}^{(ext,in)}(0:N/2-1)$ (used in P-Mode). The $\mu_{x_0}^{(out)}(0:N/2-1)$ and $\mu_{x_1}^{(out)}(0:N/2-1)$ signals arrays are wired to the $\hat{x}(0:N-1)$ output signals array.

\begin{algorithm}
\caption{ S-Mode (Steps 0 and 1)  of BP on Length $N$ $(u+v,v)$ Polar Code ($modeIn=0$) }          
\label{algo:BPSMode0}                           
\begin{description}
   \item[//STEP 0:] \hfill \\
    \ALGOSTEP  Set $c_m =  c^{(BPPE,internal)}= c^{(opMux)} = c^{(opDeMux)}= 0,  mode = 1$.

    Operate the embedded decoder in P-Mode, such that at the output of the decoder we have
     $$  {\mu_{e_1\rightarrow a_0}}(j) = f_{(=)}\left(\mu_{x_1}^{(in)}\left(j\right) ,\mu_v^{(in)}\left(j\right) \right)\,\,\,\,\,\,\, \forall j\in[N/4]_{-}.$$

    Use the auxiliary PEs array and compute
    $$  {\mu_{e_1\rightarrow a_0}}(j) = f_{(=)}\left(\mu_{x_1}^{(in)} \left(j\right) ,\mu_v^{(in)} \left(j\right) \right)\,\,\,\,\,\,\, \forall N/4 \leq j \leq N/2-1.$$
    Store these messages in their designated memory array.

    \ALGOSTEP   Set $c_m = outerCodeID =0,  c^{(BPPE,internal)}= mode = 1,  c^{(opMux)} = c^{(opDeMux)}= 2 $.

Simultaneously operate the embedded decoder (P-Mode) and the auxiliary array and store their outputs in the memory area such that
     $$  {\mu_{u}^{(out)}}(j) = f_{(+)}\left(\mu_{x_0}^{(in)}(j),\mu_{e_1\rightarrow a_0}^{(in)}\left(j\right) \right)\,\,\,\,\,\,\, \forall j\in[N/2]_{-}.$$

    Sample the first half of the frozen bits indicator $z$ by the $\tilde{z}$ register, i.e. $\tilde{z}(0:N/2-1) = z(0:N/2-1)$.
   \item[//STEP 1:] \hfill

\ALGOSTEP   Set $ mode =  outerCodeID = 0, c_m = 1  $.

Execute the embedded decoder in S-Mode on $\mu_{u}^{(out)}(0:N/2-1)$ as the LLR input and $\tilde{z}(0:N/2-1)$ as the frozen symbols indicator vector. The realization ID of the embedded decoder (denoted by $realizationID^{(N/2)}$) is calculated according to (\ref{eq:BPRealizationID}).

 \ALGOSTEP  Sample the $\tilde{ u}(0:{N/2-1})$ signals array  by the first half of $\hat{u}$, i.e.  $\hat{u}(0:{N/2-1})=\tilde{ u}(0:{N/2-1})$. Sample the $\tilde{ x}(0:{N/2-1})$ signals array  by the registers array $\mu^{(in)}_u\left(0:N/2-1\right)$, i.e.  $\mu^{(in)}_u\left(0:N/2-1\right)=\tilde{ x}(0:{N/2-1})$.

\end{description}
\end{algorithm}

\begin{algorithm}
\caption{ S-Mode (Steps 2 and 3) of BP on Length $N$ $(u+v,v)$ Polar Code ($modeIn=0$)}          
\label{algo:BPSMode2}
\begin{description}
   \item[//STEP 2:] \hfill \\
    \ALGOSTEP  Set $c_m = 0, c^{(BPPE,internal)}= c^{(opMux)} = c^{(opDeMux)}= mode = 1$.

Simultaneously operate the embedded decoder (P-Mode) and the auxiliary array and store their outputs in the memory area such that
     $$  {\mu_{a_0\rightarrow e_1}}(j) = f_{(+)}\left(\mu_{x_0}^{(in)} \left(j\right),\mu_{u}^{(in)} \left(j\right) \right)\,\,\,\,\,\,\, \forall j\in[N/2]_{-}.$$

    \ALGOSTEP  Set $c_m = c^{(BPPE,internal)}= 0,  mode = outerCodeID = 1,  c^{(opMux)} = c^{(opDeMux)}= 3$.

Simultaneously operate the embedded decoder (P-Mode) and the auxiliary array and store their outputs in the memory area such that
     $$  {\mu_{v}^{(out)}}(j) = f_{(=)}\left(\mu_{x_1}^{(in)}\left(j\right),\mu_{a_0\rightarrow e_1}(j) \right)\,\,\,\,\,\,\, \forall j\in[N/2]_{-}.$$

 Sample the second half of the frozen bits indicator $z$ by the $\tilde{z}$ register, i.e. ${\tilde{z}(0:N/2-1) = z(N/2:N-1)}$.

   \item[//STEP 3:] \hfill \\
   \ALGOSTEP   Set $c_m = mode =   0,  outerCodeID = 1  $.

Execute the embedded decoder in S-Mode on $\mu_{v}^{(out)}(0:N/2-1)$ as the LLR input and $\tilde{z}(0:N/2-1)$ as the frozen symbols indicator vector. The realization ID of the embedded decoder (denoted by $realizationID^{(N/2)}$) is calculated according to (\ref{eq:BPRealizationID}).

 \ALGOSTEP  Sample the $\tilde{ u}(0:{N/2-1})$ signals array  by the second half of $\hat{u}$, i.e.  $\hat{u}(N/2:{N-1})=\tilde{ u}(0:{N/2-1})$. Sample the $\tilde{ x}(0:{N/2-1})$ signals array  by registers array $\mu^{(in)}_v\left(0:N/2-1\right)$, i.e.  $\mu^{(in)}_v\left(0:N/2-1\right)=\tilde{ x}(0:{N/2-1})$.

       \ALGOSTEP  Set $c_m =  c^{(BPPE,internal)}= c^{(opMux)} = c^{(opDeMux)}= 0,  mode = 1$.

Simultaneously operate the embedded decoder (P-Mode) and the auxiliary array and store their outputs in the memory area such that
     $$  {\mu_{e_1\rightarrow a_0}}(j) = f_{(=)}\left(\mu_{x_1}^{(in)}(j),\mu_v^{(in)}\left(j\right) \right)\,\,\,\,\,\,\, \forall j\in[N/2]_{-}.$$

    \ALGOSTEP  Set $c_m =  0, c^{(BPPE,internal)}=  mode = 1, c^{(opMux)} = c^{(opDeMux)} = 4$.

Simultaneously operate the embedded decoder (P-Mode) and the auxiliary array and store their outputs in the memory area such that
     $$  {\mu_{x_0}^{(out)}}(j) = f_{(+)}\left(\mu_u^{(in)}(j),\mu_{e_1\rightarrow a_0}\left(j\right) \right)\,\,\,\,\,\,\, \forall j\in[N/2]_{-}.$$

        \ALGOSTEP  Set $c_m = c^{(BPPE,internal)}= 0,  mode = 1, c^{(opMux)} = c^{(opDeMux)} = 5$.

Simultaneously operate the embedded decoder (P-Mode) and the auxiliary array and store their outputs in the memory area such that
     $$  {\mu_{x_1}^{(out)}}(j) = f_{(=)}\left(\mu_v^{(in)}(j),\mu_{a_0\rightarrow e_1}\left(j\right) \right)\,\,\,\,\,\,\, \forall j\in[N/2]_{-}.$$

    //Note that $\mu_{x_0}$ and $\mu_{x_1}$ signals array are wired to the $\hat{x}(0:N-1)$ output signals array, as specified in Figure \ref{fig: BPPEPRocessor}.
 \end{description}
 \end{algorithm}

The S-Mode operation of the decoder is described in Algorithms \ref{algo:BPSMode0} and  \ref{algo:BPSMode2}.
The P-Mode procedure is described in Algorithm \ref{algo:PModeSCUVV}.

\begin{algorithm}
\caption{P-Mode of the BP Line Decoder of Length $N$ $(u+v,v)$ Polar Code ($modeIn=1$) }          
\label{algo:PModeSCUVV}
\begin{description}
    \item

    \ALGOSTEP  Set $c_m =   1,  c^{(opMux)} = c^{(opDeMux)}= 6$.

Simultaneously operate the embedded decoder (P-Mode) and the auxiliary array such that we have at the output of the decoder
$$
 \lambda^{(out)}\left(j\right) = \left\{
                 \begin{array}{ll}
                   f_{(=)}\left(\lambda(j),\lambda(j+N/2)\right), & \hbox{$c^{(BPPE,in)}=0$;} \\
                   f_{(+)}\left(\lambda(j),\lambda(j+N/2)\right), & \hbox{$c^{(BPPE,in)}=1$.}
                 \end{array}
               \right.\,\,\,\,\,\,\,\, \forall j\in \left[N/2\right]_{-}
 $$
\end{description}
\end{algorithm}

Let us, now, consider the time complexity (in terms of the number of clock cycles for running an iteration) of this design. As before, let $T(n)$ be the time complexity of  the decoder of length $N=2^n$ bits polar code. We assume that each operation of the  BP   PE requires one clock cycle. As a consequence,  we have
\begin{equation}\label{eq:recBPUV}
T(n) = 2\cdot T(n-1)+7,\,\,\, \text{for } n>1
\end{equation}
 and $T(1)=4$, resulting in $T(n)=5.5\cdot N-7=\Theta(N)$. The memory consumption, however is $\Theta(N\cdot \log N)$, because of the memory matrices for the $\mu_{v}^{(in)}$ type of messages. The number of processing elements in this design is $N/2$.
Note that our proposed PE  can be further improved to support some PE operations occuring in parallel. For example, if the  BP PE is designed such that the operation of $f_{(+)}(\cdot)$ and the operation of $f_{(=)}(\cdot)$ can be performed simultaneously in one clock cycle, we  can execute the last two  operations in step $3$ in one clock cycle. Consequently, this will reduce the free addend in (\ref{eq:recBPUV}) to $6$. Further reduction  is possible if the processor can execute $f_{(+)}(\cdot)$ and direct its output to $f_{(=)}(\cdot)$ in one clock cycle. This improvement will result in joining the two operations in step $2$, into one operation. Enabling the computation of $f_{(=)}(\cdot)$ and directing its output to $f_{(+)}(\cdot)$ in the same clock cycle,  results in consolidation of the two operations of step $0$ into one operation (actually, the latter change may also allow to consolidate the second and third computation in step $3$, leaving our first suggested change obsolete). These  changes result in $4$, as the free addend in (\ref{eq:recBPUV}) and $T(2)=2$, so $T(n)=3\cdot N-4$.

The remarks, raised on the SC line decoder recursive design at the end of Subsection \ref{sec:UVLineDecoder} also apply here. Specifically, this design also suffers from long paths hazards especially in the routing layers of P-Mode. Consequently,    more efficient designs may be applied by unfolding the recursive blocks. Furthermore, the issue of idle clock cycles for the BP PE   is also a problem of this design and the solution of Subsections \ref{sec:ParDecLine}  and \ref{sec:LimitedParDecLine} may  be adapted to this decoder too.

Note however that while in the SC decoder, the existence of inactive PEs is due to the properties of the SC algorithm, which dictates the scheduling of the message computation,  in the BP case, this is due to the scheduling we choose and not a mandatory property of the algorithm. Other types of scheduling do exist, and currently there is no evidence which scheduling is better (for example, in terms of the achieved error rate or in terms of the average number of iterations required for convergence). Hussami \textit{et al.} \cite{Hussami2009} proposed to use the Z-shape schedule, which description suggests a constant level of parallelism of $N$ PEs (of the type we considered here) operating all the time. This seems to give the Z-shape schedule an advantage over the GCC schedule if the number of processors is not limited (unless the technique of Subsection \ref{sec:ParDecLine} is applied). It is an interesting question to find out which schedule is better, when the number of processors is limited. This is a matter for further research.

\subsection{ Decoding Architectures for General Polar Codes}\label{sec:HardArchiForOthKer}

Thus far, we described decoding algorithms for the $(u+v,v)$ polar code. This notion has enabled us to restate the SC  implementation  for Arikan's construction, that were proposed by Leroux \textit{et al.} \cite{Leroux2012}. In addition, we suggested a  BP   decoding implementation employing the GCC schedule. In this subsection, we  generalize these constructions for other types of polar codes. Since we already covered  implementations for Arikan's code in some details, in this section we provide a more concise description of the implementations,  mainly emphasizing the principle differences from the designs in Subsection \ref{sec:HrdwreArikConstr}.

\subsubsection{Recursive Description of the SC Line Decoder for General Linear Kernels}\label{sec:SCLineForGeneralKernel}
Let $\mathcal{C}$ be a homogenous linear polar code over field $F$, constructed by a kernel of $\ell$ dimensions. This kernel has an $\ell\times\ell$ generating matrix, $\bf G$ associated with it. Let $f$ be the number of bits required to represent all the field elements, i.e. $f = \lceil \log_2|F| \rceil$.

\begin{figure}
\begin{subfigure}{.5\textwidth}
  \centering
  \includegraphics[scale = 0.13]{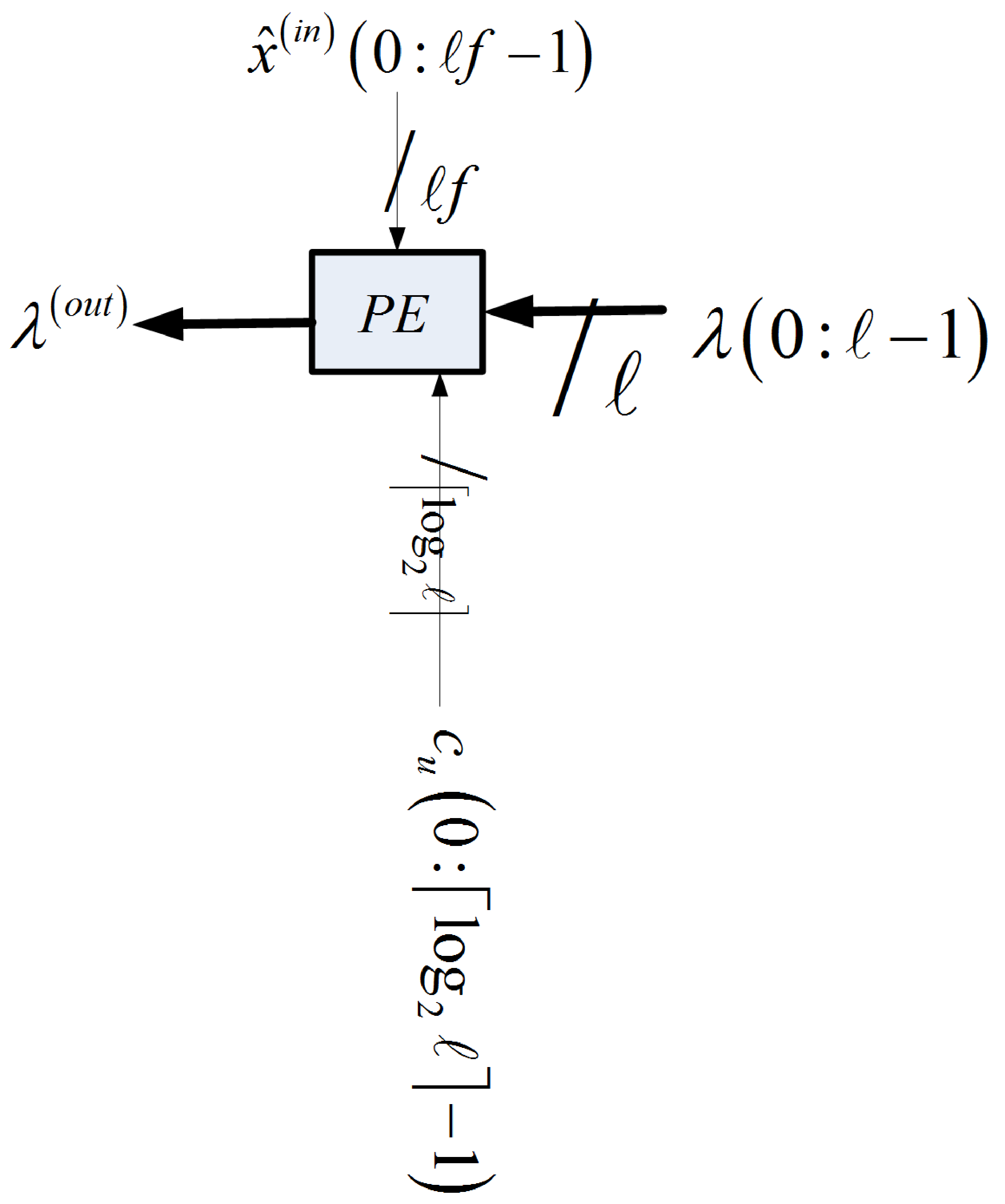}
  \caption{Processing element}
  \label{fig:GenLinKernelPE}
\end{subfigure}
\begin{subfigure}{.5\textwidth}
  \includegraphics[scale = 0.10]{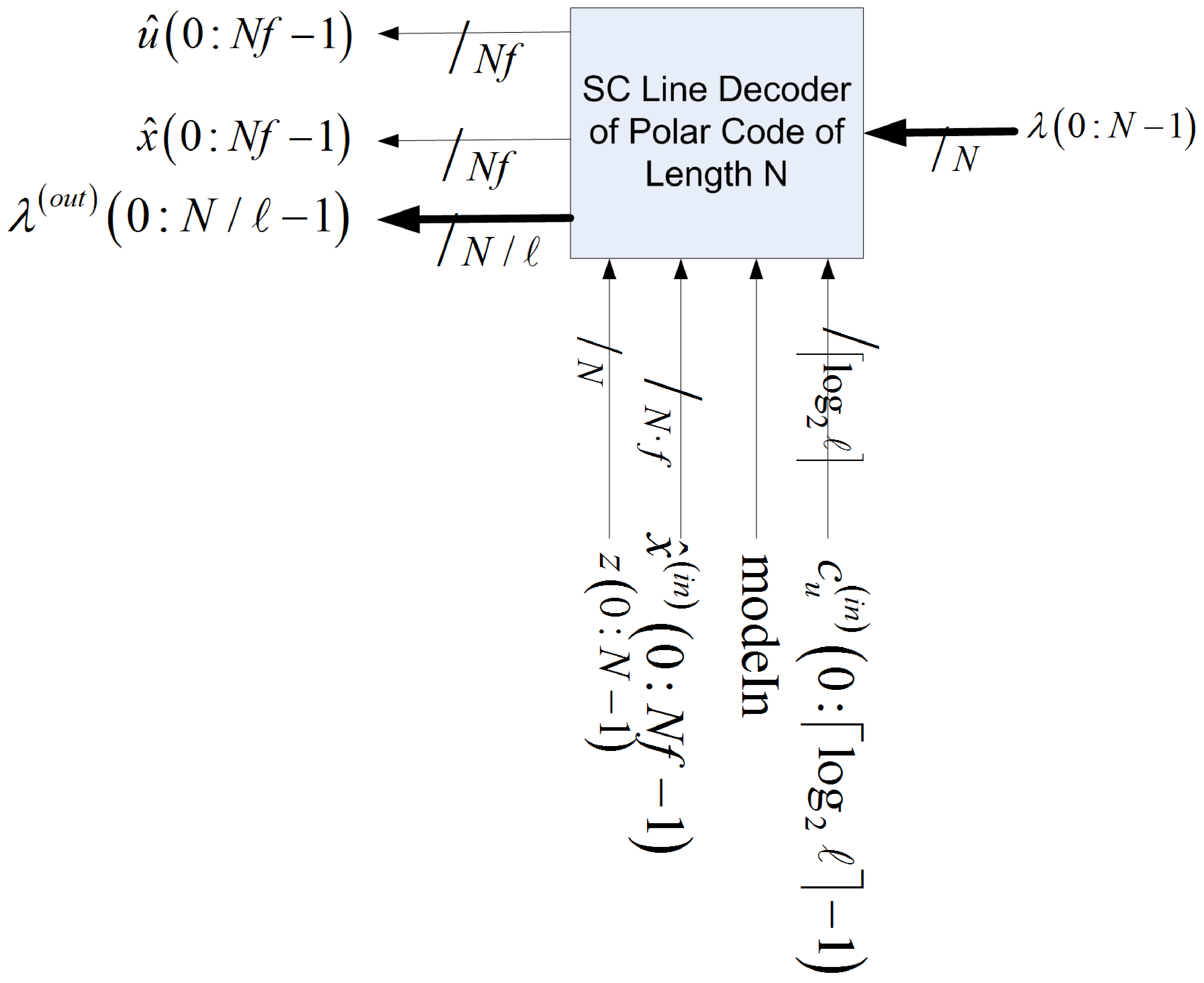}
  \caption{Decoder block definition }
  \label{fig:LineDecoderBlk}
\end{subfigure}%
\caption{Block  definitions of SC line decoder for length $N$ polar code based on a linear $\ell$ dimensions kernel  with alphabet $F$    }
\label{fig:GenLinKernelDefinitions}
\end{figure}

Figure \ref{fig:GenLinKernelPE} depicts the basic processing element (PE) of the SC line decoder.
The LLR   input  $\lambda(0:\ell-1)$ and output $\lambda^{(out)}$ are specified such that each entry $\lambda(j)$ is actually a vector of $|F|-1$ elements. These elements are the logarithms of the likelihood ratio of the zero symbol and one of the $F\backslash \{0\}$ symbols (denoted by $\lambda^{(t)}$ in (\ref{eq:llrDefinitionF}), where $t\in F\backslash\{0\}$). In our block diagrams, thick lines are used to carry these LLR signals. In other words, assuming that each aforementioned $\lambda^{(t)}$ is represented by $\beta$ bits, each thick line is composed of $\beta \cdot \left(|F|-1\right)$ bit lines. The input signal $c_u$ has $\ell$ possible values, each one corresponds to a different LLR processing step as specified in (\ref{eq:genLinKernelLLR}). The input signals array   ${\hat x}^{(in)}(0:\ell f -1)$ represents a coset vector for the currently processed kernel. This is an $\ell$ length word over $F$ and as such it is represented by $\ell\cdot f$ bits. Let $\hat{\bf x} \in F^{\ell}$ be the vector represented by this register array, furthermore let $\left[\lambda^{(t)}_i\right]_{t\in F\backslash \{0\}}$ be the LLR vector corresponding to the signal input $\lambda(i)$, where $i\in\left[\ell\right]_{-}$.  Similarly, let   $\left[\hat{\lambda}^{(t)}\right]_{t\in F\backslash \{0\}}$ be the LLR vector corresponding to the output signal $\lambda^{(out)}$. If the $c_u$ input represents the decimal value $i$ (denote it by $c_u \equiv i$), the circuit's output is defined by Equation (\ref{eq:genLinKernelLLR}).

Figure \ref{fig:LineDecoderBlk} specifies the block definition for the this general kernel line decoder.  Most of the labels of this block's input and output signals   are the same as in Figure \ref{fig:arikanLineBlock}
 and they    keep their functionality as well. There are some modifications, however, that are required in order to support the change in the kernel and the alphabet.  The signals arrays $\hat{x}^{(in)}$, $\hat{u}$ and $\hat{x}$ represent vectors of length $N$, over the $F$ alphabet. As a consequence, each entry in them is represented by $f$ bits. The input signal $c_u^{(in)}\left(0:\lceil\log_{2}\ell\rceil-1\right)$, used in P-Mode ($modeIn = 1$), has $\ell$ possible values, each one corresponds to a different LLR processing step as specified in (\ref{eq:genSCRule2}) in Algorithm \ref{algo:GenDetDescSC}. Since the maximum number of PEs employed simultaneously is $N/{\ell}$, the line decoder is designed to have $N/{\ell}$ length LLR output signals array. The functionality of the decoder in P-Mode is that for all $j\in \left[N/{\ell} \right]_{-}$  we have $\lambda^{(out)}(j)$ be the output of a PE that is given as inputs the LLR array $\lambda\left(j\cdot\ell:(j+1)\cdot \ell -1\right)$, the coset vector $\hat{x}^{(in)}\left(j\cdot\ell:(j+1)\cdot \ell -1\right)$ and $c_u = c_u^{(in)}$. In S-Mode ($modeIn = 0$) the decoder outputs its estimations for the information word $\hat{u}\left(0:Nf-1\right)$ and its corresponding codeword $\hat{x}\left(0:Nf-1\right)$ given the LLR input signals array $\lambda\left(0:N-1\right)$ and the frozen indicator vector $z\left(0:N-1\right)$.

The generalization of the $(u+v,v)$ block diagram in Figure \ref{fig:lineArikanLimPar} and its corresponding algorithms can be easily completed using the above $PE$ description and Algorithms \ref{algo:GenDetDescSC} and \ref{algo:SCDecodingOutput}. We leave the details for the reader.

\subsubsection{Recursive Description of the BP Line Decoder for General  Kernels}\label{sec:BPLineForGeneralKernel}
Subsection \ref{sec:SCLineForGeneralKernel} considered  the adaptation of the $(u+v,v)$ line decoder for supporting general kernels. Designing a BP line decoder for general polar codes entails similar difficulties. In this subsection we only highlight the principal necessary  modifications to the  BP decoder in Subsection \ref{sec:UVLineDecoderBP} in order to  adjust it to the case of $\ell$ dimensions kernel over alphabet $F$.
\begin{itemize}
\item The LLR inputs, internal signals and memories should be extended to support LLRs over $F$. See Subsection \ref{sec:SCLineForGeneralKernel} for more details.
\item The routing layers OP-MUX and OP-De-MUX need to be extended in order to support all the different messages calculated by the PE.
\item The Memory Region in Figure \ref{fig: BPLineDecodeUV} needs to include registers array to support each of the algorithm's possible messages.  Messages that are required to be kept beyond the iteration boundary have to be stored in a matrix, such that each row corresponds to a different realization of the code. The number of LLRs in each row of these matrices is $N/{\ell}$, the outer-code length in $F$ symbols. On the other hand, messages that in each iteration, their values are calculated before being used for the first time (in the iteration) requires only registers arrays of length $N/{\ell}$. See Subsection \ref{sec:UVLineDecoderBP} for more details on the distinction between these two types of messages.
\item Algorithms \ref{algo:BPSMode0} and \ref{algo:BPSMode2} are replaced by $\ell$ pairs of steps each one is dedicated to a different outer-code $\mathcal{C}_i$. See Algorithms \ref{algo:BPGenLengthN} and \ref{algo:BPIterationStepsForCi} for further details.
\end{itemize}

\subsubsection{Decoders for Mixed-Kernels and General Concatenated Codes}\label{sec:MixedKernelsHW}
So far, we considered decoders for homogenous polar codes  over   alphabet $F$. These codes have the attractive property, that the outer-codes in their GCC structure are themselves (shorter) polar codes from the same family. Therefore, we were able to use a single embedded decoder of a code of length $N/{\ell}$ symbols within the decoder of the code of length $N$ symbols. This embedded decoder is used  $\ell$ times, each time on different inputs (i.e. indices of the frozen symbols  and the input messages). Unfortunately, this property no longer applies when mixed-kernels polar codes are used.

Let us consider the  $\ell=4$ dimensions mixed-kernels polar code described in Example \ref{ex:MxdKernelsExample}. In the decoder for length $N=4^n$ bits   code, we need to have an embedded decoder of the mixed-kernels code of length $N/4$ bits and an additional embedded decoder for the  $RS4$ polar code of  length $N/4$ quaternary symbols. Note, however, that even here, a reuse of circuits is still possible, as the decoder for the $RS4$ code of length $N/4$,  requires an embedded  decoder for the $RS4$ code of length $N/16$  within it. The latter decoder (and its embedded decoders) can be shared with the decoder for the mixed-kernels code of length $N/4$ (that requires an embedded $RS4$ decoder of the same length).

\section*{Summary and Conclusions}
We considered the recursive GCC structures of polar codes which led to recursive descriptions of their encoding and decoding algorithms. Specifically,  known algorithms (SC, SCL and BP) were formalized in a recursive fashion, and then were generalized for arbitrary kernels.  Moreover, recursive  architectures for these algorithms were considered. We restated known architectures, and generalized them for arbitrary kernels.

In our discussion, we preferred for brevity, to give somewhat abstract descriptions of the subjects, emphasizing the main properties while neglecting some of the technical details. However, a complete  design requires a full treatment of all of these specifics (see e.g. Leroux \textit{et al.} for the $(u+v,v)$ case \cite{Leroux2012}).

A subject that requires a more careful attention, is the study of BP decoder and specifically the proposed GCC schedules. A comparison between this schedule and other proposed schedules (e.g. the $Z$ shaped schedule) is an intriguing question.  Furthermore, a comparison of the BP decoder versus SCL decoder for general kernels taking into account error-correction performance and the decoder's complexity  is also an interesting topic.  These questions are subjects for further research.


\bibliographystyle{IEEEtran}
\bibliography{IEEEabrv,bibTexPolar}

\end{document}